\documentclass[final,twoside,a4paper]{IEEEtran}  
\RequirePackage{ifxetex}  
\usepackage{scalerel}
\RequirePackage[table,svgnames]{xcolor} 
\usepackage{tikz}
\usetikzlibrary{svg.path}

\definecolor{orcidlogocol}{HTML}{A6CE39}
\tikzset{
  orcidlogo/.pic={
    \fill[orcidlogocol] svg{M256,128c0,70.7-57.3,128-128,128C57.3,256,0,198.7,0,128C0,57.3,57.3,0,128,0C198.7,0,256,57.3,256,128z};
    \fill[white] svg{M86.3,186.2H70.9V79.1h15.4v48.4V186.2z}
                 svg{M108.9,79.1h41.6c39.6,0,57,28.3,57,53.6c0,27.5-21.5,53.6-56.8,53.6h-41.8V79.1z M124.3,172.4h24.5c34.9,0,42.9-26.5,42.9-39.7c0-21.5-13.7-39.7-43.7-39.7h-23.7V172.4z}
                 svg{M88.7,56.8c0,5.5-4.5,10.1-10.1,10.1c-5.6,0-10.1-4.6-10.1-10.1c0-5.6,4.5-10.1,10.1-10.1C84.2,46.7,88.7,51.3,88.7,56.8z};
  }
}   

\makeatletter
\newif\ifhbonecolumn
\@ifclasswith{IEEEtran}{onecolumn}{\hbonecolumntrue}{\hbonecolumnfalse}
\makeatother

\newcommand\orcidicon[1]{\,\,\href{https://orcid.org/#1}{\mbox{\scalerel*{
\begin{tikzpicture}[yscale=-1,transform shape]
\pic{orcidlogo};
\end{tikzpicture}
}{|}}}}

\usepackage[normalem]{ulem}  
\usepackage{multirow}  

\usepackage{graphicx}      
\DeclareRobustCommand*{\IEEEauthorrefmark}[1]{\raisebox{0pt}[0pt][0pt]
{\textsuperscript{\footnotesize\ensuremath{\ifcase#1\or 1\or 2\or 3\or%
    4\or 5\or 6\or 7\or 8  
\else\textsuperscript{\expandafter\romannumeral#1}\fi}}}}

\graphicspath{{Figs/}}

\usepackage{flushend} 

\xdefinecolor{TTYblue}{rgb}{0.0, 0.4980, 0.7725}   
\colorlet{spotColor}{gray!100!black}

\usepackage{amsmath} 
\usepackage{amssymb}
\usepackage{stix}
\usepackage[cal=boondoxo,bb=boondox]{mathalfa}

\usepackage[  
draft=false,
colorlinks=true,  
breaklinks=true, 
allcolors=DarkBlue,
urlbordercolor={1 0 0}, 
pdfborderstyle={/S/U/W 0.5},
]{hyperref}
\hypersetup{pdflinkmargin=-0.5pt} 

\hypersetup{%
  pdfsubject={This paper investigates the application of fast convolution (FC) filtering schemes for flexible and effective waveform generation and processing in 5th generation (5G) systems.},
  pdfkeywords={5G, physical layer, 5G New Radio, 5G-NR, multicarrier, waveforms, filtered-OFDM, fast-convolution},
  pdfauthor={Juha Yli-Kaakinen, AlaaEddin Loulou, Toni Levanen, Arto Palin, Markku Renfors, and Mikko Valkama}%
}
\hypersetup{pdflinkmargin=0.5pt}

\usepackage[version-1-compatibility]{siunitx}
\DeclareSIUnit\prb{\text{PRB}}
\DeclareSIUnit\prbs{\text{PRBs}}

\usepackage{tabularx,booktabs} 
\usepackage{cite}
 
\providecommand{\iu}{\ensuremath{{\mathop{\mspace{1mu}\mathrm{j}\mspace{0.5mu}}\nolimits}}}

\usepackage{mathtools}
\DeclarePairedDelimiter{\diagpars}{(}{)}
\DeclarePairedDelimiter{\conjpars}{(}{)}

\newcommand{\diag}{\operatorname{diag}\diagpars}
\newcommand{\conj}{\operatorname{conj}\conjpars}

\makeatletter
\newcommand*{\transpose}{%
  {\mathpalette\@transpose{}}%
}
\newcommand*{\@transpose}[2]{%
  \raisebox{\depth}{$\m@th#1\intercal$}%
}

\makeatother
\usepackage[nolist,nohyperlinks]{acronym} 

\makeatletter
\DeclareMathSizes{\@xpt}{\@xpt}{7}{5}
\makeatother
   
\newlength{\figwidth}
\newlength{\figwidthSC}
\ifhbonecolumn
\else
\fi

\begin{document}
\title{\vspace{0.5em}Frequency-Domain Signal Processing for Spectrally-Enhanced CP-OFDM Waveforms\\ in 5G New Radio}
\author{Juha Yli-Kaakinen\orcidicon{0000-0002-4665-9332}, %
AlaaEddin Loulou\orcidicon{0000-0001-9715-5647}, %
Toni Levanen\orcidicon{0000-0002-9248-0835}, \\ %
Kari Pajukoski, 
Arto Palin\orcidicon{0000-0001-8567-5549}, %
Markku Renfors\orcidicon{0000-0003-1548-6851}, and %
Mikko Valkama\orcidicon{0000-0003-0361-0800}
\thanks{This work was partially supported by Business Finland (formerly known as the Finnish Funding Agency for Innovation, Tekes) and Nokia Bell Labs, under the projects ``5G Radio Systems Research'', ``Wireless for Verticals (WIVE)'', and ``5G-FORCE'', in part by Nokia Networks, and in part by the Academy of Finland under the projects no. 284694, no. 284724 and no. 319994.}
\thanks{Juha Yli-Kaakinen, Markku Renfors, and Mikko Valkama are with the Department of Electrical Engineering, Tampere University, FI-33101 Tampere, Finland (e-mail: $\lbrace$juha.yli-kaakinen; markku.renfors; mikko.valkama$\rbrace$@tuni.fi)} 
\thanks{AlaaEddin Loulou, Toni Levanen, and Arto Palin are with Nokia Networks, Finland (e-mail: $\lbrace$alaaeddin.loulou; toni.a.levanen; arto.palin$\rbrace$@nokia.com)}
\thanks{Kari Pajukoski is with Nokia Bell Labs, Finland (e-mail: kari.pajukoski@nokia-bell-labs.com)}
\thanks{Early stage results of this paper have been published in \emph{Proc. Asilomar Conf. Signals, Syst., Comput. (ACSSC)}, Pacific Grove, California, USA \cite{C:Yli-Kaakinen:ASILOMAR2018}. 
}    
\thanks{This article contains multimedia material, available at \url{http://yli-kaakinen.fi/DiscontinuousSymbolSynchronizedFastConvolution/}}
\vspace{-2em}}

\sloppy

\maketitle
\begin{abstract} 
Orthogonal frequency-division multiplexing (OFDM) has been selected as the basis for the \ac{5g-nr} waveform developments. However, effective signal processing tools are needed for enhancing the OFDM spectrum in various advanced transmission scenarios. In earlier work, we have shown that \ac{fc} processing is a very flexible and efficient tool for filtered-OFDM signal generation and receiver-side subband filtering, e.g., for the mixed-numerology scenarios of the \ac{5g-nr}. \ac{fc} filtering approximates linear convolution through effective \ac{fft}-based circular convolutions using partly overlapping processing blocks. However, with the continuous overlap-and-save and overlap-and-add processing models with fixed block-size and fixed overlap, the \ac{fc}-processing blocks cannot be aligned with all \acs{ofdm} symbols of a transmission frame. Furthermore, \ac{5g-nr} numerology does not allow to use transform lengths shorter than 128 because this would lead to non-integer cyclic prefix (CP) lengths. In this article, we present new \ac{fc}-processing schemes which solve or avoid the mentioned limitations. These schemes are based on dynamically adjusting the overlap periods and extrapolating the CP samples, which make it possible to align the \ac{fc} blocks with each \acs{ofdm} symbol, even in case of variable CP lengths. This reduces complexity and latency, e.g., in mini-slot transmissions and, as an example, allows to use 16-point transforms in case of a 12-subcarrier-wide subband allocation, greatly reducing the implementation complexity. On the receiver side, the proposed scheme makes it possible to effectively combine cascaded inverse and forward \ac{fft} units in \ac{fc}-filtered OFDM processing. Transform decomposition is used to simplify these computations, leading to significantly reduced implementation complexity in various transmission scenarios. A very extensive set of numerical results is also provided, in terms of the radio-link performance and associated processing complexity.
\end{abstract}


\section{Introduction}  
\label{sec:introduction} 
\IEEEPARstart{O}{rthogonal} frequency-division multiplexing (OFDM)\acused{ofdm} is the dominating multicarrier modulation scheme and it is extensively deployed in modern radio access systems. 
\ac{ofdm} offers high flexibility and efficiency in allocating spectral resources to different users through the division of subcarriers, simple and robust way of channel equalization due to the inclusion of \acf{cp}, as well as simplicity of combining multi-antenna schemes with the core physical-layer processing \cite{B:Dahlman2018}.  The main drawback is the limited spectrum localization, especially in challenging new spectrum use scenarios like asynchronous multiple access, as well as mixed-numerology cases aiming to use adjustable symbol and \ac{cp} lengths, \acp{scs}, and frame structures depending on the service requirements \cite{J:20145GNOW,J:2014BanelliModFormatsAndWaveformsFor5G,J:Memisoglu2020,J:Chen2020,J:Mao2020}. 

\begin{figure}[t!]                    
  \centering    
  \includegraphics[width=\figwidthSC]{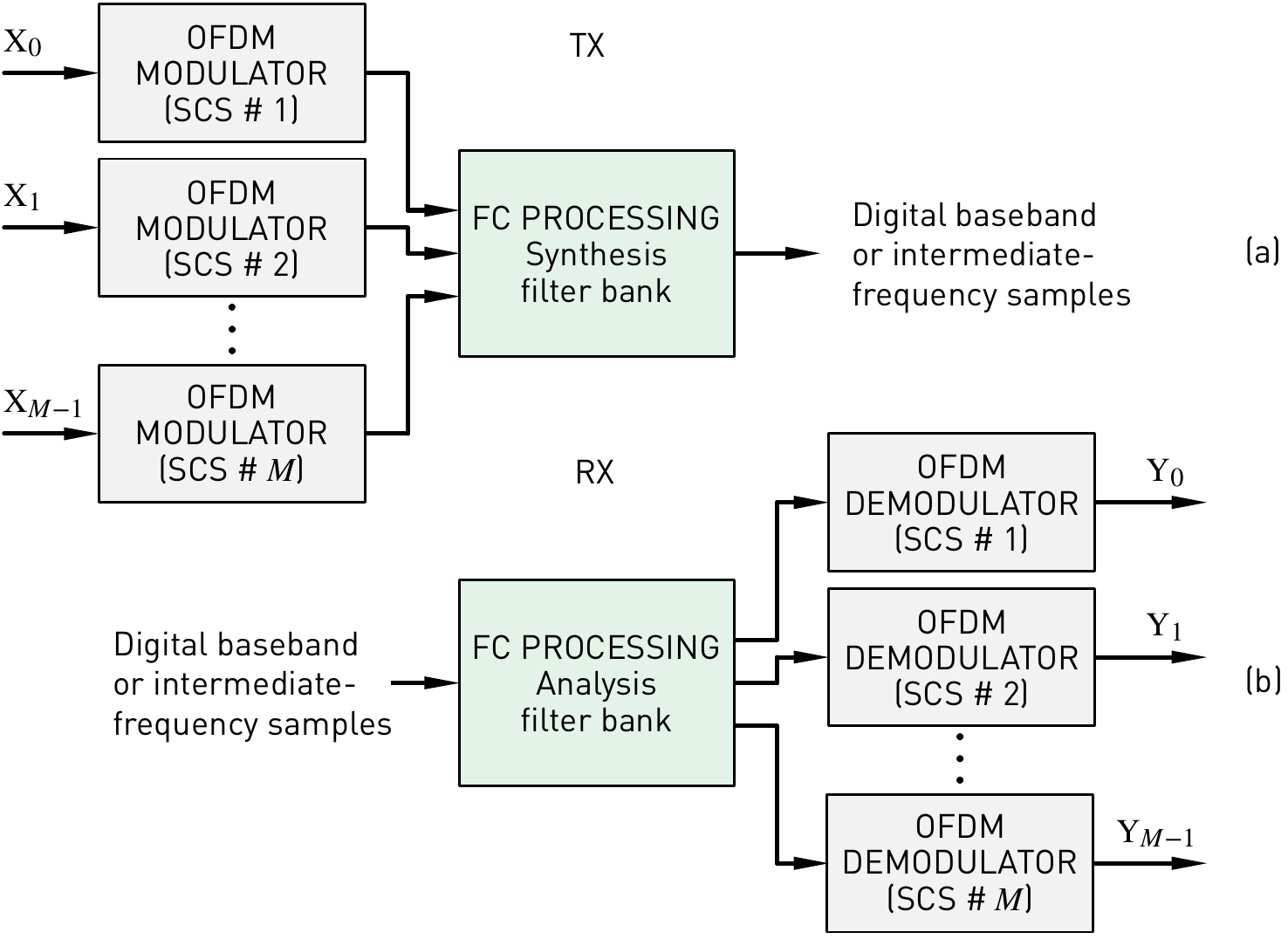}
  \caption{In \acf{fc}-based filtered-OFDM, filtering is applied at subband level, which means one or multiple contiguous \acfp{prb} with same \acf{scs}, while utilizing normal \ac{cp-ofdm} waveform for the \acsp{prb}. (a) Transmitter processing using the \acs{fc} synthesis filter bank of $M$ subbands. (b) Receiver processing using \acs{fc} analysis filter bank.}  
  \label{fig:OFDMpros}    
\end{figure}

Initial studies on filtered \ac{ofdm} were based on time-domain filtering \cite{C:2016_Coexistence_UFOFDM_CPOFDM,C:2015_Zhang_f-OFDM_for_5G,C:Ahmed19:ofdm_enhan_based_filter_window}, and later also polyphase filter-bank-based solutions have been presented \cite{J:Li2014:RB-F-OFDM,J:Zakaria12:_novel_filter_bank_multic_schem,C:Gerzaguet17:block_filter_OFDM,J:Zayani2018}. Fast-convolution-based filtered \ac{ofdm}\acused{fc-f-ofdm} (see Fig.~\ref{fig:OFDMpros}) has been presented in \cite{C:Renfors2015:fc-f-ofdm, C:Renfors16:adjustableCP, C:Yli-Kaakinen17:EUCNC, J:Yli-Kaakinen:JSAC2017}.  Especially in \cite{J:Yli-Kaakinen:JSAC2017}, the flexibility, good performance, and low computational complexity of \ac{fc-f-ofdm} was clearly demonstrated in the \ac{5g-nr} context.  These schemes typically apply filtering in continuous manner over a frame of \ac{cp-ofdm} (or zero-prefix-\ac{ofdm}) symbols.

\begin{figure*}[h!]
  \centering    
  \includegraphics[clip,width=0.95\textwidth]{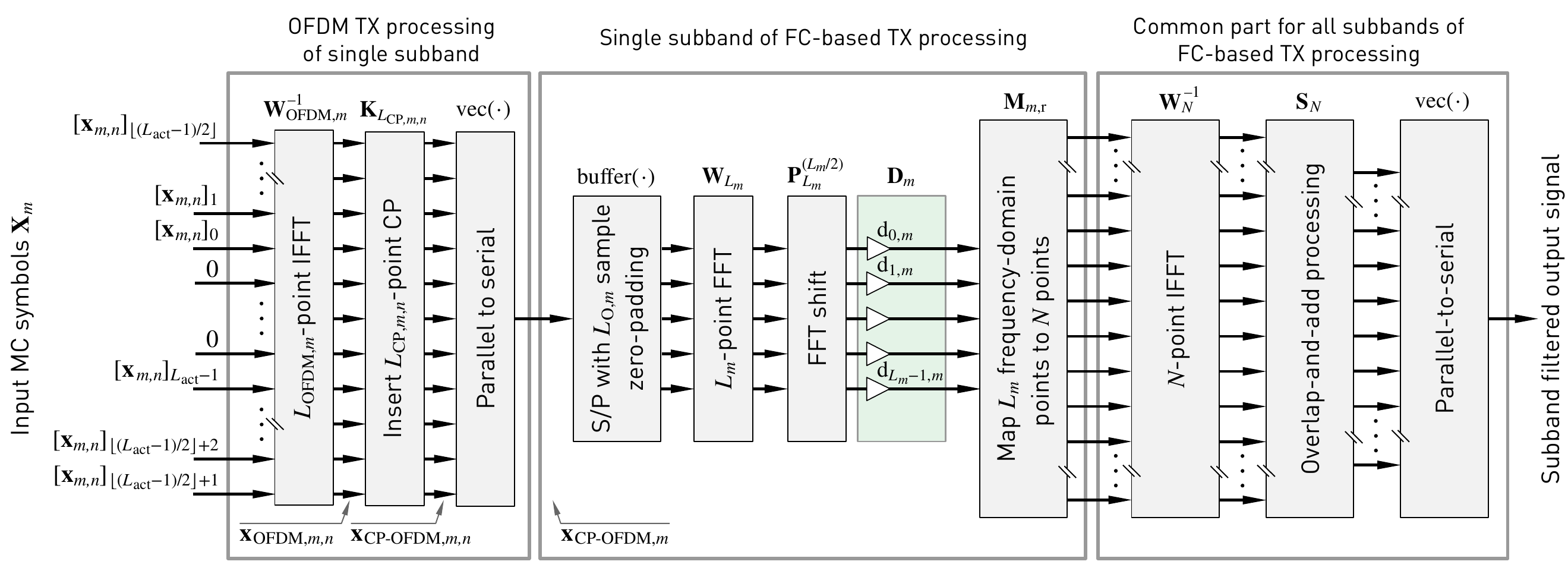}
  \caption{Block diagram for basic FC-F-OFDM transmitter processing using the \ac{ola} model.}
  \label{fig:FC-F-OFDM}      
\end{figure*}      
 
The problematic or inconvenient aspect of the conventional time-domain filtering-based schemes is their high complexity. In \cite{Ch:Yli-Kaakinen:5GRef}, it was shown that the complexity of time-domain filtering based solutions in mixed-numerology case is $70$ times the complexity of \ac{cp-ofdm} processing. This complexity can be reduced using classical (e.g., polyphase) filter-bank models, however, these solutions have somewhat reduced flexibility in adjusting the subband center frequencies and bandwidths. Since \ac{fc} is a block-wise processing scheme with fixed block length, the position of the useful parts of the \ac{ofdm} symbols vary within a frame of transmitted \ac{ofdm} symbols. With the usual continuous \ac{fc}-processing model as described in \cite{C:Renfors2015:fc-f-ofdm,J:Yli-Kaakinen:JSAC2017}, it is necessary that the \ac{cp} lengths and useful symbol durations correspond to an integer number of samples at the lower sampling rate used for transmitter \ac{ofdm} processing at each subband. In case of narrow-band allocations, this limits the choice of the transform lengths, significantly increasing the computational complexity. With 3GPP \ac{lte} and \acs{5g-nr} numerologies, the shortest possible transform length is \num{128}, while length-\num{16} transform would be sufficient when a subband contains one physical-layer resource block (\ac{prb}) (\num{12} subcarriers) only. This restriction applies to both time-domain filtering and \ac{fc}-based solutions with continuous processing model. On the receiver side, the proposed scheme makes it possible to effectively combine cascaded inverse and forward \ac{fft} units in \ac{fc}-filtered \ac{ofdm} processing. Transform decomposition is used to simplify these computations, leading to significantly reduced implementation complexity in various transmission scenarios. 

Building on our early work in \cite{C:Yli-Kaakinen:ASILOMAR2018}, this article proposes symbol-synchronized discontinuous \ac{fc} processing targeting at increased flexibility and reduced complexity of \ac{fc-f-ofdm}. It is shown that the proposed processing supports more flexible parametrization of the \ac{fc} engine, resulting in reduced complexity and latency with narrow subband allocations and in mini-slot transmission. Important use cases are seen, e.g., in spectrally well-contained \ac{nb-iot} transmission with one \ac{prb} allocation and in \ac{urllc} for generating short transmission bursts, so-called mini-slots, to reduce the radio link latency. However, the proposed schemes can be used for wide-band allocations as well.

The proposed scheme allows to reduce the complexity in the case of (\emph{i}) short transmissions (e.g.  mini-slot), (\emph{ii}) in multiplexing multiple relatively narrow subbands (e.g., gateway for \ac{mmtc} communications), and (\emph{iii}) \ac{ue} side \ac{tx} processing, assuming that only one numerology is transmitted.  Moreover, in case of parallelized hardware implementations, it is a benefit that each \ac{ofdm} symbol can be generated and filtered independently of the others.  This also minimizes the \ac{tx} signal processing latency.

In this article, we develop and describe discontinuous symbol-synchronized \ac{fc}-based processing techniques. The main contributions of the article can be listed as follows:
\begin{itemize}
\item[$\smallblacktriangleright$] Mathematical models for discontinuous symbol-synchronized \ac{fc}-based \ac{tx} and \ac{rx} processing are described. Both \acf{ola} and \ac{ols} variants are discussed.
\item[$\smallblacktriangleright$] Extrapolating \ac{tx} \ac{fc} processing is suggested for reducing the required \ac{ifft} size for \ac{ofdm} modulation by relaxing the \ac{cp} length related constraints on the \ac{ifft} size.
\item[$\smallblacktriangleright$] Model for simplifying the \ac{ifft} computations in \ac{rx} processing is proposed. The computational savings are achieved by effectively combining the computations of the \ac{ifft} of the \ac{fc} module and the \ac{fft} of subband \ac{ofdm} processing through transform decomposition.
\item[$\smallblacktriangleright$] Extensive numerical results are provided, verifying the validity of the proposed models and illustrating the benefits of the proposed \ac{fc}-based filtered-\ac{ofdm} processing.
\item[$\smallblacktriangleright$] Complexity evaluations are given quantifying the savings achieved using the proposed techniques.
\item[$\smallblacktriangleright$] Discontinuous \ac{fc} processing is proposed as an additional useful element in the toolbox for frequency-domain waveform processing, and it can be expected to find applications also in other areas of digital signal processing.
\end{itemize}

The remainder of this paper is organized as follows. Section~\ref{sec:cont-fast-conv}, first shortly reviews the continuous \ac{fc}-based filtered-\ac{ofdm} processing. Then, the proposed discontinuous TX \ac{fc}-processing model is described with implementation alternatives resulting to the reduced complexity and latency in Section~\ref{sec:symb-synchr-fast} for \ac{tx} and in Section~\ref{sec:symb-synchr-fast-rx} for \ac{rx}. Section~\ref{sec:impl-compl} presents an analysis of the computational complexity of considered alternative FC schemes. In Section \ref{sec:numerical-results}, the performance of the discontinuous processing is analyzed in terms of uncoded \ac{ber} in different interference/multiplexing scenarios and channel conditions, while also numerical results for the complexity for alternative \ac{fc} schemes are provided. Finally, the conclusions are drawn in Section \ref{sec:conclusions}. 

\section{Continuous Fast-Convolution Processing}
\label{sec:cont-fast-conv}
The block diagram for the basic continuous, symbol-nonsychronized \ac{ola}-based \ac{fc-f-ofdm} \ac{tx} processing for subband $m$ is shown in Fig.~\ref{fig:FC-F-OFDM}.  First, the \ac{cp-ofdm} signal for subband $m$ is generated by using the smallest \ac{ifft} length equal to or larger than $L_{\text{act},m}$ supporting an integer length \ac{cp}. Let us denote this transform (\ac{ifft}) size by $L_{\text{OFDM},m}$. Then, low-rate \ac{cp} of length $L_{\text{CP},m,n}$ is added to each of the $B_{\text{OFDM},m}$ \ac{ofdm} symbols for $n=0,1,\dots,B_{\text{OFDM},m}-1$ and the signal is converted to serial format.  These are all operations equivalent to basic \ac{cp-ofdm} \ac{tx} processing.  
 
The actual \ac{fc} processing per subband starts by partitioning the time-domain input sample stream to \ac{fc} blocks, as illustrated in Fig.~\ref{fig:FC-F-OFDM}.  Note that the exact number of \ac{fc} processing blocks depends on the input sequence length, overlap factor, and \ac{fft} length $L_m$.  Next, we take $L_m$-point \ac{fft} of each processing block and apply \ac{fft}-shift operation which essentially places the \acs{dc}-carrier in the middle of each vector.  Then, a frequency-domain window $\mathbf{D}_m$ is applied to implement the designed filter response.  After frequency-domain windowing the given subband is placed at the allocated \ac{fft} bins with transition-band values possibly exceeding the nominal allocation range.  The overlapping transition-band bins of adjacent subbands are added together. 
 
The $N$-point \ac{ifft} is common part for all subbands. It converts all the low-rate frequency-division multiplexed subband signals to time domain per \ac{fc} block.  In addition, it provides the sampling-rate conversion by the factor of $I_m=N/L_m$. Next, \ac{ola} processing is used to concatenate the high-rate \ac{fc}-blocks in order to construct the filtered time-domain representation of the transmitted signal. Alternatively, \ac{fc} processing can be realized using \ac{ols} scheme. In this case, the zero padding in block partitioning is replaced by the straightforward segmentation into the overlapping blocks and the \ac{ola} after the last transform is replaced by the discarding of the overlapping output segments. More detailed description of the \ac{fc} filtering process can be found, e.g., from \cite{J:Yli-Kaakinen:JSAC2017,J:Yli-Kaakinen:TSP2018}.

\begin{figure}[t!] 
  \centering
  \ifhbonecolumn
  \includegraphics[clip,width=0.53\textwidth]{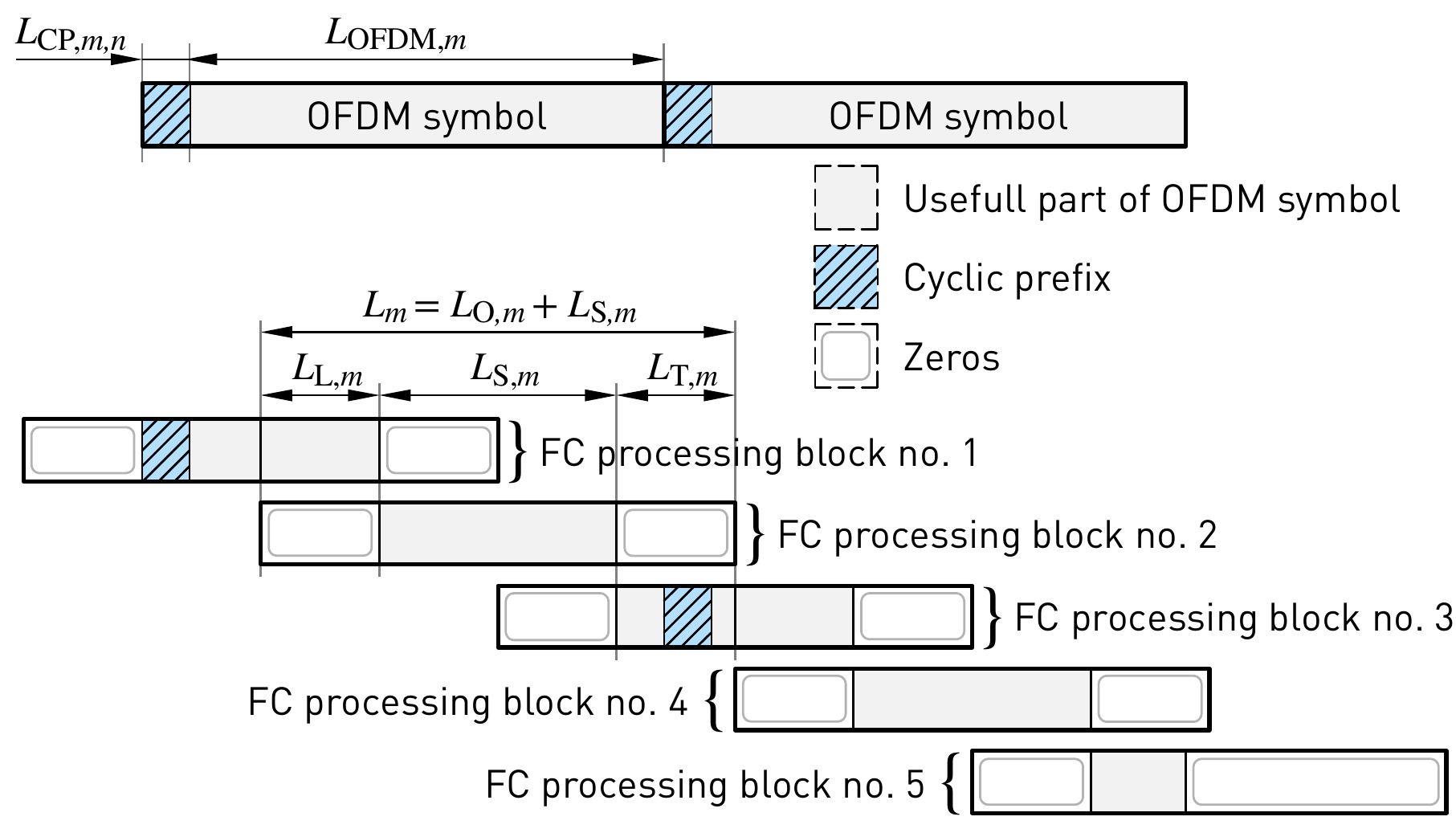}
  \else
  \includegraphics[clip,width=0.48\textwidth]{BlockProcessingTX}
  \fi
  \caption{\ac{fc}-processing block partitioning in basic continuous \ac{ola}-based \ac{fc-f-ofdm}.  \ac{fc} blocks are not synchronized to \ac{cp-ofdm} symbols.  Five \ac{fc}-processing blocks are needed for two \ac{ofdm} symbols.}
  \label{fig:BlockProc}        
\end{figure}       

The basic continuous \ac{fc}-processing flow of \ac{fc-f-ofdm} transmitter for $L_{\text{OFDM},m}=L_m$ is illustrated in Fig.~\ref{fig:BlockProc}.  The assumed overlap between processing blocks is \SI{50}{\%} (the overlap factor is $\lambda=0.5$). From Fig.~\ref{fig:BlockProc} we can observe how the \ac{fc} processing is continuous by collecting $L_{\text{OFDM},m}/2=L_m/2$ samples from the input sample stream to each \ac{fc} processing block. Also, the overlap factor is constant over all \ac{fc} processing blocks.

\section{Symbol-Synchronized Discontinuous FC-based Filtered-OFDM TX Processing} 
\label{sec:symb-synchr-fast}
\begin{figure}[t!]           
  \centering      
  \includegraphics[clip,width=\figwidthSC]{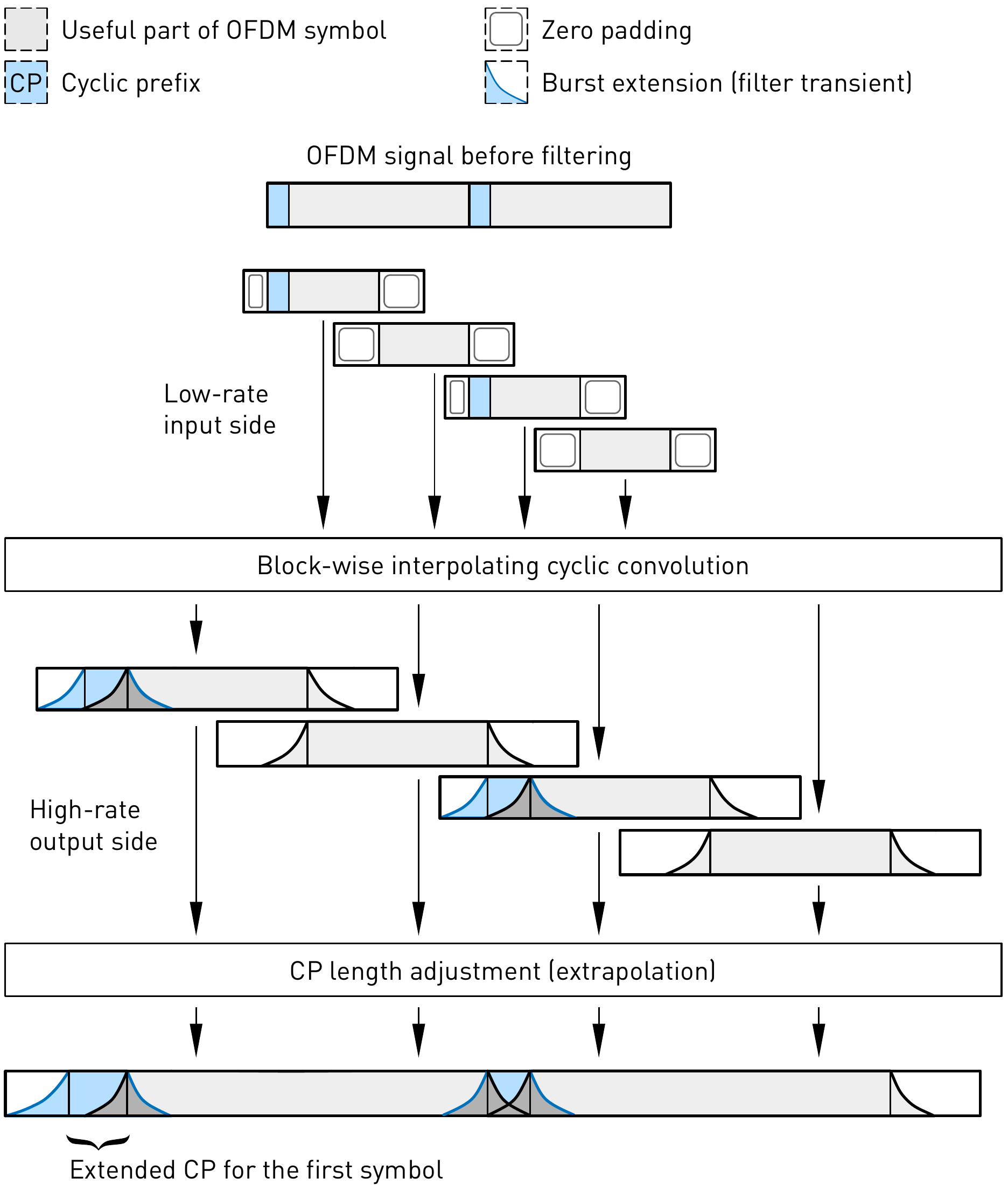}
  \caption{Discontinuous \ac{ola}-based block partitioning for \ac{fc-f-ofdm} \ac{tx} processing.  \ac{fc} processing blocks are synchronized to the \ac{ofdm} symbols.  Four \ac{fc}-processing blocks are needed for two \ac{ofdm} symbols.  The \ac{fc} overlap factor is dynamic: $0.5-N_{\text{CP},n}/N$ for the first block and \num{0.5} for the second block of each \ac{ofdm} symbol.}
  \label{fig:SymbolwiseProc}      
\end{figure}

Fig.~\ref{fig:SymbolwiseProc} illustrates the proposed discontinuous \ac{tx} \ac{fc-f-ofdm} processing flow for a mini-slot of two \ac{ofdm} symbols ($B_{\text{OFDM},m}=2$).  It can be observed that in discontinuous processing, two \ac{fc} processing blocks are synchronized to each \ac{ofdm} symbol, where the first \ac{fc} block contains the first half of the \ac{ofdm} symbol and the second \ac{fc} block contains the second half of the \ac{ofdm} symbol.  In addition, the first \ac{fc} processing block contains the low-rate \ac{cp} samples.  This reduces the overlap in the beginning of the first \ac{fc} block, that is, the overlap factor becomes $\lambda=0.5-L_{\text{CP},m,n}/L_m$. Here, $L_{\text{CP},m,n}$ is the \ac{cp} length of the $n$th symbol on subband $m$.  In practice, this reduction is relatively small, causing only minor increase in the related distortion effects. For discontinuous processing,  only four \ac{fc} processing blocks are used, instead of five in the continuous processing model (see Fig.~\ref{fig:BlockProc}), resulting in reduced complexity.  
 
\begin{figure*}[!t]          
  \centering      
  \includegraphics[clip,width=0.75\textwidth]{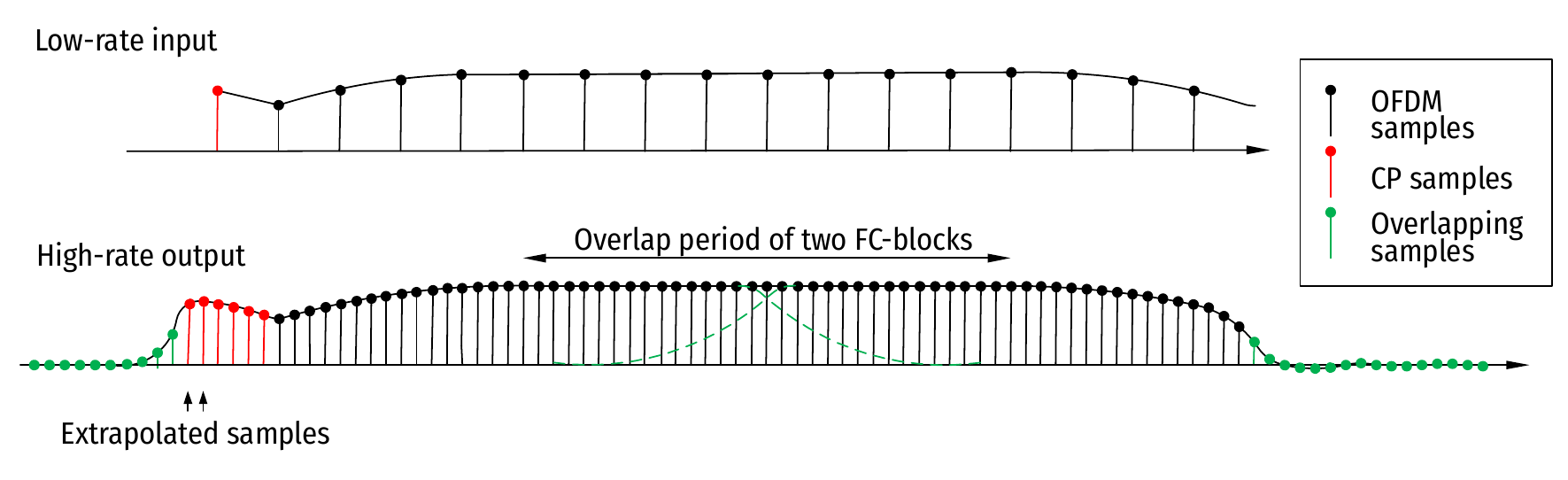}
  \caption{Discontinuous \ac{ola} processing for a single \ac{cp-ofdm} symbol with \SI{50}{\%} \ac{fc} overlap, interpolation factor of four ($I_m=4$), and high-rate \ac{ofdm} symbol duration of \num{64} samples.  \ac{cp} length is one sample at low rate and six samples at high rate (corresponding to one and half low-rate samples)}
  \label{fig:Extrapol}      
\end{figure*} 

This scheme is particularly beneficial in cases like Fig.~\ref{fig:SymbolwiseProc} where the \ac{fc} block length is equal to the \ac{ofdm} symbol duration (a common assumption in earlier studies of \ac{fc-f-ofdm}) and the two halves of the basic \ac{ofdm} symbol are processed in two consecutive \ac{fc} blocks.  Then the \ac{fc} blocks are synchronized to the \ac{ofdm} symbols, and the \ac{cp} is also processed within the first \ac{fc} block.  Such discontinuous \ac{fc}-based \ac{tx} filter processing reduces computational complexity through dynamically adjustable overlap of consecutive \ac{cp-ofdm} symbols. 

Fig.~\ref{fig:Extrapol} shows a detailed example of the sample-level interpolation and extrapolation process.  
The \ac{cp} part is included in the leading overlap section of the first \ac{fc} block of each \ac{ofdm} symbol and the \ac{cp} length is fine tuned in the \ac{ola} processing for consecutive \ac{ofdm} symbols at high rate.  The time resolution in adjusting the \ac{cp} length is equal to the sampling interval at high rate (as in traditional \ac{cp-ofdm}).
 
In generic setting, the discontinuous \ac{fc-f-ofdm} process can be formulated as follows.  First the \ac{cp-ofdm} symbols are generated at the minimum feasible sampling rate for each subband and, if needed, the \ac{cp} length is truncated to the highest integer number of low-rate samples which does not exceed the \ac{cp} length of the transmitted signal.  \ac{fc}-based filtering (or interpolation) is applied to each \ac{cp-ofdm} symbol individually to generate filtered symbols at the high (output) sampling rate.  The \ac{cp-ofdm} signal for a transmission slot is constructed from the generated individual symbols using the \ac{ola} principle.  When combining the individual symbols, their spacing in time direction is adjusted (with the precision of the output sampling interval) to correspond to the precise \ac{cp} duration.

In basic form, the proposed scheme is suitable for scenarios where the overall symbol durations of all subband signals to be transmitted have equal lengths and the symbols are synchronized.  It is notable that different durations (e.g., different \ac{cp} lengths) are allowed for different \ac{cp-ofdm} symbol intervals within a transmission slot.

\subsection{CP-OFDM Processing with Variable CP Lengths} 
\label{sec:cp-ofdm-processing}
Let $L_{\text{OFDM},m}$ and $L_{\text{CP},m,n}$ be the \ac{ofdm} symbol length and the \ac{cp} length of the $n$th symbol, respectively, on $m$th subband for $m=0,1,\dots,M-1$, where $M$ is the number of subbands.  The \ac{cp-ofdm} \ac{tx} processing, as illustrated in left-hand side block in Fig.~\ref{fig:FC-F-OFDM}, is formally expressed as
\begin{subequations}
  \label{eq:TX_OFDMproc}
  \begin{equation}
    \label{eq:TX_OFDMmod}
    \mathbf{x}_{\text{OFDM},m,n}=  
    \sqrt{L_{\text{OFDM},m}}\mathbf{W}_{L_{\text{OFDM},m}}^{-1}\mathbf{x}_{m,n}
  \end{equation}
  and
  \begin{equation}
    \label{eq:TX_CPins}
    \mathbf{x}_{\text{CP-OFDM},m,n}=  
    \mathbf{K}_{L_{\text{CP},m,n}} 
    \mathbf{x}_{\text{OFDM},m,n},
  \end{equation}
\end{subequations}
where $\mathbf{x}_{m,n}\in\mathbb C^{L_{\text{OFDM},m}\times 1}$ is the vector containing the incoming \ac{qam} symbols on $L_{\text{act},m}$ active subcarriers, $\mathbf{W}^{-1}_{L_{\text{OFDM},m}}\in\mathbb{C}^{L_{\text{OFDM},m}\times L_{\text{OFDM},m}}$ is the \ac{idft} matrix, and $\mathbf{K}_{L_{\text{CP},m,n}}\in\mathbb Z^{(L_{\text{OFDM},m}+L_{\text{CP},m,n})\times L_{\text{OFDM},m}}$ is the \ac{cp} insertion matrix.  In general, the \ac{cp} length could be different for each symbol, while for \ac{5g-nr} and \ac{lte}, two \ac{cp} lengths are used for normal \ac{cp} configuration \cite{S:3GPP:TS36.104v162,S:3GPP:TS38.104v160} such that the first symbol of a half subframe (\SI{0.5}{ms}) is longer than the others.

\subsection{FC-based Synthesis Filter Bank with Overlap-and-Add or Overlap-and-Save Processing} 
\label{sec:fc-based-filtering}
\ac{fc}-based filtering carries out the processing in overlapping blocks. In the \ac{sfb} case, the input block length of the $m$th subband is $L_m$ and the output block length is $N$. The overlap between the input blocks is determined by the number of overlapping input samples $L_{\text{O},m}$. The number of non-overlapping input samples is given as $L_{\text{S},m}=L_m-L_{\text{O},m}$ while the overlap factor is expressed using these values as
\begin{equation}
  \lambda=L_{\text{O},m}/L_m = (L_m-L_{\text{S},m})/L_m.
\end{equation}
The number of overlapping input samples can be further divided into leading and tailing overlapping parts as follows:
\begin{equation}
  L_{\text{L},m}=\lceil L_{\text{O},m}/2 \rceil
  \quad\text{and}\quad
  L_{\text{T},m}=\lfloor L_{\text{O},m}/2 \rfloor.
\end{equation}
The corresponding number of overlapping and non-overlapping output samples are determined as $N_{\text{O}}=\lambda N$ and $N_{\text{S}}=(1-\lambda)N$, respectively. Similarly, the number of overlapping output samples are divided into leading and tailing parts, $N_\text{L}$ and $N_\text{T}$, respectively.

The \ac{fc} processing increases the sampling rate by the factor of 
\begin{equation}
  \label{eqn:Rk} 
  I_m=N/L_{m},   
\end{equation}  
resulting in \ac{ofdm} symbol and \ac{cp} durations of $N_{\text{OFDM},m}=I_m L_{\text{OFDM},m}$ and $N_{\text{CP},m,n}=I_m L_{\text{CP},m,n}$, respectively.
Here $L_{\text{OFDM},m}$ and $L_{\text{CP},m,n}$ have integer values. It is convenient, but not necessary, that $N_{\text{OFDM},m}$ and $N_{\text{CP},m,n}$ have integer values as well.

In continuous \ac{fc} \ac{sfb}, the filtering of the $m$th \ac{cp-ofdm} subband signal for the generation of the high-rate waveform $\mathbf{w}_{m}$ can be represented as
\begin{subequations}
  \label{eq:BDM}    
  \begin{equation}  
    \mathbf{w}_{m} =
    \sqrt{I_m}
    \mathbf{F}_m
    \begin{bmatrix}
      \mathbf{0}_{S_{\text{L},m}\times 1}   \\
      \mathbf{x}_{\text{CP-OFDM},m} \\
      \mathbf{0}_{S_{\text{L},m}\times 1}
    \end{bmatrix},
  \end{equation} 
  where $\mathbf{F}_m$ is the block diagonal transform matrix of the form
  \begin{align}
    \label{eq:tx_matrix1}
    \mathbf{F}_m &= \diag*{
                   \mathbf{F}_{m,0}(\varphi_{m,0}),
                   \mathbf{F}_{m,1}(\varphi_{m,1}),
                   \dots,
                   \mathbf{F}_{m,R_m-1}(\varphi_{m,R_m-1})
                   }
  \end{align}
\end{subequations}   
with $R_m$ overlapping blocks $\mathbf{F}_{m,r}(\varphi_{m,r})\in\mathbb C^{N\times L_m}$ for $r=0,1,\dots,R_m-1$. Here, $\mathbf{x}_{\text{CP-OFDM},m}$ is the column vector formed by concatenating the $\mathbf{x}_{\text{CP-OFDM},m,n}$ for $n=0,1,\dots,B_{\text{OFDM},m}-1$. The zero-padding before and after the \ac{cp-ofdm} symbols is typically selected to be $S_{\text{L},m}=L_m-L_{\text{S},m}$.
The overall high-rate waveform to be transmitted is then obtained by combining all the subband waveforms as follows:
\begin{equation}
  \mathbf{x}_{\text{FC-F-OFDM}}=\sum_{m=0}^{M-1} \mathbf{w}_m.
\end{equation}  

The multirate version of the \ac{fc} \ac{sfb} can be represented either using the \ac{ola} block processing by decomposing the $\mathbf{F}_{m,r}(\varphi_{m,r})$'s as the following matrix
\begin{subequations}
  \begin{equation}
    \label{eq:ola_synthesis}
    \mathbf{F}^{\text{(OLA)}}_{m,r}(\varphi_{m,r}) = 
    \mathbf{W}^{-1}_{N}
    \mathbf{M}_{m}(\varphi_{m,r})
    \mathbf{D}_m 
    \mathbf{P}^{(L_m/2)}_{L_m}
    \mathbf{W}_{L_m}
    \mathbf{A}_{m,r}
  \end{equation}
  or \ac{ols} block processing when the $\mathbf{F}_{m,r}(\varphi_{m,r})$'s are decomposed as 
  \begin{equation}
    \label{eq:ols_synthesis}
    \mathbf{F}^{\text{(OLS)}}_{m,r}(\varphi_{m,r}) = 
    \mathbf{S}_{m,r} 
    \mathbf{W}^{-1}_{N}
    \mathbf{M}_{m}(\varphi_{m,r})
    \mathbf{D}_m 
    \mathbf{P}^{(L_m/2)}_{L_m}
    \mathbf{W}_{L_m}.
  \end{equation}
\end{subequations}
Here, $\mathbf{W}_{L_m}\in\mathbb{C}^{L_m\times L_m}$ and $\mathbf{W}_N^{-1}\in\mathbb{C}^{N\times N}$ are \ac{dft} and \ac{idft} matrices, respectively. 
The \ac{dft} shift matrix $\mathbf{P}^{(L_m/2)}_{L_m}\in\mathbb{N}^{L_m\times L_m}$ is circulant permutation matrix
expressed as
\begin{equation}
  \mathbf{P}^{(L_m/2)}_{L_m} = 
  \begin{bmatrix}
    \mathbf{0}_{{\lfloor L_m/2 \rfloor}\times {\lceil L_m/2 \rceil}} &
    \mathbf{I}_{\lfloor L_m/2 \rfloor} \\ 
    \mathbf{I}_{\lceil L_m/2 \rceil} &
    \mathbf{0}_{{\lceil L_m/2 \rceil}\times {\lfloor L_m/2 \rfloor}}
  \end{bmatrix}
\end{equation}
while $\mathbf{D}_m\in\mathbb{R}^{L_m\times L_m}$ is diagonal matrix with diagonal elements being the weights of the subband $m$.  The frequency-domain mapping matrix $\mathbf{M}_{m}(\varphi_{m,r})\in\mathbb{N}^{N\times L_m}$ maps $L_m$ frequency-domain bins of the input signal to frequency-domain bins of the output signal as follows:
\begin{subequations}
  \begin{equation}
    [\mathbf{M}_m(\varphi_{m,r})]_{q,p} = 
    \begin{cases}
      \varphi_{m,r}, &
      \text{if $\Xi(p) = q$} \\
      0, & \text{otherwise}
    \end{cases}
  \end{equation}
  with
  \begin{equation}
    \Xi(p)=(c_m-\lceil{{L_m}/{2}}\rceil+p-1 \bmod N)+1,
  \end{equation}
  where $c_m$ is the center bin of the subband $m$. The phase rotation needed to maintain the phase continuity between the consecutive overlapping processing blocks is given as
  \begin{equation}
    \varphi_{m,r}=\exp(\iu 2 \pi r c_m L_{\text{S},m}/L_m).
  \end{equation}
\end{subequations}
For further details, see \cite{J:Yli-Kaakinen:JSAC2017}.

For \ac{ola} processing, the time-domain analysis window matrix $\mathbf{A}_{m,r}\in\mathbb{N}^{L_m\times L_m}$ is a diagonal weighting matrix as given by
\begin{subequations}
  \begin{equation}
    \label{eq:ola_proc}
    \mathbf{A}_{m,r}=\diag*{\mathbf{a}_{m,r}}
    \quad\text{with}\quad
    \mathbf{a}_{m,r}=
    \begin{bmatrix}
      \mathbf{0}_{L_{\text{L},m}\times 1} \\
      \mathbf{1}_{L_{\text{S},m}\times 1} \\
      \mathbf{0}_{L_{\text{T},m}\times 1} 
    \end{bmatrix}.
  \end{equation} 
  For \ac{ols} processing, the time-domain synthesis window matrix $\mathbf{S}_{m,r}\in\mathbb{N}^{N\times N}$ is given by
  \begin{equation}
    \mathbf{S}_{m,r}=\diag*{\mathbf{s}_{m,r}}
    \quad\text{with}\quad
    \mathbf{s}_{m,r}=
    \begin{bmatrix}
      \mathbf{0}_{N_{\text{L}}\times 1} \\
      \mathbf{1}_{N_{\text{S}}\times 1} \\
      \mathbf{0}_{N_{\text{T}}\times 1} 
    \end{bmatrix}.
  \end{equation}  
\end{subequations}

\begin{figure}[!t]        
  \centering  
  \includegraphics[width=\figwidthSC]{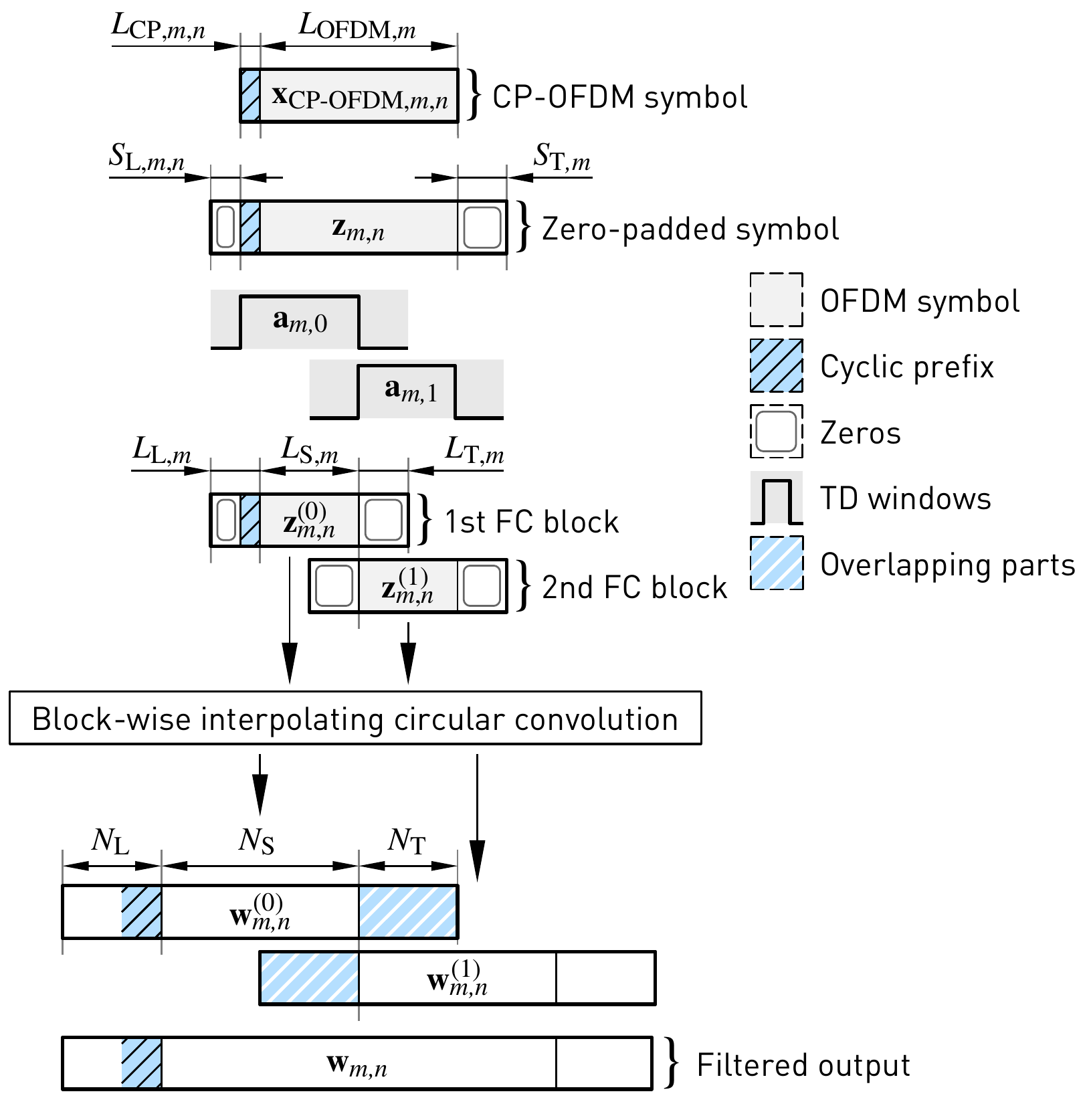} 
  \caption{Overlap-and-add (OLA)-based block processing of one \ac{cp-ofdm} symbol using two FC-processing blocks.}
  \label{fig:OLAbuffering}        
\end{figure}         

\subsection{Symbol-Synchronized TX FC Processing for One \acs{cp}-\acs{ofdm} Symbol} 
\label{sec:symb-synchr-fc} 
Fig.~\ref{fig:OLAbuffering} illustrates the \ac{ola}-based \ac{fc} processing of one \ac{cp-ofdm} symbol in two processing blocks corresponding to overlap factor of $\lambda=0.5$.  Here, it is assumed that the \ac{ofdm} modulation \ac{ifft} lengths and the \ac{fc}-processing short transform (FFT) lengths are the same, that is, $L_{\text{OFDM},m}=L_m$ for $m=0,1,\dots,M-1$. 

In symbol-synchronized processing, the incoming symbol is first zero padded in the beginning and the end by $S_{\text{L},m,n}=L_{{\text{L}},m}-L_{\text{CP},m,n}$ and $L_{\text{T},m}$ zeros, respectively, to form a zero-padded symbol ${\mathbf{z}}_{m,n}$ of length $3/2L_{m}$ as follows
\begin{equation}
  \label{eq:padding}
  {\mathbf{z}}_{m,n} = 
  \begin{bmatrix}
    \mathbf{0}_{S_{\text{L},m,n}\times (L_{\text{OFDM},m}+L_{\text{CP},m,n})} \\
    \mathbf{I}_{L_{\text{OFDM},m}+L_{\text{CP},m,n}} \\
    \mathbf{0}_{L_{\text{T},m}\times (L_{\text{OFDM},m}+L_{\text{CP},m,n})} \\
  \end{bmatrix}
  \mathbf{x}_{\text{CP-OFDM},m,n}.
\end{equation}
Now, $\mathbf{F}_{m,r}(\varphi_{m,r})$'s for $r=0,1$ essentially process (filter and possibly interpolate) two overlapping segments of length $L_m$ from the zero-padded symbol. Let us denote these segments by ${\mathbf{z}}^{(r)}_{m,n}\in \mathbb{C}^{L_{m}\times 1}$ for $r=0,1$ and the samples belonging to these processing segments are given by
\begin{subequations}  
\begin{equation}  
  {\mathbf{z}}^{(r)}_{m,n}=
  (\mathbf{R}_{L_m}^{(r)})^\transpose
  {\mathbf{z}}_{m,n}
\end{equation} 
for $r=0,1$ where
\begin{equation}
  \mathbf{R}_{L_m}^{(0)}=
  \begin{bmatrix}
    \mathbf{I}_{L_m} \\  
    \mathbf{0}_{L_m/2\times L_m}
  \end{bmatrix}
  \qquad\text{and}\qquad 
  \mathbf{R}_{L_m}^{(1)}=
  \begin{bmatrix}
    \mathbf{0}_{L_m/2\times L_m} \\
    \mathbf{I}_{L_m}  
  \end{bmatrix}.
\end{equation}   
\end{subequations}    
 
The effective overlap factor for the first processing block is reduced to $0.5-L_{\text{CP},m,n}/L_m$ due to inclusion of \ac{cp} and, therefore, the first time-domain analysis window has to be redefined to 
\begin{equation} 
  \mathbf{a}_{m,0}= 
  \begin{bmatrix} 
    \mathbf{0}_{S_{\text{L},m,n} \times 1} \\
    \mathbf{1}_{L_{\text{S},m}+L_{\text{CP},m,n} \times 1} \\
    \mathbf{0}_{L_{\text{T},m} \times 1} 
  \end{bmatrix}
\end{equation}
as well. Let $\mathbf{w}_{m,n}^{(r)}\in\mathbb{C}^{N\times 1}$ for $r = 0,1$ denote the product of the processing blocks by the transform matrices as expressed by
\begin{subequations}    
\begin{equation}
  \mathbf{w}_{m,n}^{(r)} = 
  \mathbf{F}_{m,r}(\varphi_{m,r}\theta_n)
  {\mathbf{z}}^{(r)}_{m,n}
\end{equation}
for $r=0,1$. Here, an additional phase rotation as given by
\begin{equation}
  \theta_n = \exp(\iu 2 \pi c_m \varphi_n)
\end{equation}
with
\begin{equation}
  \varphi_n = \frac{1}{N}\sum_{q=1}^{n}N_{\text{CP},m,q},
\end{equation}
\end{subequations}    
is included to compensate the truncation of the \ac{cp} length to interger samples on the low-rate side.

The filtered high-rate subband waveform of length $3/2N$ corresponding the $n$th symbol on the $m$th subband can finally obtained be combining these filtered blocks as
\begin{subequations}
  \begin{equation}
    \mathbf{w}_{m,n} = \sum_{r=0,1} \mathbf{R}^{(r)}_{N}{\mathbf{w}}_{m,n}^{(r)}
  \end{equation}
  where  
  \begin{equation}
    \mathbf{R}_N^{(0)}=
    \begin{bmatrix}
      \mathbf{I}_{N} \\ 
      \mathbf{0}_{N/2\times N}
    \end{bmatrix}\quad\text{and}\quad 
    \mathbf{R}_N^{(1)}=
    \begin{bmatrix}
      \mathbf{0}_{N/2\times N}\\
      \mathbf{I}_{N}
    \end{bmatrix}.
  \end{equation} 
\end{subequations} 

\begin{figure}[!t]        
  \centering  
  \includegraphics[width=\figwidthSC]{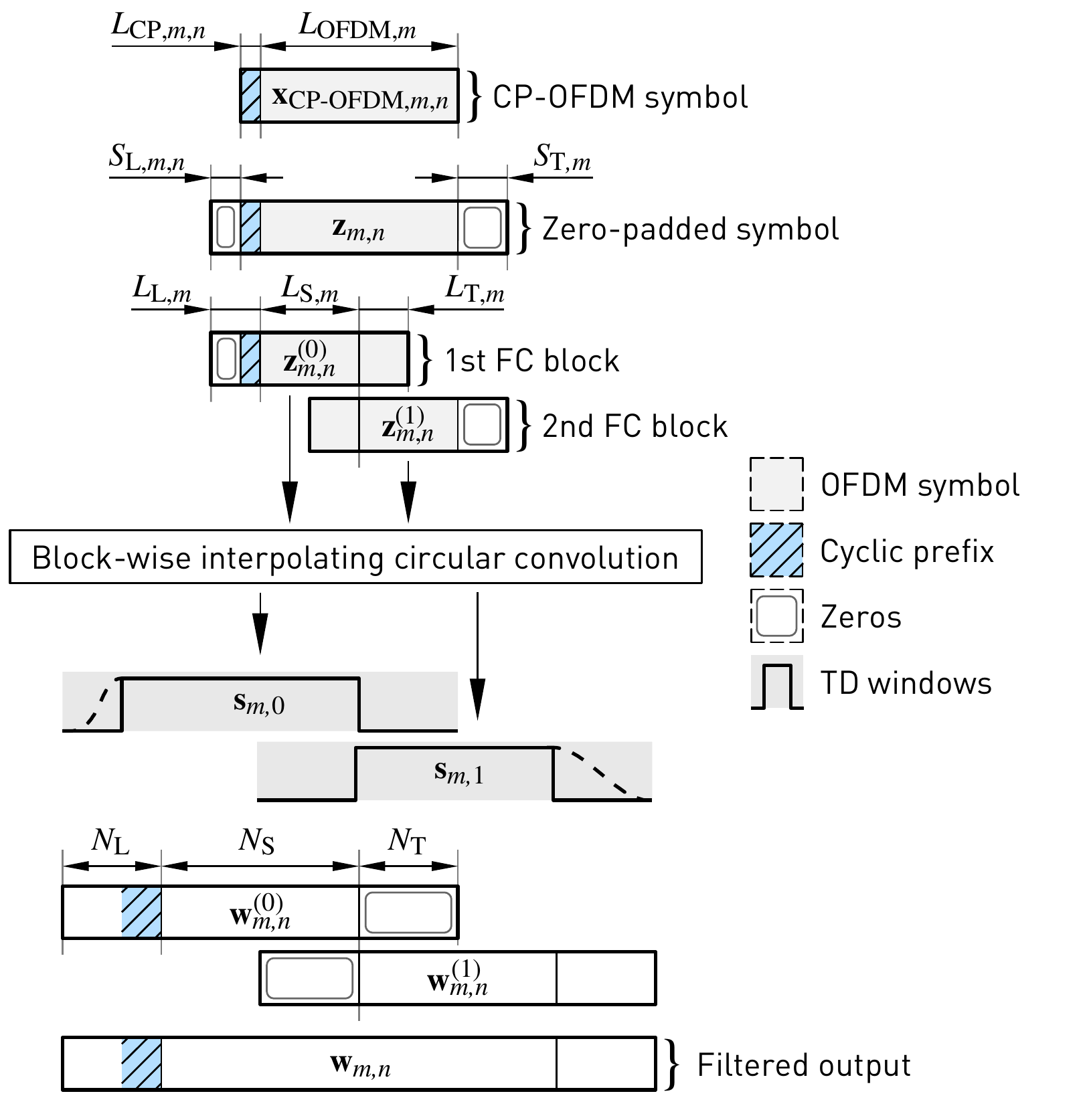} 
  \caption{Overlap-and-save (OSA)-based block processing of one \ac{cp-ofdm} symbol using two FC-processing blocks.}
  \label{fig:OSAbuffering}        
\end{figure}        

Alternative to \ac{ola} scheme, the above processing can also be carried out following the \ac{ols} approach as illustrated in Fig.~\ref{fig:OSAbuffering}. The basic difference is that now the time-domain windowing is realized after the convolution and only one of the filtered blocks is non-zero in the overlapping regions. The abrupt truncation of output waveform at the edges of the filtered symbol can be avoided by smoothly tapering the raising edge of the first time-domain synthesis window $\mathbf{s}_{m,0}$ and the falling edge of the second time-domain window $\mathbf{s}_{m,1}$, as illustrated using the dashed line in Fig.~\ref{fig:OSAbuffering}. Discontinuities in the output waveform give raise to a high spectral leakage and, therefore, the \ac{ola} scheme is preferable on the \ac{tx} side in general.

\subsection{Symbol-Synchronized TX FC Processing for Multiple Symbols}
\label{sec:symb-synchr-fc-1}
\begin{figure}[!t]         
  \centering  
  \includegraphics[width=\figwidthSC]{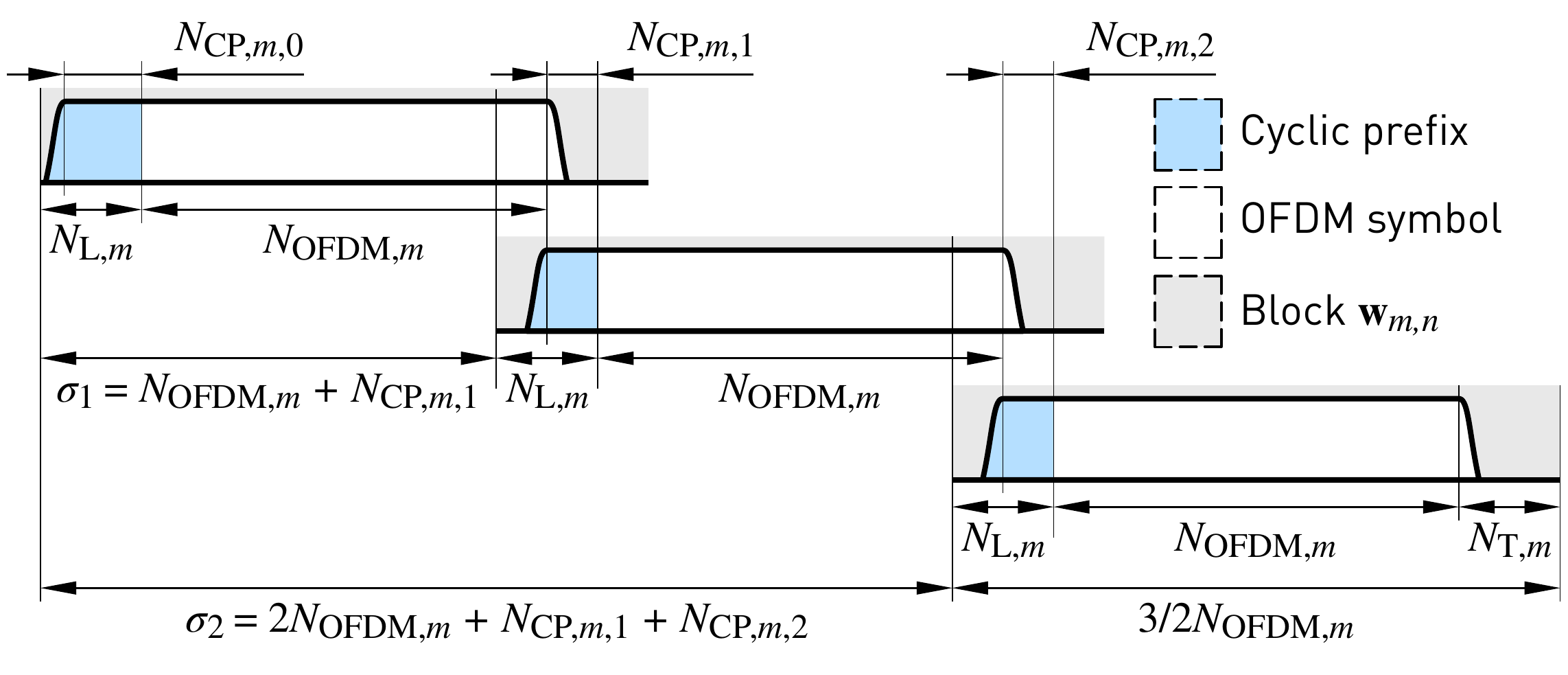} 
  \caption{Symbol-wise overlap-and-add processing for concatenating the filtered \ac{cp-ofdm} symbols.}
  \label{fig:concatenation}        
\end{figure}

In the case of multiple \ac{ofdm} symbols, the high-rate filtered symbols are combined with symbol-wise \ac{ola} processing as follows
\begin{subequations}
  \label{eq:multi-symb}
  \begin{equation}
    \mathbf{w}_{m} = \sum_{n=0}^{B_{\text{OFDM},m}-1}
    \boldsymbol{\Gamma}_{\sigma_n} \mathbf{w}_{m,n}
  \end{equation}
  with
  \begin{equation}
    \sigma_n = nN_{\text{OFDM},m}+\sum_{q=1}^{n} 
    N_{\text{CP},m,q}
  \end{equation}
  being the starting index of the $n$th filtered block $\mathbf{w}_{m,n}$ of length $3/2N_{\text{OFDM},m}$ in $\mathbf{w}_{m}$ as illustrated in Fig.~\ref{fig:concatenation}. Here, $Q$-by-$3/2N$ matrix 
  \begin{equation}
    \label{eq:Gamma}
    \boldsymbol{\Gamma}_{p}= 
    \begin{bmatrix}
      \mathbf{0}_{p\times 3/2N}\\ 
      \mathbf{I}_{3/2N} \\
      \mathbf{0}_{(Q-3/2N-p)\times 3/2N} 
    \end{bmatrix} 
  \end{equation}
with 
\begin{equation}
  \label{eq:waveformLen}
  Q = N_{\text{L},m} + N_{\text{T},m} +
  B_{\text{OFDM},m}N_{\text{OFDM},m}+
  \sum_{n=1}^{B_{\text{OFDM},m}-1}N_{\text{CP},m,n}
\end{equation}
\end{subequations}
aligns the filtered symbols to their desired time-domain locations at the high-rate output sequence.

\subsection{CP Extrapolation by TX FC Processing}
\label{sec:cp-extrapolation-fc}
Suppose that $L_{\text{OFDM},m}$ is chosen such that the \ac{cp} length on the low-rate side is not an integer, that is, $L_{\text{OFDM},m}<128$ for \ac{5g-nr} and \ac{lte} numerologies.  In this case, \ac{cp} length can be rounded to next smaller integer as given by
\label{eq:ext_multi-symb}
\begin{equation}
  L_{\text{CP},m,n} = \lfloor N_{\text{CP},m,n}/I_m \rfloor
\end{equation}
and the \ac{fc}-based filtering with the accompanying symbol-wise overlap-and-add processing as given by \eqref{eq:multi-symb} inherently extrapolates at the high-rate side $N_{\text{CP},m,n}-N/L_m L_{\text{CP},m,n}$ samples corresponding to fractional part of the low-rate \ac{cp}.

As an example, in the \SI{10}{MHz} \ac{5g-nr} or \ac{lte} case with \SI{15}{kHz} \ac{scs} and normal \ac{cp} length, the output sampling rate is $f_\text{s,out}=\SI{15.36}{MHz}$, the useful \ac{ofdm} symbol duration is $N_{\text{OFDM},m}=1024$ high-rate samples, and the \ac{cp} length is $N_{\text{CP},m,n}=80$ high-rate samples for the first symbol (for $n \bmod 7 = 0$) of each slot of \num{7} symbols, and $N_{\text{CP},m,n}=72$ samples for the others (for $n \bmod 7 \neq 0$).  Then in the continuous processing model, the smallest possible \ac{ofdm} \ac{ifft} length is $L_{\text{OFDM},m}=128$, corresponding to $f_\text{s,in}=\SI{1.92}{MHz}$ input sampling rate, and the \ac{cp} lengths are $L_{\text{CP},m,n}=10$ and $L_{\text{CP},m,n}=9$ low-rate samples for the first and other symbols, respectively.  However, using the discontinuous \ac{fc} processing model with narrow subband allocations, like \num{12}, \num{24}, or \num{48} \acp{subc}, the \ac{ofdm} \ac{ifft} length $L_{\text{OFDM},m}$ can be reduced to \num{16}, \num{32}, or \num{64}, respectively. These transform lengths correspond at the low-rate side to \ac{cp} lengths $L_{\text{CP},m,n}$ of \num{1.25}, \num{2.5}, and \num{5.0} for the first symbol or \num{1.125}, \num{2.25}, and \num{4.5} for the others.  The same transform (\ac{ifft}) lengths are used for the subband \ac{ofdm} signal generation. One important case is \ac{nb-iot} using \SI{180}{kHz} transmission bandwidth corresponding to a single \ac{prb} (\num{12} SCs) and needs \ac{ifft} length of \num{128} and sampling rate of \SI{1.92}{MHz} in traditional implementation. Discontinuous processing allows to generate the signal by using the \ac{fft} size of \num{16} for the \ac{ofdm} symbol generation and \ac{fc} processing at the sampling rate of \SI{240}{kHz}. 

\section{Symbol-Synchronized Discontinous FC-based Filtered-OFDM RX Processing}
\label{sec:symb-synchr-fast-rx}
\begin{figure}[!t]
  \centering
  \includegraphics[clip,width=\figwidthSC]{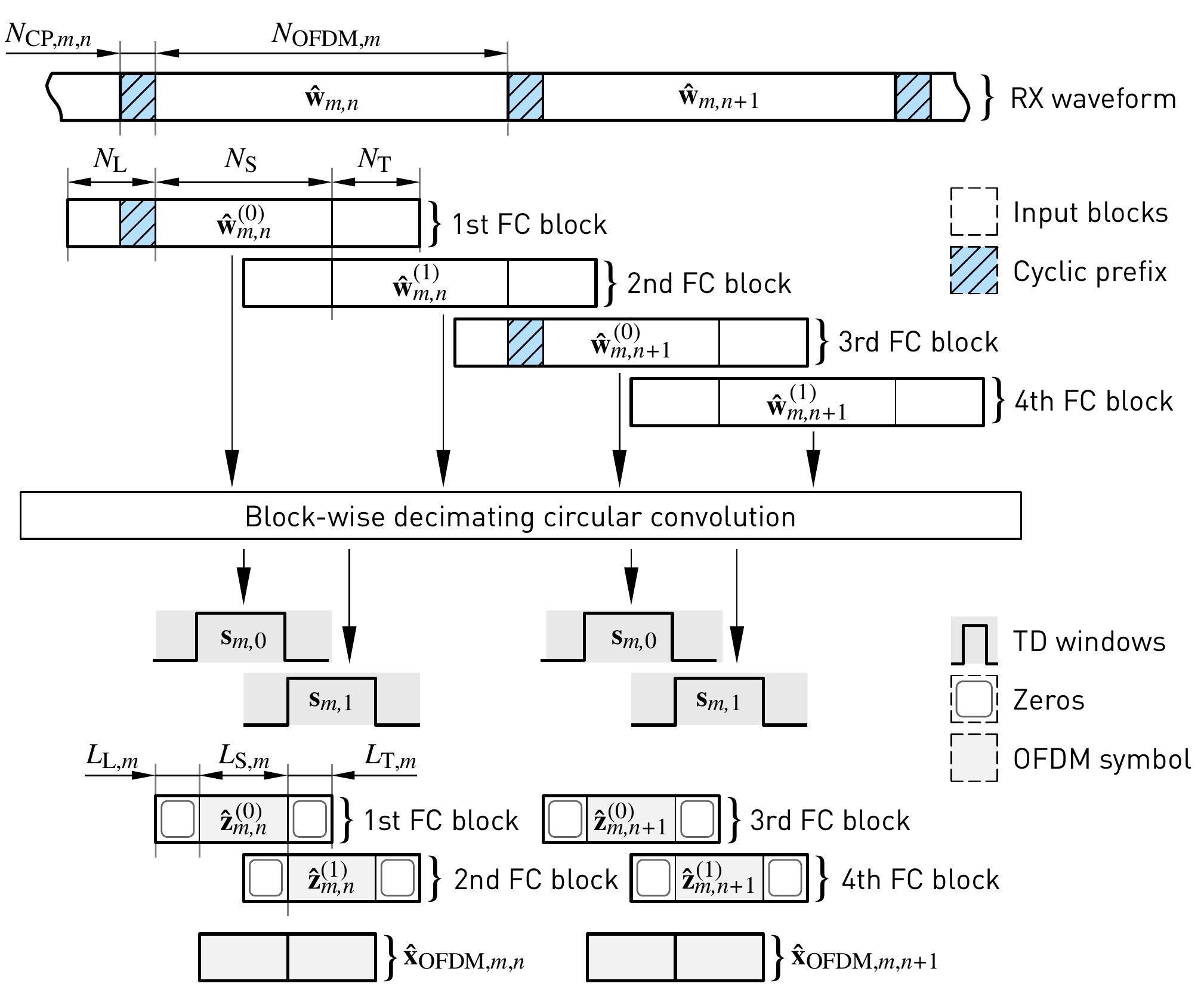} 
  \caption{Discontinuous \ac{ols} processing for \ac{fc-f-ofdm} receiver  with overlap factor of $\lambda=0.5$. \ac{fc} processing blocks are synchronized to the \ac{ofdm} symbols. Four \ac{fc}-processing blocks are needed for two \ac{ofdm} symbols.}
  \label{fig:BlockProcRX}       
\end{figure}

As for the \ac{tx} side, the basic continuous \ac{fc}-processing flow of \ac{fc-f-ofdm} \ac{rx} requires five \ac{fc}-processing blocks in order to process two \ac{cp-ofdm} symbols. In this case, the \acp{cp} are discarded after the \ac{fc} filtering as part of normal \ac{rx} \ac{cp-ofdm} reception. With continuous processing, the \ac{fc}-processing chain needs to wait for varying number of samples belonging to the second \ac{ofdm} symbol before it can start processing the third \ac{fc}-processing block in order to obtain final samples in the output for the first \ac{ofdm} symbol. In discontinuous processing, as seen in Fig.~\ref{fig:BlockProcRX}, the \ac{rx} waits only for the samples belonging to the first \ac{ofdm} symbol and desired amount of overlapping samples from the beginning of the second symbol, after which it can start processing the second \ac{fc} processing block, providing at the output the last filtered samples of the first \ac{ofdm} symbol.
     
In the case of maximal timing-adjustment flexibility, the number of samples collected from the following \ac{cp-ofdm} symbol time corresponds to the number of overlapping samples at the end of \ac{fc} processing block. In addition, in discontinuous processing, two first \ac{fc} processing blocks can be processed independently from the two following \ac{fc} processing blocks, as they represent different \ac{ofdm} symbols. In continuous processing, two \ac{ofdm} symbols are linked through common samples in the third \ac{fc} processing block.

Because the content (\ac{fc} block contains first or second half of \ac{ofdm} symbol) and processing of even and odd \ac{fc} processing blocks remain constant over the whole \ac{rx} signal, we can process even and odd blocks separately from each other. This allows for an implementation where even and odd \ac{fc} processing blocks are processed in parallel \ac{fc} processing chains, allowing to minimize the latency of the \ac{rx} implementation.

In the \ac{rx} case, the transform matrices for \ac{fc} processing with \ac{ola} and \ac{ols} schemes are given as
\begin{subequations}
\begin{align}
  \label{eq:ola_analysis}
  \mathbf{G}^{\text{(OLA)}}_{m,r}(\varphi_{m,r})
  &= \mathbf{W}^{-1}_{L_m}
  \mathbf{P}^{(-L_m/2)}_{L_m}
  \mathbf{D}_m 
  \mathbf{M}_{m,r}(\varphi_{m,r})^\transpose 
  \mathbf{W}_{N}
  \mathbf{{A}}_{m,r} \nonumber\\
  &= \conj{\mathbf{F}^{\text{(OLS)}}_{m,r}(\varphi_{m,r})}^\transpose
\end{align}
and
\begin{align}
  \label{eq:ols_analysis}
  \mathbf{G}^{\text{(OLS)}}_{m,r}(\varphi_{m,r})
  &= \mathbf{{S}}_{m,r} 
  \mathbf{W}^{-1}_{L_m}
  \mathbf{P}^{(-L_m/2)}_{L_m}
  \mathbf{D}_m 
  \mathbf{M}_{m,r}(\varphi_{m,r})^\transpose  
  \mathbf{W}_{N} \nonumber\\
  &= \conj{\mathbf{F}^{\text{(OLA)}}_{m,r}(\varphi_{m,r})}^\transpose,
\end{align}
\end{subequations}
respectively, while the time-domain analysis and synthesis window matrices are now given as
\begin{subequations}
\begin{equation}
  \mathbf{A}_{m,r}=\label{eq:1}
  \diag*{
  \begin{bmatrix}
    \mathbf{0}_{N_{\text{L}}\times 1} \\
    \mathbf{1}_{N_{\text{S}}\times 1} \\
    \mathbf{0}_{N_{\text{T}}\times 1} 
  \end{bmatrix}
  }
  \quad\text{and}\quad
  \mathbf{S}_{m,r}=
  \diag*{
  \begin{bmatrix}
    \mathbf{0}_{L_{\text{L},m}\times 1} \\
    \mathbf{1}_{L_{\text{S},m}\times 1} \\
    \mathbf{0}_{L_{\text{T},m}\times 1} 
  \end{bmatrix}
  },
\end{equation}
\end{subequations}
respectively.

\begin{figure}[!t] 
  \centering   
  \includegraphics[clip,width=\figwidthSC]{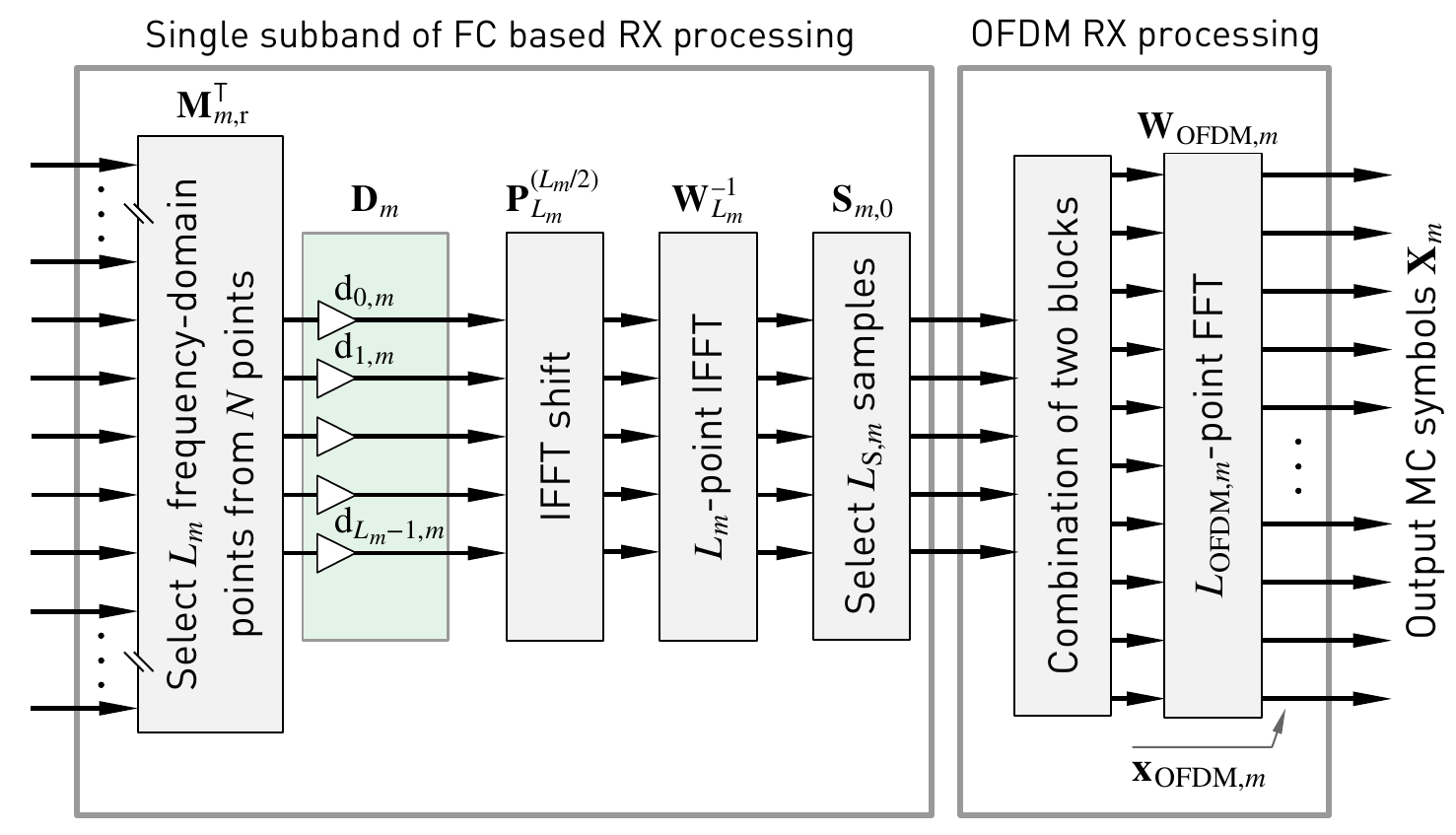}
  \caption{Block diagram for discontinuous FC-F-OFDM receiver processing using the overlap-save model.}
  \label{fig:DISCO_RX}  
\end{figure}     

The \ac{rx} side discontinuous \ac{fc} processing starts by segmenting the received high-rate waveform $\mathbf{\hat{x}}_{\text{FC-F-OFDM}}$ into the blocks of length $3/2N$ as
\begin{subequations}
  \begin{equation}
    \mathbf{\hat{w}}_{m,n} =
    \sum_{n=0}^{B_{\text{OFDM},m}-1}
    \boldsymbol{\Gamma}_{\sigma_n}^\transpose\mathbf{\hat{x}}_{\text{FC-F-OFDM}}.
  \end{equation}
  Here, it is assumed for simplicity that the length of the received waveform is $Q$ as given by \eqref{eq:waveformLen} and that the waveform is synchronized such that it contains $I_mS_{\text{L},m,n}$ samples before the first data symbol. The high-rate symbols $\mathbf{\hat{w}}_{m,n}$ are further divided into the overlapping blocks of length $N$ as
\begin{equation}
  \mathbf{\hat{w}}_{m,n}^{(r)} =
  (\mathbf{R}_N^{(r)})^\transpose\mathbf{\hat{w}}_{m,n},
\end{equation}
for $r=0,1$. These blocks are processed by $\mathbf{G}_{m,r}(\varphi_{m,r})$'s as 
\begin{equation}
  \mathbf{\hat{z}}_{m,n}^{(r)} =
  \mathbf{G}_{m,r}(\varphi_{m,r})
  {\mathbf{\hat{w}}}^{(r)}_{m,n}
\end{equation}
for $r=0,1$ to obtain filtered low-rate \ac{fc} blocks corresponding to 
first and second half of the $n$th symbol on subband $m$. Finally, the filtered blocks are concatenated and the extensions are removed as
\begin{align} 
  \mathbf{\hat{x}}_{\text{OFDM},m,n} =
  \begin{bmatrix}
    \mathbf{0}_{S_{\text{T},m,n}\times L_{\text{OFDM},m}} \\
    \mathbf{I}_{L_{\text{OFDM},m}} \\
    \mathbf{0}_{S_{\text{T},m}\times L_{\text{OFDM},m}} 
  \end{bmatrix}^\transpose
  \sum_{r=0,1}
  \mathbf{R}^{(r)}_{L_m}{\mathbf{\hat{z}}}_{m,n}^{(r)},
\end{align}
and the resulting low-rate \ac{ofdm} symbol is converted back to frequency domain as a part of the \ac{ofdm} demodulation process as follows:
\begin{equation} 
  \mathbf{\hat{x}}_{m,n} =  
  \frac{1}{\sqrt{L_{\text{OFDM},m}}}
  \mathbf{W}_{L_{\text{OFDM},m}}
  \mathbf{\hat{x}}_{\text{OFDM},m,n}. 
\end{equation}
Alternatively, the concatenation and the removal of the extensions can be combined in
\begin{equation}
  \label{eq:concat2}
  \mathbf{\hat{x}}_{\text{OFDM},m,n} = 
  \mathbf{P}_{L_m}^{(L_m/4)}\mathbf{\hat{z}}_{m,n}^{(0)} + 
  \mathbf{P}_{L_m}^{(-L_m/4)}\mathbf{\hat{z}}_{m,n}^{(1)},
\end{equation}
where
\begin{align} 
  \label{eq:perm1}
  \mathbf{P}_{L_m}^{(L_m/4)} &= 
  \begin{bmatrix}
    \begin{matrix}
      \mathbf{0}_{L_m/2\times L_m/4} & \mathbf{I}_{L_m/2} & \mathbf{0}_{L_m/2\times L_m/4}
    \end{matrix} \\
    \mathbf{0}_{L_m/2\times L_m}
  \end{bmatrix}\\ 
  \intertext{and} 
  \label{eq:perm2}
  \mathbf{P}_{L_m}^{(-L_m/4)} &= 
  \begin{bmatrix}
    \mathbf{0}_{L_m/2\times L_m} \\
    \begin{matrix}
      \mathbf{0}_{L_m/2\times L_m/4} & \mathbf{I}_{L_m/2} & \mathbf{0}_{L_m/2\times L_m/4}
    \end{matrix} 
  \end{bmatrix}.
\end{align}
\end{subequations}
This latter form is beneficial when finding the simplified implementations for the discontinuous \ac{rx} processing, as described in next section.

\section{Implementation Complexity}  
\label{sec:impl-compl}
The \ac{fc} processing complexity can be divided into high-rate side complexity, $C_\text{high-rate}$, corresponding to long \ac{fc}-processing transform and low-rate (or subband-wise) complexity, $C_{\text{low-rate},m}$, corresponding to short \ac{fc}-processing transform, frequency-domain windowing, and \ac{ofdm} (de)modulation transform. Let us denote the \ac{fc} \ac{rx} processing long transform (\ac{fft}) and short transform (\ac{ifft}) complexities given in terms of number of real multiplications by $\mu(N)$ and $\mu(L_m)$, respectively, and \ac{ofdm} transform (\ac{fft}) complexity by $\mu(L_{\text{OFDM},m})$.  

The number of real multiplications per received data symbol can now be evaluated as
\begin{subequations}
  \label{eq:complexity}
  \begin{equation} 
    C_\text{mult} = \frac{
      C_\text{high-rate} + \sum_{m=0}^{M-1} C_{\text{low-rate},m}
    }{
      \sum_{m=0}^{M-1} L_{\text{act},m}
    },
  \end{equation}
  where the high-rate and low-rate complexities per subband are defined, respectively, as
  \begin{align}
    \label{eq:highrate}
    C_\text{high-rate} &= \alpha \mu(N) \\ 
    \intertext{and}
    \label{eq:lowrate} 
    C_{\text{low-rate},m} &= \alpha\beta \mu(L_m) + 6\alpha N_{\text{tb},m} + \mu(L_{\text{OFDM},m}). %
  \end{align} 
\end{subequations}
Here, $N_{\text{tb},m}$ is the number of transition-band bins per transition band, $\alpha=2$ is the number of \ac{fc} blocks per \ac{ofdm} symbol, and $\beta\leq1$ is the implementation related factor as described in the next subsection. Factor \num{6} in \eqref{eq:lowrate} is due to fact that two transition bands are needed for each subband and one complex non-trivial transition-band bin requires three real multiplications in general.

\begin{figure}[t!] 
\centering    \includegraphics[clip,width=0.85\figwidth]{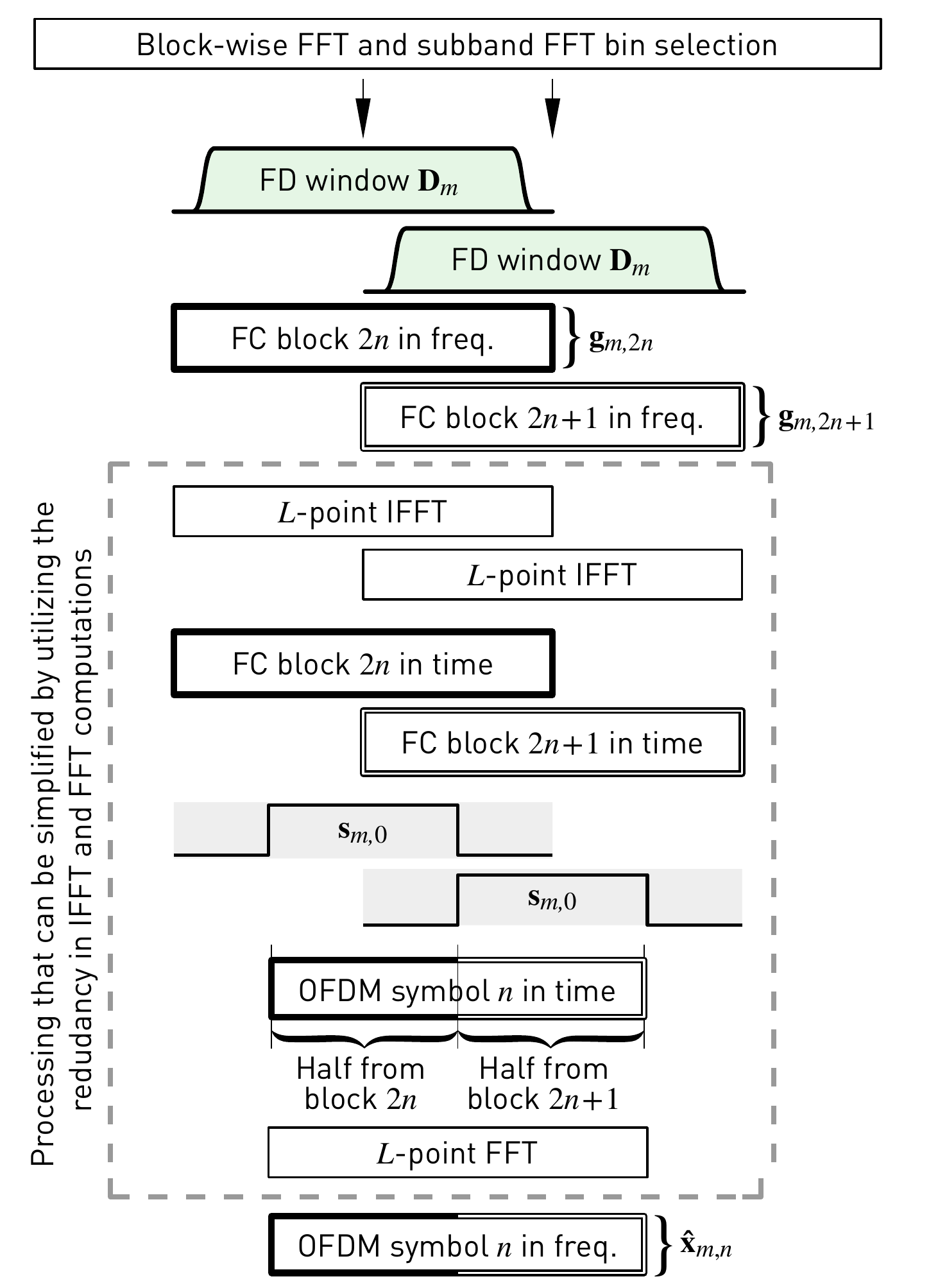} 
  \caption{Discontinuous processing for one output OFDM symbol. Dashed block represents the processing that can be simplified by sharing the \ac{ifft} computations.}
  \label{fig:SimplBlockProcRX}         
\end{figure}       

For \acs{5g-nr} and \ac{lte} numerologies, the minimum allocation size is one \ac{prb} corresponding to \num{12} subcarriers. In this case, the minimum usable \ac{fft} size for continuous processing is $L_{\text{OFDM},m}=L_m=128$ whereas for discontinuous processing, \ac{fft} of size $L_{\text{OFDM},m}=L_m=16$ can be used. Assuming that an efficient implementation (using split-radix algorithm) requires 
\begin{equation} 
  \mu(L)=L\log_2(L)-3L+4
\end{equation}
real multiplications for transform of size $L$, each transform of size \num{128} requires $\mu(128)=516$ real multiplications and each transform of size \num{16} requires $\mu(16)=20$ real multiplications.  These complexities are evaluated according to scheme requiring three real multiplications per complex multiplication as described in \cite{J:Sorensen86:SP-FFT}.

\subsection{Simplified Implementation}
The basic idea for joint processing of the \acp{ifft} of the \ac{fc}-based filter and the \ac{fft} of \ac{ofdm} receiver is shown in Fig.~\ref{fig:SimplBlockProcRX}. This structure is made possible by the symbol-synchronized discontinuous \ac{rx} processing scheme. 

Here, it is assumed that the length of the subband-wise \ac{ofdm} symbol and the size of the \ac{ifft} used in discontinuous \ac{rx} \ac{fc} processing are the same $L\equiv L_{\text{OFDM},m}=L_m$. $L$ is selected such that it contains all active \acp{subc} per subband and transition-band bins used in the frequency-domain windowing performed in the ``Block-wise \ac{fft}, subband \ac{fft} bin selection and weighting'' block. 
 
For simplicity, we denote the time-domain synthesis window matrix $\mathbf{S}_{m,r}$ by $\mathbf{S}$. Now, the processing of the $n$th symbol from the frequency-domain \ac{fc} output blocks $\mathbf{g}_{m,2n}$ and $\mathbf{g}_{m,2n+1}$ can be written as
\begin{align}
  \label{eq:FCcomb}    
  \mathbf{\hat{x}}_{m,n} & = 
  \mathbf{W}_{L} 
  \begin{bmatrix}
    \mathbf{P}_L^{(L/4)}\mathbf{S}\mathbf{W}_{L}^{-1}\mathbf{g}_{m,2n} +
    \mathbf{P}_L^{(-L/4)}\mathbf{S}\mathbf{W}_{L}^{-1}\mathbf{g}_{m,2n+1}
  \end{bmatrix}.
\end{align}
Here, $\mathbf{P}_L^{(L/4)}$ and $\mathbf{P}_L^{(-L/4)}$ are given by \eqref{eq:perm1} and \eqref{eq:perm2}, respectively. Alternatively, \eqref{eq:FCcomb} can be represented as
\begin{subequations} 
\begin{equation}
  \mathbf{\hat{x}}_{m,n} = \mathbf{F}_0\mathbf{g}_{m,2n} + \mathbf{F}_1\mathbf{g}_{m,2n+1},
\end{equation}
where
\begin{align}
  \mathbf{F}_0 = \mathbf{W}_{L}\mathbf{P}_L^{(L/4)}\mathbf{S}\mathbf{W}_{L}^{-1} 
  \quad\text{and}\quad
  \mathbf{F}_1 = \mathbf{W}_{L}\mathbf{P}_L^{(-L/4)}\mathbf{S}\mathbf{W}_{L}^{-1}
\end{align}
\end{subequations}
are circular convolution matrices. 

After some manipulations \eqref{eq:FCcomb} can be reformulated as
\begin{subequations}
  \label{eq:reduc}
  \begin{equation}
  \mathbf{\hat{x}}_{m,n} = 
    \boldsymbol{\Omega}_{-L/4}[\mathbf{F}(\mathbf{g}_{m,2n} - 
    \boldsymbol{\Omega}_{L/2}\mathbf{g}_{m,2n+1}) +
    \boldsymbol{\Omega}_{L/2}\mathbf{g}_{m,2n+1}],
\end{equation}
where 
\begin{equation}
  \label{eq:reduc_sub}
  \mathbf{F} = 
  \boldsymbol{\Omega}_{L/4}
  \mathbf{C}
  \boldsymbol{\Omega}_{-L/4}
 \end{equation} 
 with
 \begin{equation}
   \label{eq:halfband}
   \mathbf{C}=
   \mathbf{W}_L
   \mathbf{\hat{S}}  
   \mathbf{W}_L^{-1}
 \end{equation} 
 and
\begin{equation}
  \boldsymbol{\Omega}_\phi = \diag*{
  \begin{bmatrix}
    W_L^0 & W_L^\phi & \cdots & W_L^{\phi(L-1)}
  \end{bmatrix}^\transpose}
\end{equation}
with $W_{L}=\exp(-\iu{2}\pi/L)$. Here, $\mathbf{\hat{S}}\in\mathbb{Z}^{L\times L}$ as given by
\begin{equation}
  \mathbf{\hat{S}} = 
  \mathbf{P}_L^{(L/4)}  
  \mathbf{S} 
  \mathbf{P}_L^{(-L/4)}  
\end{equation}
\end{subequations}
is the time-domain synthesis window matrix circularly left and up shifted by $L/4$ samples.

\begin{figure}[!t]         
  \centering   
  \includegraphics[width=\figwidthSC]{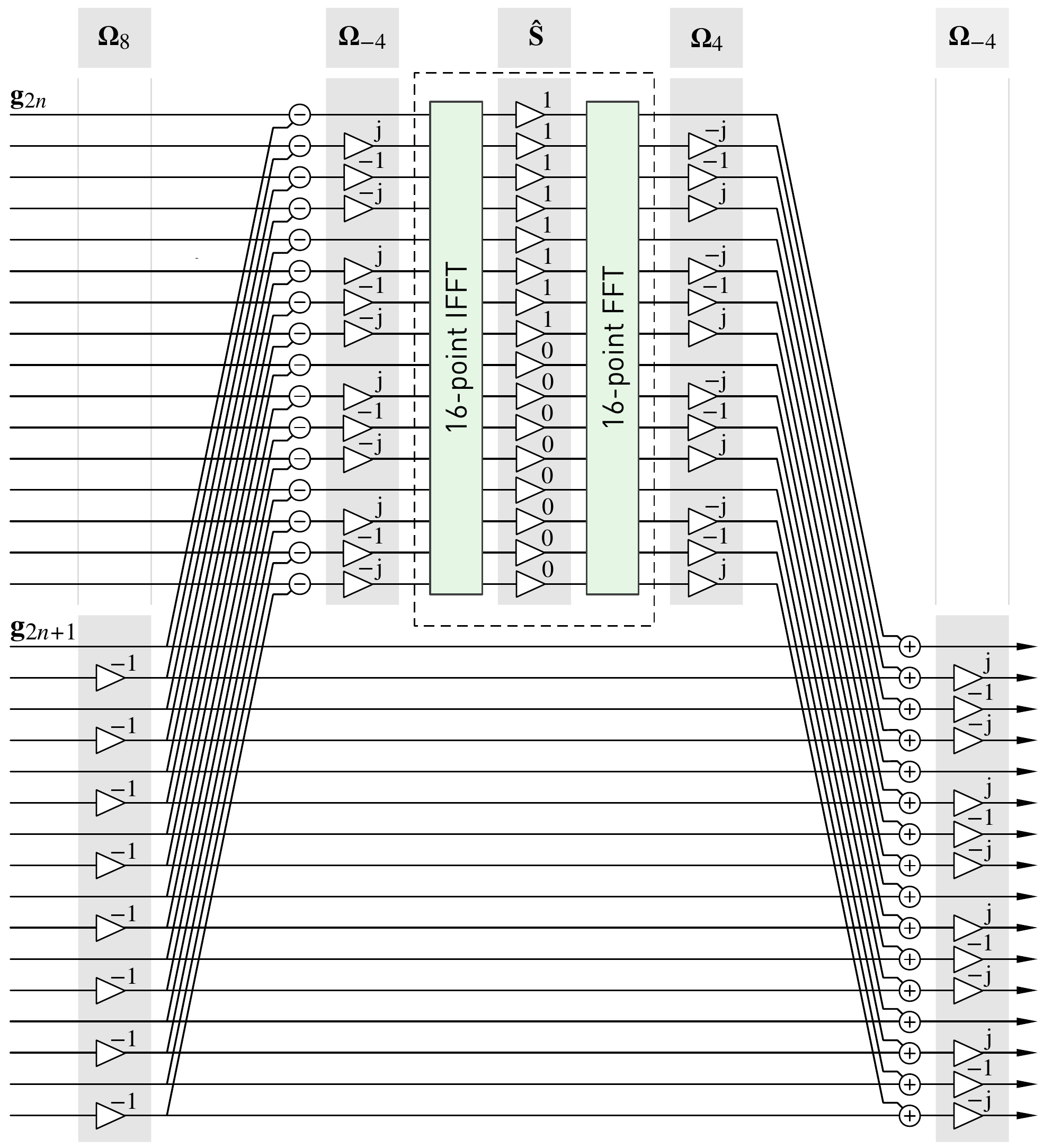} 
  \caption{Simplified discontinuous processing  for $L=16$ as described by \eqref{eq:reduc}. The dashed region can be further simplified by decomposing transforms into smaller transforms.}
  \label{fig:SimpliefiedL16}
\end{figure}

The above formulation relies on rotating the $\mathbf{g}_{m,2n+1}$ and the corresponding window by $L/2$ samples. In this case, $\mathbf{F}_0$ and $\mathbf{F}_1$ form a lowpass-highpass filter pair essentially meaning that $\mathbf{F}_0\mathbf{F}_1=\mathbf{I}_L$.  The low-rate complexity of \eqref{eq:reduc} per \ac{ofdm} symbol is one \ac{ifft} and one \ac{fft} of length $L$ as depicted in Fig. \ref{fig:SimpliefiedL16}. The direct approach, as illustrated in Fig.~\ref{fig:SimplBlockProcRX}, requires two \acp{ifft} and one \ac{fft} of length $L$, that is, the saving in number of real multiplications for this simplified processing is \SI{33.3}{\%}. In \eqref{eq:complexity}, $\beta=1/2$ gives the complexity of the simplified processing whereas $\beta=1$ gives the complexity of the direct approach.  The half-band filtering provided by \eqref{eq:halfband} can be further decomposed into smaller transforms to achieve some savings in implementation, however, these decompositions are beyond the scope of this paper.

\section{Numerical results} 
\label{sec:numerical-results} 
In this section, we will analyze the performance of the discontinuous \ac{fc} processing in terms of uncoded \ac{ber} in different interference and channel conditions, and also show complexity comparison between continuous and discontinuous \ac{fc} processing. Here we assume the overlap factor of $\lambda=0.5$. Continuous \ac{fc} processing with the overlap of $\lambda=0.25$ marginally degrades the spectral containment and \ac{evm} performance compared to the overlap of $\lambda=0.5$ with the benefit of somewhat lower implementation complexity.
  
\ifhbonecolumn 
  \begin{table*}[t]
  \caption{Narrow-band transmission scenarios and filtering configurations}
  \label{tab:conf-narr}
  \centering
  \vspace{-2.8em}
  \footnotesize{
   \begin{tabular}[t]{@{}ccccccccccc@{}}  
     \toprule
     \multicolumn{1}{@{}m{3cm}}{\raisebox{2.95em}{\emph{Channel model}}}   & 
     \multicolumn{5}{m{6.1cm}}{\raisebox{2.95em}{Additive white Gaussian noise (AWGN)}} &
     \multicolumn{5}{m{6.1cm}@{}}{Tapped-delay line (TDL)-C with \SI{300}{ns} and \SI{1000}{ns} root mean squared (RMS) channel delay spread} \\
     \midrule
     \multicolumn{1}{@{}m{3cm}}{\raisebox{4.75em}{\emph{Synchronicity}}} & 
     \multicolumn{5}{m{6.1cm}}{\raisebox{4.75em}{\parbox[t]{6cm}{Quasi-synchronous: No timing offset and no frequency offset between different uplink signals}}} &
     \multicolumn{5}{m{6.1cm}@{}}{Asynchronous: Timing offset of 256 samples ($L_{\text{OFDM},m}/4$) between the target subband for \ac{ber} evaluation and adjacent subbands on both sides} \\
     \midrule
     \multicolumn{1}{@{}m{3cm}}{\raisebox{2.95em}{\emph{Allocated subband width}}} &     
     \multicolumn{5}{m{6.1cm}}{\SI{1}{\prb}, 12 \acp{sc}: 8 active \acp{sc} and 4-\ac{sc} guardbands between adjacent active subbands} & 
     \multicolumn{5}{m{6.1cm}@{}}{\raisebox{2.95em}{\parbox[t]{6cm}{\SI{4}{\prbs}, 48 \acp{sc}: 44 active \acp{sc} and 4-\ac{sc} guardband between adjacent active subbands}}} \\ 
     \midrule 
     \multicolumn{1}{@{}m{3cm}}{\raisebox{1.6em}{\emph{Filtering configuration}}} &
     \multicolumn{2}{m{4.0cm}}{\raisebox{1.6em}{\parbox[t]{3.80cm}{1) No filtering on \ac{tx} and \ac{rx} sides}}} &
     \multicolumn{2}{m{4.0cm}}{\raisebox{1.6em}{\parbox[t]{3.80cm}{2) No \ac{tx} filtering, \ac{rx} filtering with continuous \ac{fc} model with $L_{\text{RX}}=128$}}} &
     \multicolumn{2}{m{4.0cm}}{\raisebox{1.6em}{\parbox[t]{3.80cm}{3) Continuous \ac{tx} filtering, continuous \ac{rx} filtering, $L_{\text{TX}}=L_{\text{RX}}=128$}}} \\[2pt]
     \noalign{\smallskip}
     \cline{2-11}
     & 
     \multicolumn{2}{m{4.0cm}}{\raisebox{0.0em}{\parbox[t]{3.80cm}{4) Discontinuous \ac{tx} filtering, continuous \ac{rx} filtering, $L_{\text{TX}}=L_{\text{RX}}=128$}}} &
     \multicolumn{2}{m{4.0cm}}{\parbox[t]{3.80cm}{5) Discontinuous \ac{tx} filtering, continuous \ac{rx} filtering, $L_{\text{TX}}=16$ with \SI{1}{\prb}, $L_{\text{TX}}=64$  with \SI{4}{\prbs}, and $L_{\text{RX}}=128$}} &
     \multicolumn{2}{m{4.0cm}}{\parbox[t]{3.80cm}{6) Discontinuous \ac{tx} and \ac{rx} filtering, $L_{\text{TX}}=16$ with \SI{1}{\prb}, $L_{\text{TX}}=64$  with \SI{4}{\prbs}, and $L_{\text{RX}}=128$}}
     \\[2pt]
     \bottomrule
   \end{tabular}}
  \end{table*}
\else
  \begin{table*}[t]
  \caption{Narrow-band transmission scenarios and filtering configurations}
  \label{tab:conf-narr}
  \centering
  \vspace{-1.8em}
  \footnotesize{
   \begin{tabular}[t]{@{}ccccccccccc@{}}  
     \toprule
     \multicolumn{1}{@{}m{3cm}}{\raisebox{1.15em}{\emph{Channel model}}}   & 
     \multicolumn{5}{m{7.1cm}}{\raisebox{1.15em}{Additive white Gaussian noise (AWGN)}} &
     \multicolumn{5}{m{7.1cm}@{}}{Tapped-delay line (TDL)-C with \SI{300}{ns} and \SI{1000}{ns} root mean squared (RMS) channel delay spread} \\
     \midrule
     \multicolumn{1}{@{}m{3cm}}{\raisebox{2.4em}{\emph{Synchronicity}}} & 
     \multicolumn{5}{m{7.1cm}}{\raisebox{2.4em}{\parbox[t]{7cm}{Quasi-synchronous: No timing offset and no frequency offset between different uplink signals}}} &
     \multicolumn{5}{m{7.1cm}@{}}{Asynchronous: Timing offset of 256 samples ($L_{\text{OFDM},m}/4$) between the target subband for \ac{ber} evaluation and adjacent subbands on both sides} \\
     \midrule
     \multicolumn{1}{@{}m{3cm}}{\raisebox{1.15em}{\emph{Allocated subband width}}} &     
     \multicolumn{5}{m{7.1cm}}{\SI{1}{\prb}, 12 \acp{sc}: 8 active \acp{sc} and 4-\ac{sc} guardbands between adjacent active subbands} & 
     \multicolumn{5}{m{7.1cm}@{}}{\SI{4}{\prbs}, 48 \acp{sc}: 44 active \acp{sc} and 4-\ac{sc} guardband between adjacent active subbands} \\ 
     \midrule 
     \multicolumn{1}{@{}m{3cm}}{\raisebox{1.6em}{\emph{Filtering configuration}}} &
     \multicolumn{2}{m{4.6cm}}{\raisebox{1.6em}{\parbox[t]{4.55cm}{1) No filtering on \ac{tx} and \ac{rx} sides}}} &
     \multicolumn{2}{m{4.6cm}}{\raisebox{1.6em}{\parbox[t]{4.55cm}{2) No \ac{tx} filtering, \ac{rx} filtering with continuous \ac{fc} model with $L_{\text{RX}}=128$}}} &
     \multicolumn{2}{m{4.6cm}}{\raisebox{1.6em}{\parbox[t]{4.55cm}{3) Continuous \ac{tx} filtering, continuous \ac{rx} filtering, $L_{\text{TX}}=L_{\text{RX}}=128$}}} \\[2pt]
     \cline{2-11}
     \noalign{\smallskip}
     & 
     \multicolumn{2}{m{4.6cm}}{\raisebox{0.0em}{\parbox[t]{4.55cm}{4) Discontinuous \ac{tx} filtering, continuous \ac{rx} filtering, $L_{\text{TX}}=L_{\text{RX}}=128$}}} &
     \multicolumn{2}{m{4.6cm}}{\parbox[t]{4.55cm}{5) Discontinuous \ac{tx} filtering, continuous \ac{rx} filtering, $L_{\text{TX}}=16$ with \SI{1}{\prb}, $L_{\text{TX}}=64$  with \SI{4}{\prbs}, and $L_{\text{RX}}=128$}} &
     \multicolumn{2}{m{4.6cm}}{\parbox[t]{4.55cm}{6) Discontinuous \ac{tx} and \ac{rx} filtering, $L_{\text{TX}}=16$ with \SI{1}{\prb}, $L_{\text{TX}}=64$  with \SI{4}{\prbs}, and $L_{\text{RX}}=128$}}
     \\
     \bottomrule
   \end{tabular}}
  \end{table*}
\fi

\begin{figure*}[!ht]            
  \centering      
  \includegraphics[width=0.49\textwidth]{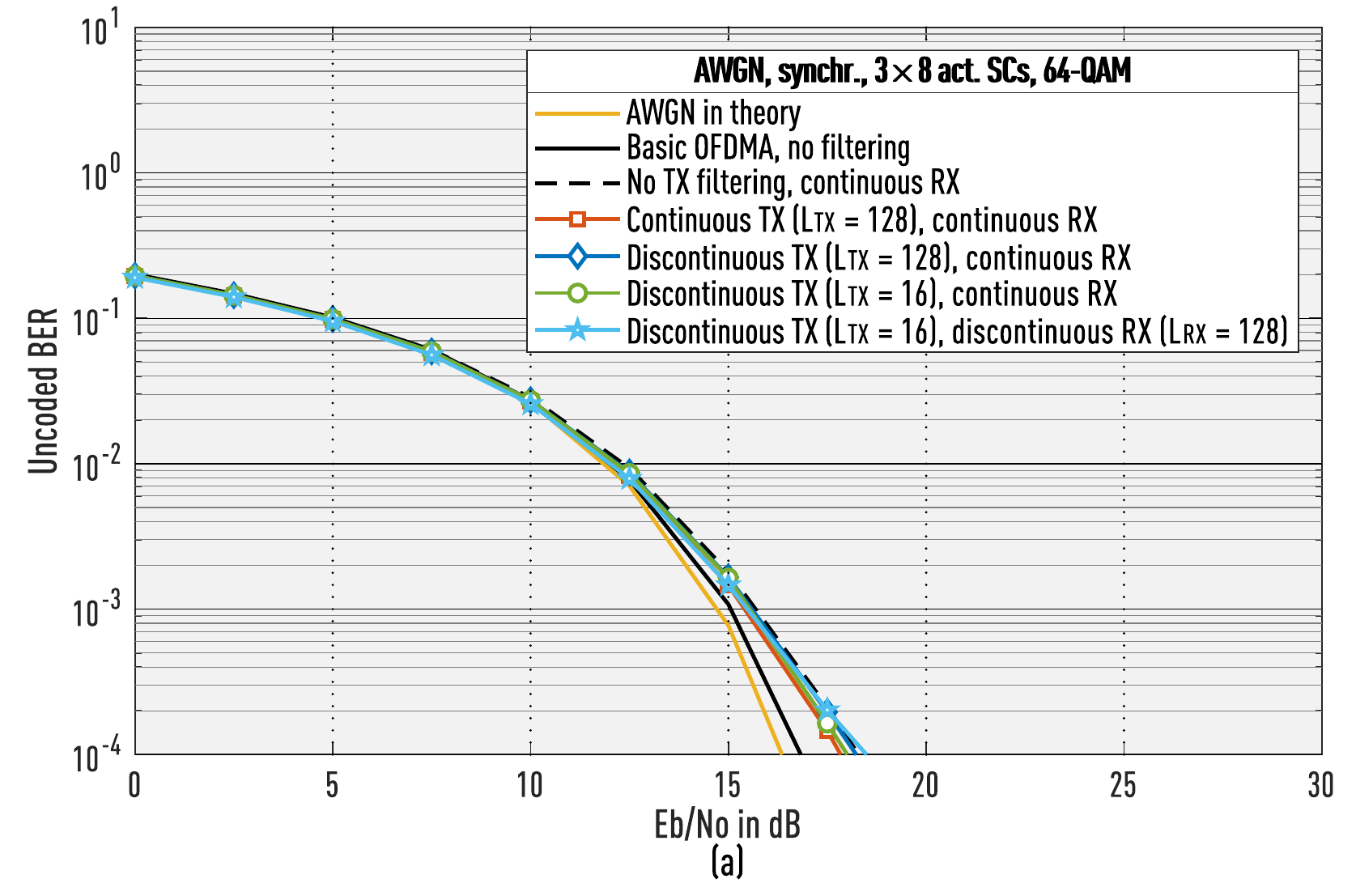} 
  \includegraphics[width=0.49\textwidth]{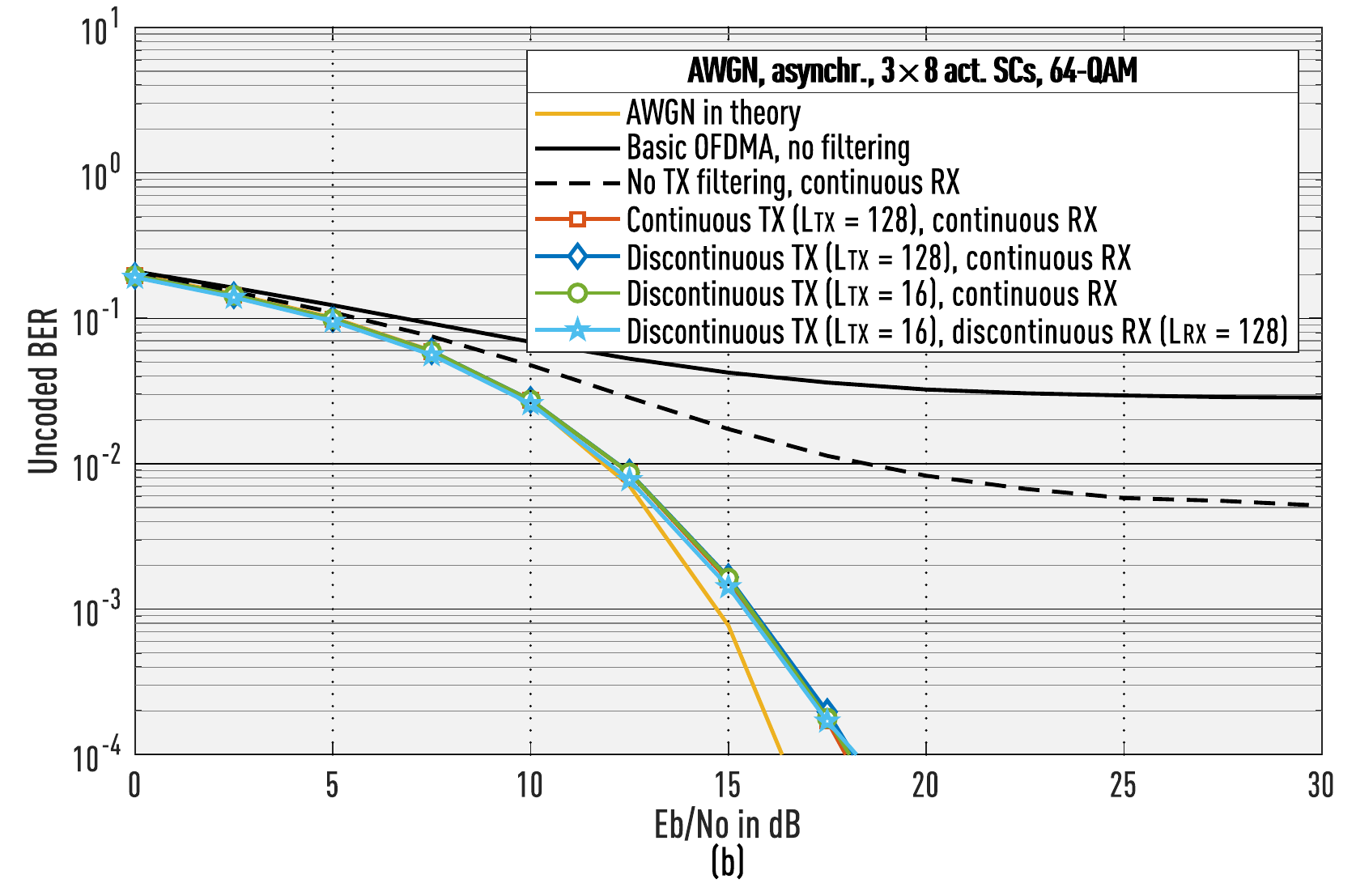}
  \caption{Simulation results with \acs{awgn} channel and 64-\ac{qam} modulation. 
(a) Quasi-synchronous cases and (b) asynchronous cases (quarter-symbol timing offset). 
Single \ac{prb} allocated for each of the three subbands (12 \acp{scs}, 8 active \acp{sc}, and 4 \acp{sc} for guardband).}
  \label{fig:simResAWGN_narr}        
\end{figure*}        

\subsection{Bit-Error Rate Performance in Narrow-band Allocations}   
\label{sec:bit-error-rate}
\begin{figure*}[!ht]          
  \centering   
  \includegraphics[width=0.49\textwidth]{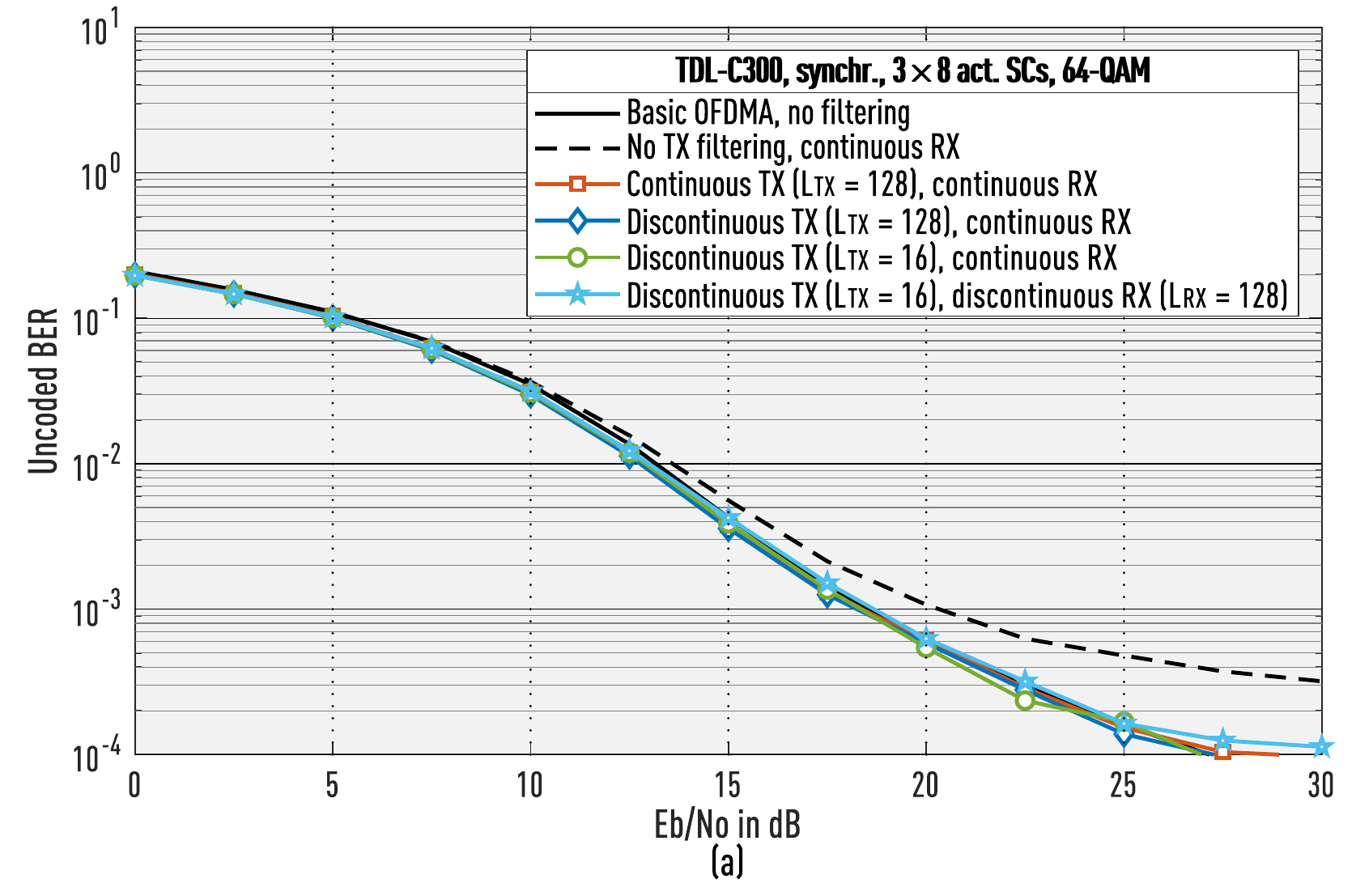}
  \includegraphics[width=0.49\textwidth]{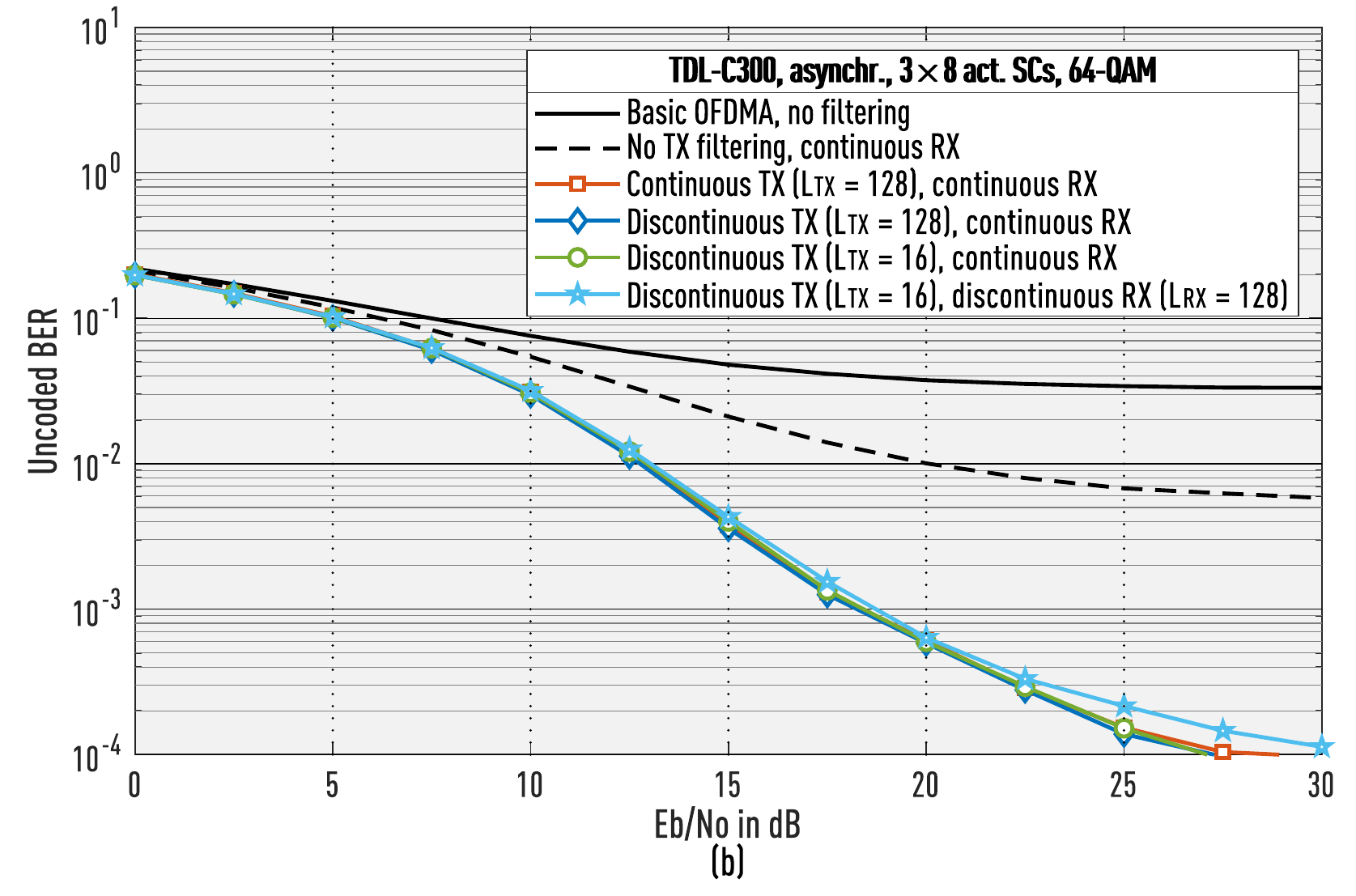}
  \caption{Performance simulation results with \ac{tdl}-C \SI{300}{ns} channel and 64-\ac{qam} modulation. 
(a) Quasi-synchronous cases and (b) asynchronous cases (quarter-symbol timing offset).
Single \ac{prb} allocated for each subband (12 \acp{scs}, 8 active \acp{sc}, and 4 \acp{sc} for guardband).}
  \label{fig:simResTDL-C300_narr} 
\end{figure*}

\begin{figure*}[!ht]          
  \centering   
  \includegraphics[width=0.49\textwidth]{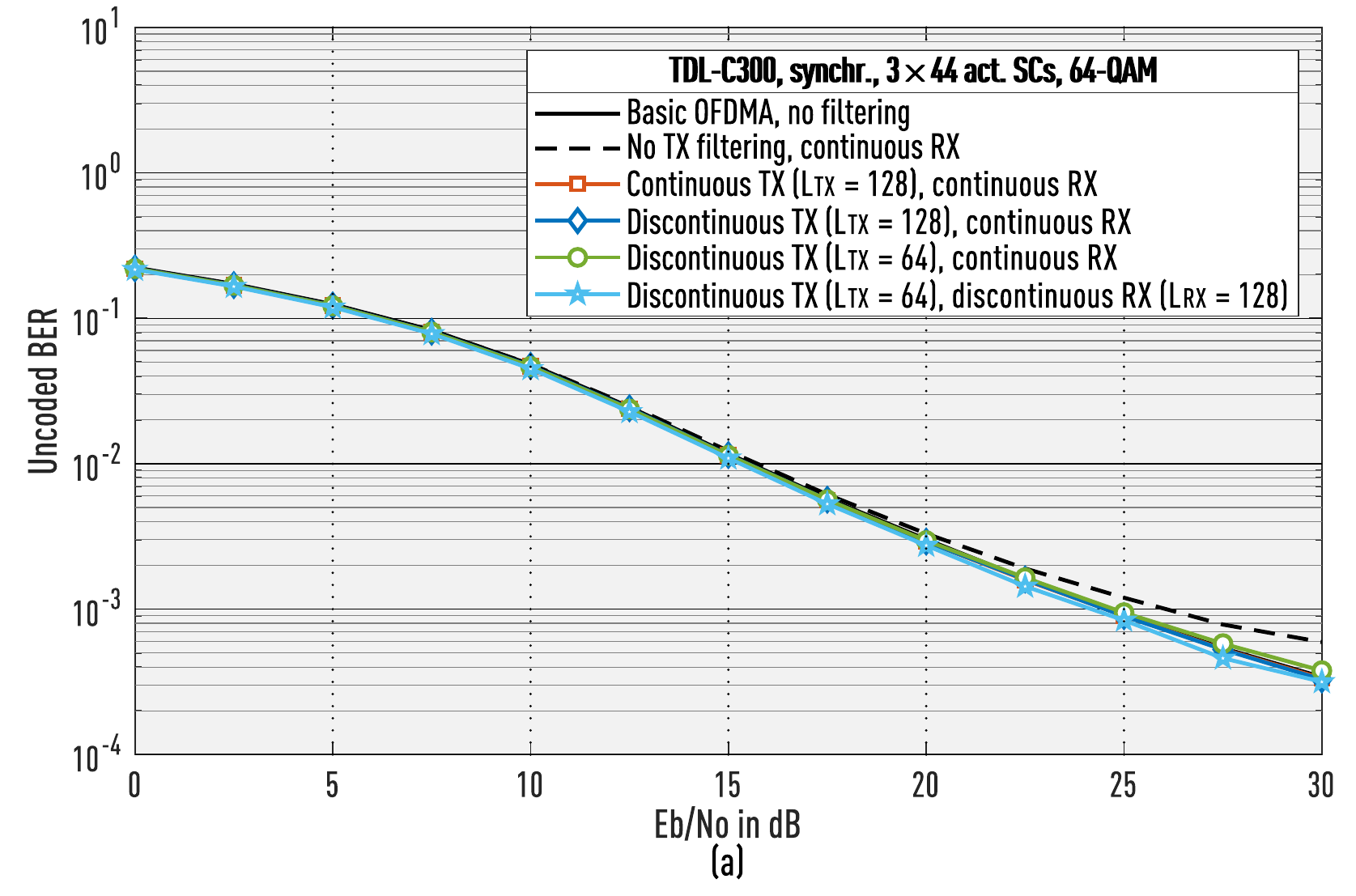}
  \includegraphics[width=0.49\textwidth]{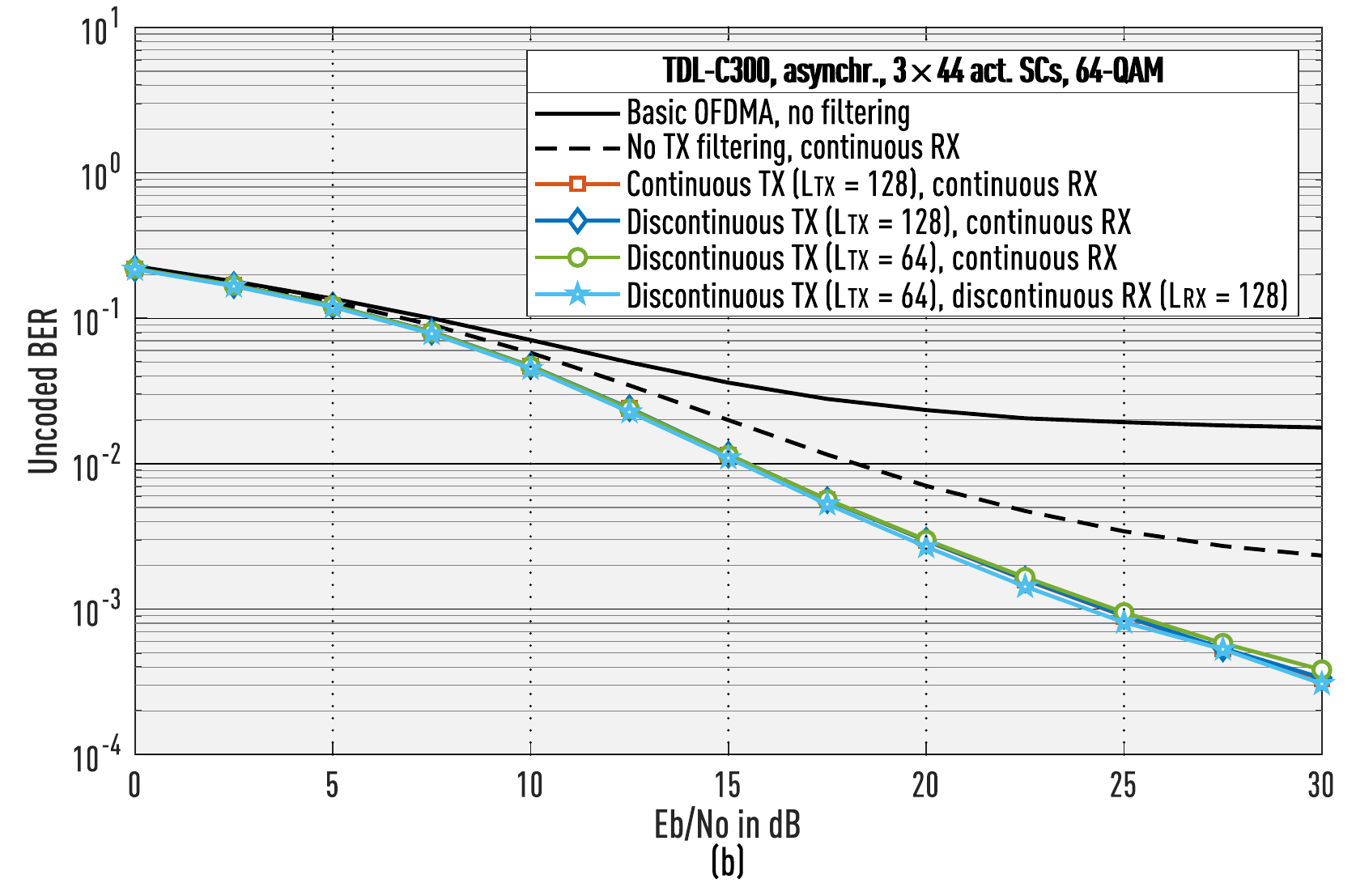}
  \caption{Performance simulation results with \ac{tdl}-C \SI{300}{ns} channel and 64-\ac{qam} modulation. 
(a) Quasi-synchronous cases and (b) asynchronous cases (quarter-symbol timing offset).
Four \acp{prb} allocated for each subband (48 \acp{sc}, 44 active \acp{sc}, and 4 \acp{sc} for guardband).}
  \label{fig:simResTDL-C300_wide}       
\end{figure*}

\begin{figure*}[!ht]          
  \centering   
  \includegraphics[width=0.49\textwidth]{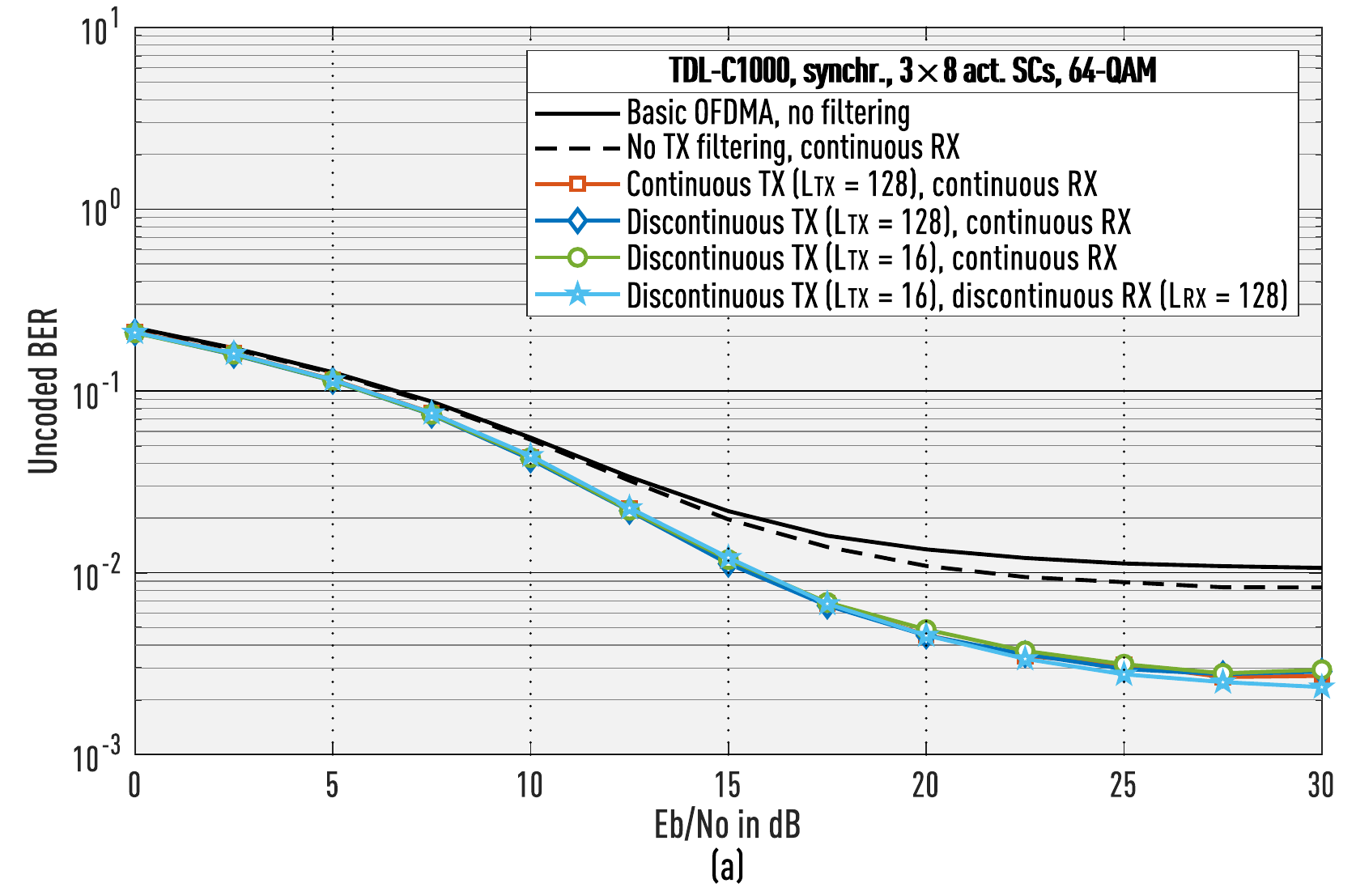}
  \includegraphics[width=0.49\textwidth]{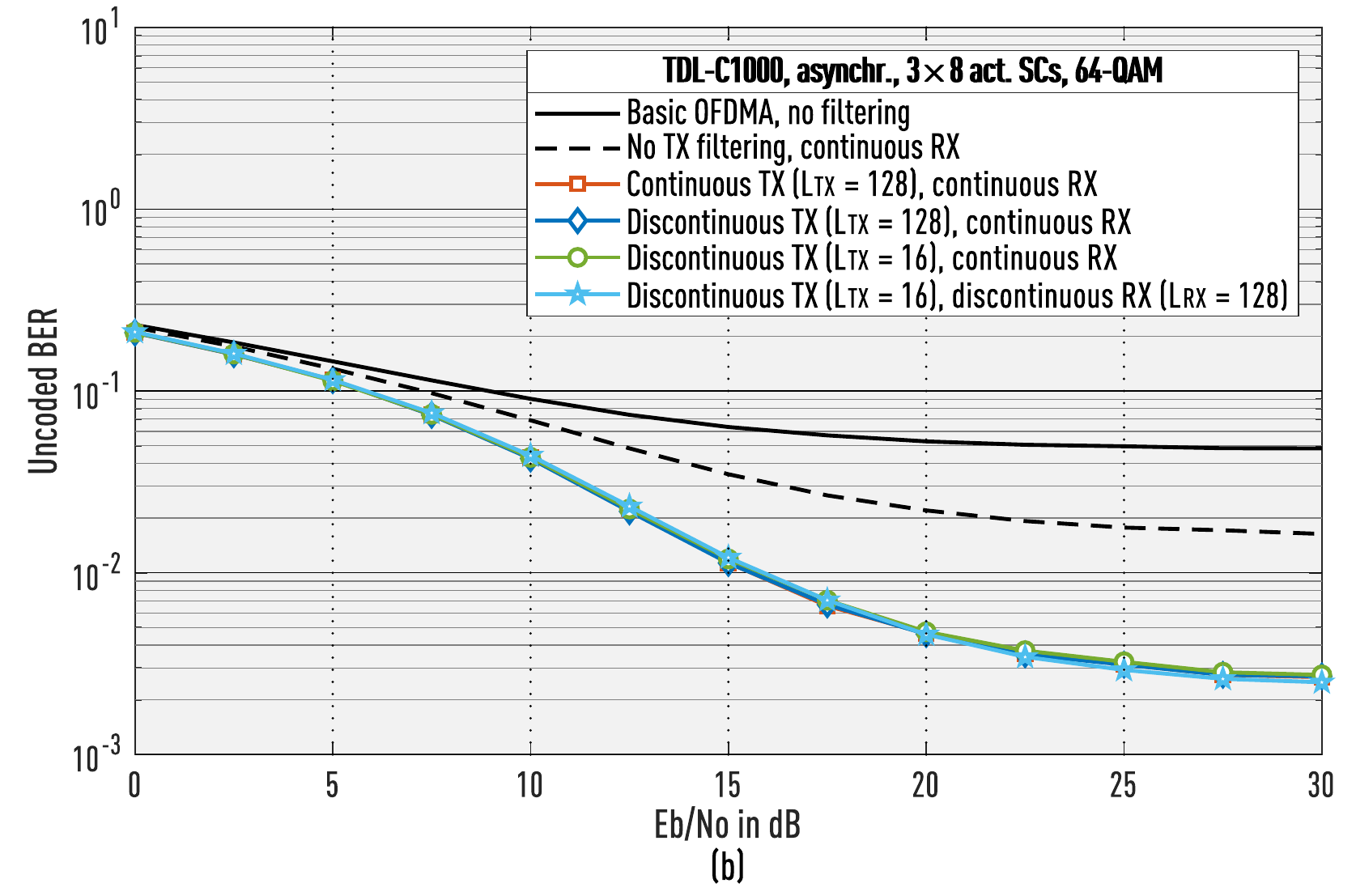}
  \caption{Performance simulation results with \ac{tdl}-C \SI{1000}{ns} channel and 64-\ac{qam} modulation. 
(a) Quasi-synchronous cases and (b) asynchronous cases (quarter-symbol timing offset).
Single \ac{prb} allocated for each subband (12 \acp{sc}, 8 active \acp{sc}, and 4 \acp{sc} guardband)}
  \label{fig:simResTDL-C1000_narr}      
\end{figure*}

\begin{figure*}[!ht]           
  \centering   
  \includegraphics[width=0.49\textwidth]{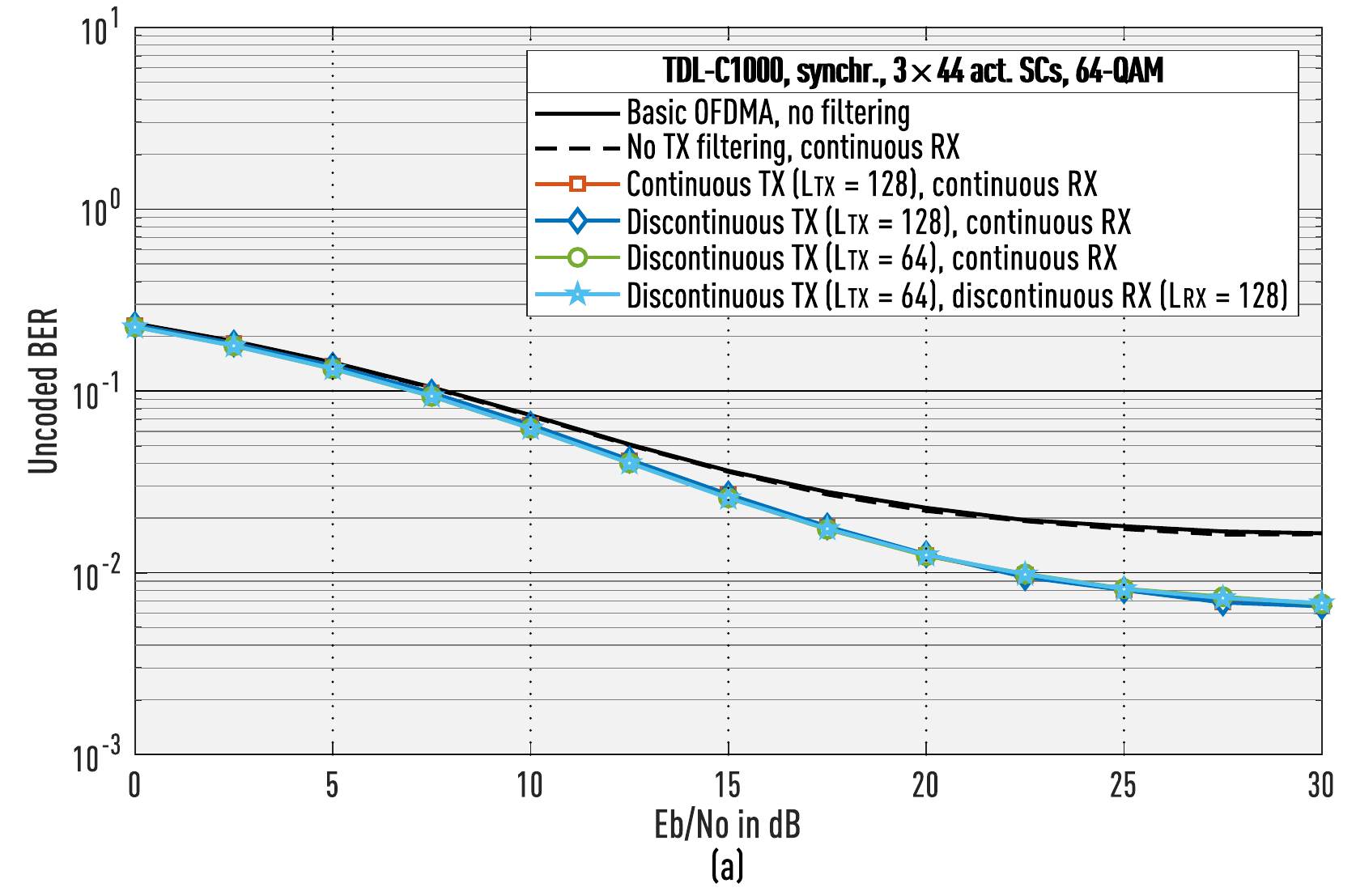}
  \includegraphics[width=0.49\textwidth]{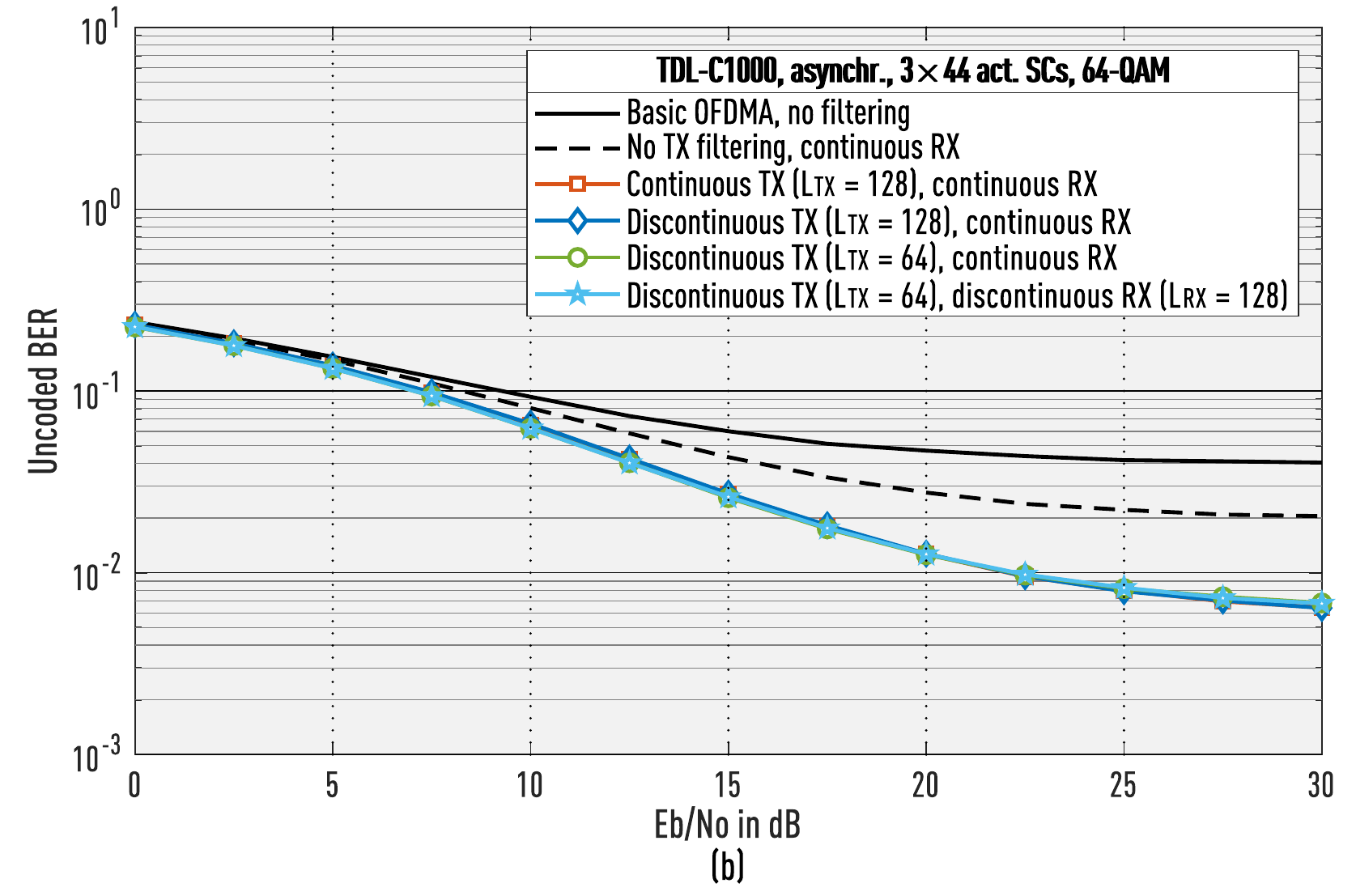}
  \caption{Performance simulation results with \ac{tdl}-C \SI{1000}{ns} channel and 64-\ac{qam} modulation. 
(a) Quasi-synchronous cases and (b) asynchronous cases (quarter-symbol timing offset).
Single \ac{prb} allocated for each subband (48 \acp{sc}, 44 active \acp{sc}, and 4 \acp{sc} for guardband)}
  \label{fig:simResTDL-C1000_wide}           
\end{figure*}

Figs.~\ref{fig:simResAWGN_narr}--\ref{fig:simResTDL-C1000_wide} compare the simulated uncoded \ac{ber} performance of different \ac{cp-ofdm} configurations, with or without subband filtering in \SI{10}{MHz} \ac{5g-nr} uplink scenario, with  high-rate \ac{ifft} size of $N=1024$ and \ac{scs} of \SI{15}{kHz}. Table~\ref{tab:conf-narr} shows details of  the considered scenarios and filtering configurations. The used channel models are \ac{awgn} and \ac{tdl}-C, which is one of the channel models considered in the \ac{5g-nr} development \cite{S:3GPP:TR38.900v150}. Two different values of the \ac{rms} channel delay spread, \SI{300}{ns} and \SI{1000}{ns}, are used for \ac{tdl}-C channel. Two subband configurations are considered: single \ac{prb} or four \acp{prb} of \num{12} subcarriers. In both cases, four deactivated subcarriers are used as for guard bands. Focusing on the asynchronous up-link scenario, different instances of the channel model are always used for the three adjacent subbands included in the simulations. Perfect power control is assumed in such a way that the three adjacent subbands are always received at the same power level and constant \acs{snr} for all channel instances.

Fig.~\ref{fig:simResAWGN_narr}(a) shows the \ac{ber} simulation results for \ac{awgn} channel in synchronous scenario with single \ac{prb} per subband. In this case, the performance of the \ac{tx} filtered continuous and discontinuous approaches and the \ac{rx}-filtering-only scheme are practically the same.  When compared with the basic synchronous \ac{ofdm}, the filtered schemes have a minor performance loss.  In asynchronous scenario, as depicted in Fig.~\ref{fig:simResAWGN_narr}(b), the performance of \ac{tx} filtered schemes remains close to theoretical one whereas the \ac{ber} floor for \ac{rx}-filtering-only and basic \ac{ofdm} schemes is about $\SI{0.5}{\%}$ and $\SI{3}{\%}$, respectively. 
Fig.~\ref{fig:simResTDL-C300_narr}(a) shows the \ac{ber} simulation results for \ac{tdl}-C \SI{300}{ns} channel in synchronous scenario with single \ac{prb} per subband. In this case, the channel maximum delay spread (about \SI{2.6}{\mics}) is well below the \ac{cp} length (about \SI{4.7}{\mics}). When comparing with basic synchronous \ac{ofdm}, we can see minor performance degradation of the schemes with filtering at both ends, while the degradation of the \ac{rx}-filtering-only scheme is more visible. In the asynchronous case, the benefits of subband-filtered \ac{ofdm} are clearly visible as illustrated in Fig.~\ref{fig:simResTDL-C300_narr}(b).

Fig.~\ref{fig:simResTDL-C300_wide}(a) compares the performance for \ac{tdl}-C \SI{300}{ns} channel in synchronous scenario with four \acp{prb} per subband. Now the performance difference between the schemes has been decreased due to the fact that, on the average, the wider subband suffer less from the interference between the subbands. The same trend can also be seen from simulation results of asynchronous scenario as shown in Fig.~\ref{fig:simResTDL-C300_wide}(b) where the performance degradation of \ac{rx}-filtering-only and basic \ac{ofdm} schemes is less obvious.
 
Fig.~\ref{fig:simResTDL-C1000_narr}(a) shows the \ac{ber} simulation results for \ac{tdl}-C \SI{1000}{ns} channel with single \ac{prb} per subband. In this case the channel maximum delay spread (about \SI{8.7}{\mics}) exceeds the \ac{cp} duration (about \SI{4.7}{\mics}), resulting in higher error floor in all configurations. The same conclusions can be made as above, except that the \ac{tx} filtered schemes have now better performance than the basic \ac{ofdm}. The performance degradation of basic \ac{ofdm} scheme is due to the \ac{ici} induced by the increased frequency dispersion of the channel whereas, for \ac{tx} filtered \ac{ofdm}, the better spectral containment provides also better protection against the \ac{ici} \cite{J:Strohmer03,J:Zhao16:PulseShapingDesign}. For asynchronous scenario, as illustrated in Fig.~\ref{fig:simResTDL-C1000_narr}(b), the performance of \ac{tx} filtered schemes are approximately the same as in synchronous scenario whereas the basic \ac{ofdm} and \ac{rx}-filtering-only schemes have considerably higher error floors than the synchronous cases. 
 
The \ac{ber} simulation results for \ac{tdl}-C \SI{1000}{ns} channel with four \acp{prb} per subband are shown in Figs.~\ref{fig:simResTDL-C1000_wide}(a) and (b) for synchronous and asynchronous scenarios, respectively. As seen for these figures, the performance improvement of the \ac{fc} filtered waveforms remain consistent with other results.

\ifhbonecolumn 
\begin{table}[t]  
  \caption{Wide-band transmission scenarios and filtering configurations}
  \label{tab:conf-wide}
  \centering 
  \vspace{-2.8em}   
  \footnotesize{ 
   \begin{tabular}[t]{@{}ccccc@{}}    
     \toprule
     \multicolumn{1}{@{}m{3.4cm}}{\raisebox{0em}{\emph{Channel models}}}   & 
     \multicolumn{2}{m{10.8cm}@{}}{\raisebox{0em}{\parbox[t]{9.75cm}{%
     AWGN and TDL-C with \SI{1000}{ns} RMS channel delay spread}}} \\
     \midrule 
     \multicolumn{1}{@{}m{3.4cm}}{\raisebox{0.0em}{\parbox[t]{3.25cm}{%
           \emph{Allocated subband width}}}} &     
     \multicolumn{2}{m{10.8cm}@{}}{\raisebox{0em}{\parbox[t]{9.75cm}{%
       \SI{52}{\prbs}, 624 active \acp{sc} and 8-\ac{sc} guardband around the active subband
     }}} \\
     \midrule 
     \multicolumn{1}{@{}m{3.4cm}}{\raisebox{0.0em}{\parbox[t]{3.25cm}{\emph{Filtering configuration}}}} & 
     \multicolumn{2}{m{10.8cm}@{}}{\raisebox{0em}{\parbox[t]{9.75cm}{
           1) Continuous filtering on \ac{tx} and \ac{rx} sides\\
           2) Continuous \ac{tx} and discontinuous \ac{rx} filtering\\
           3) Discontinuous \ac{tx} and continuous \ac{rx} filtering\\
           4) Discontinuous filtering on \ac{tx} and \ac{rx} sides
         }}}  \\
     \bottomrule
   \end{tabular}} 
\end{table} 
\else
\begin{table}[t]  
  \caption{Wide-band transmission scenarios and filtering configurations}
  \label{tab:conf-wide}
  \centering 
  \vspace{-1.8em}   
  \footnotesize{ 
   \begin{tabular}[t]{@{}ccccc@{}}    
     \toprule
     \multicolumn{1}{@{}m{2.4cm}}{\raisebox{0em}{\emph{Channel models}}}   & 
     \multicolumn{2}{m{5.8cm}@{}}{\raisebox{0em}{\parbox[t]{5.75cm}{%
     AWGN and TDL-C with \SI{1000}{ns} RMS channel delay spread}}} \\
     \midrule
     \multicolumn{1}{@{}m{2.4cm}}{\raisebox{1.15em}{\emph{Synchronicity}}} & 
     \multicolumn{2}{m{5.8cm}@{}}{\raisebox{1.15em}{\parbox[t]{5.75cm}{
           Quasi-synchronous: No timing offset and no frequency offset between different uplink signals
         }}} \\ 
     \midrule 
     \multicolumn{1}{@{}m{2.4cm}}{\raisebox{0.0em}{\parbox[t]{2.25cm}{%
           \emph{Allocated subband width}}}} &     
     \multicolumn{2}{m{5.8cm}@{}}{\raisebox{0em}{\parbox[t]{5.75cm}{%
       \SI{52}{\prbs}, 624 active \acp{sc} and 8-\ac{sc} guardband around the active subband
     }}} \\
     \midrule 
     \multicolumn{1}{@{}m{2.4cm}}{\raisebox{0.0em}{\parbox[t]{1.85cm}{\emph{Filtering configuration}}}} & 
     \multicolumn{2}{m{5.8cm}@{}}{\raisebox{0em}{\parbox[t]{5.75cm}{
           1) Continuous filtering on \ac{tx} and \ac{rx} sides\\
           2) Continuous \ac{tx} and discontinuous \ac{rx} filtering\\
           3) Discontinuous \ac{tx} and continuous \ac{rx} filtering\\
           4) Discontinuous filtering on \ac{tx} and \ac{rx} sides
         }}}  \\
     \bottomrule
   \end{tabular}} 
\end{table} 
\fi

\subsection{Bit-Error Rate Performance in Wide-band Allocations} 
\begin{figure*}[!ht]             
  \centering     
  \includegraphics[width=0.48\textwidth]{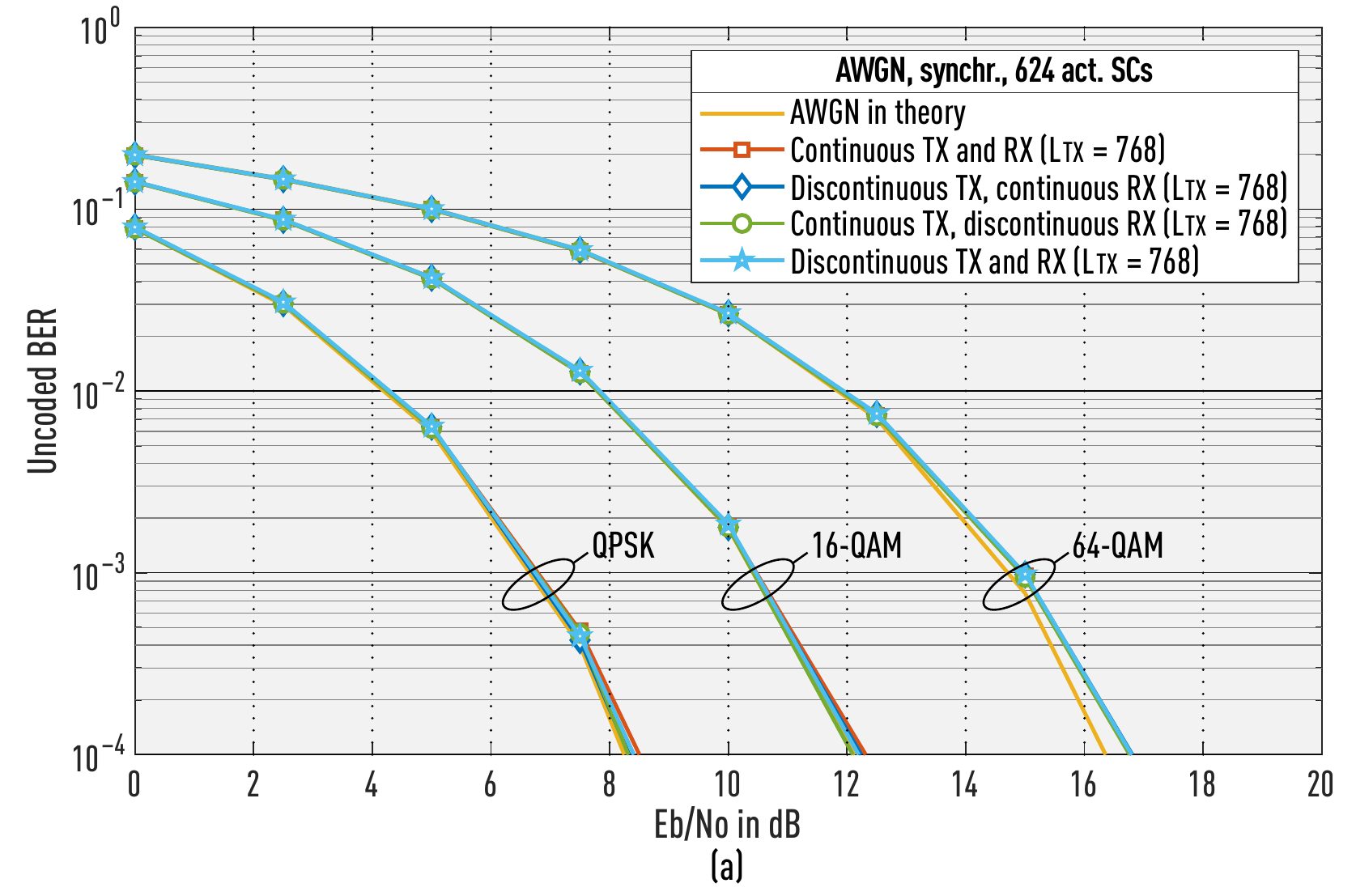}
  \raisebox{-3pt}{\includegraphics[width=0.50\textwidth]{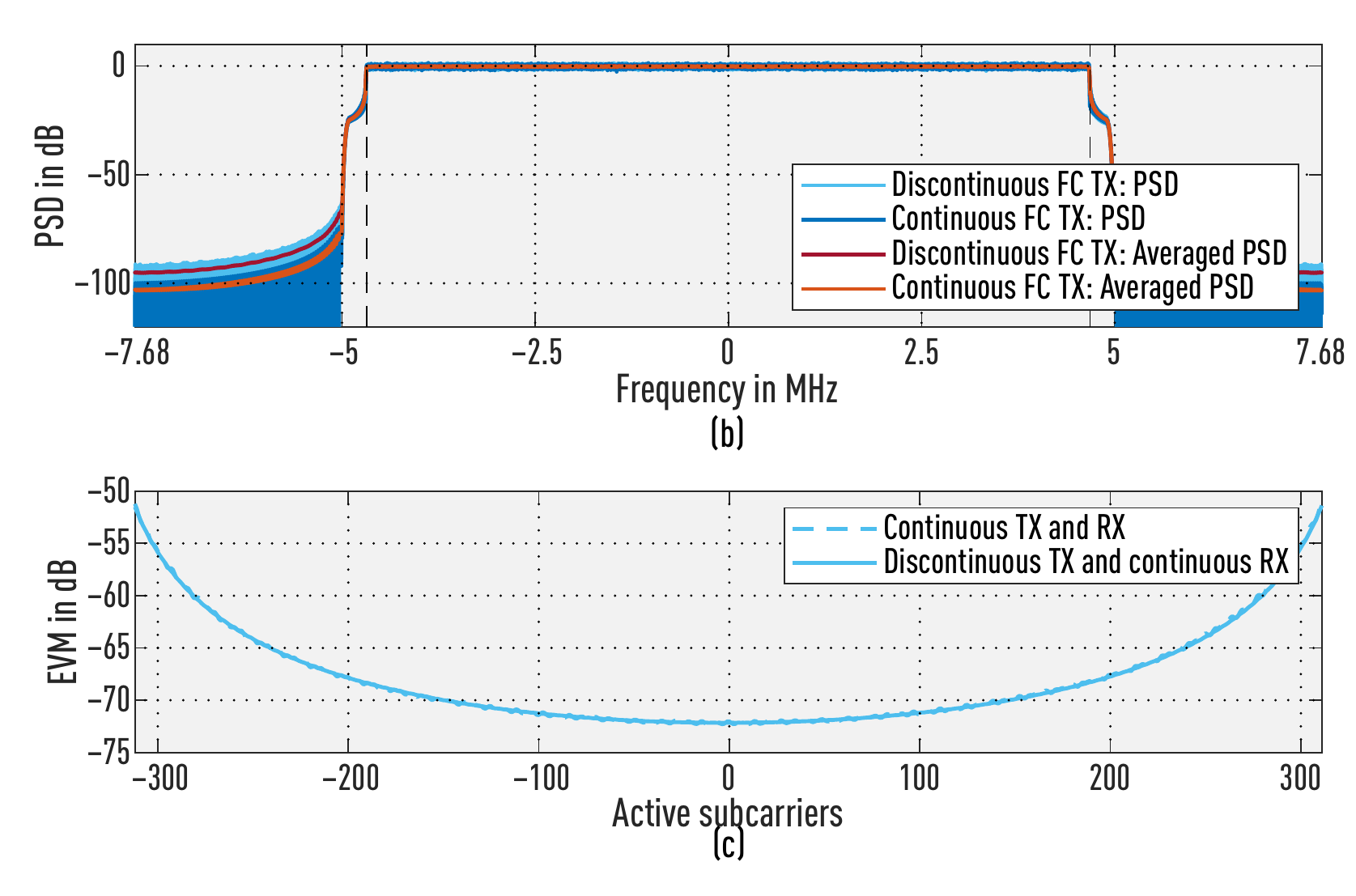}} 
  \caption{(a) Performance simulation results with \ac{awgn} channel and \acs{qpsk}, \num{16}-\ac{qam}, and \num{64}-\ac{qam} modulations. The number of \acp{prb} allocated for single subband is \num{52} (624 active \acp{sc} and 8 \acp{sc} for guardband). In-band region (active subcarriers) is denoted by the vertical dashed lines. (b) Power spectral densities for continuous and discontinuous processing. (c) Error-vector magnitude on active subcarriers with 64-\acs{qam} modulation.} 
  \label{fig:simResAWGN_624}      
\end{figure*}

\begin{figure*}[!ht]           
  \centering   
  \includegraphics[width=0.49\textwidth]{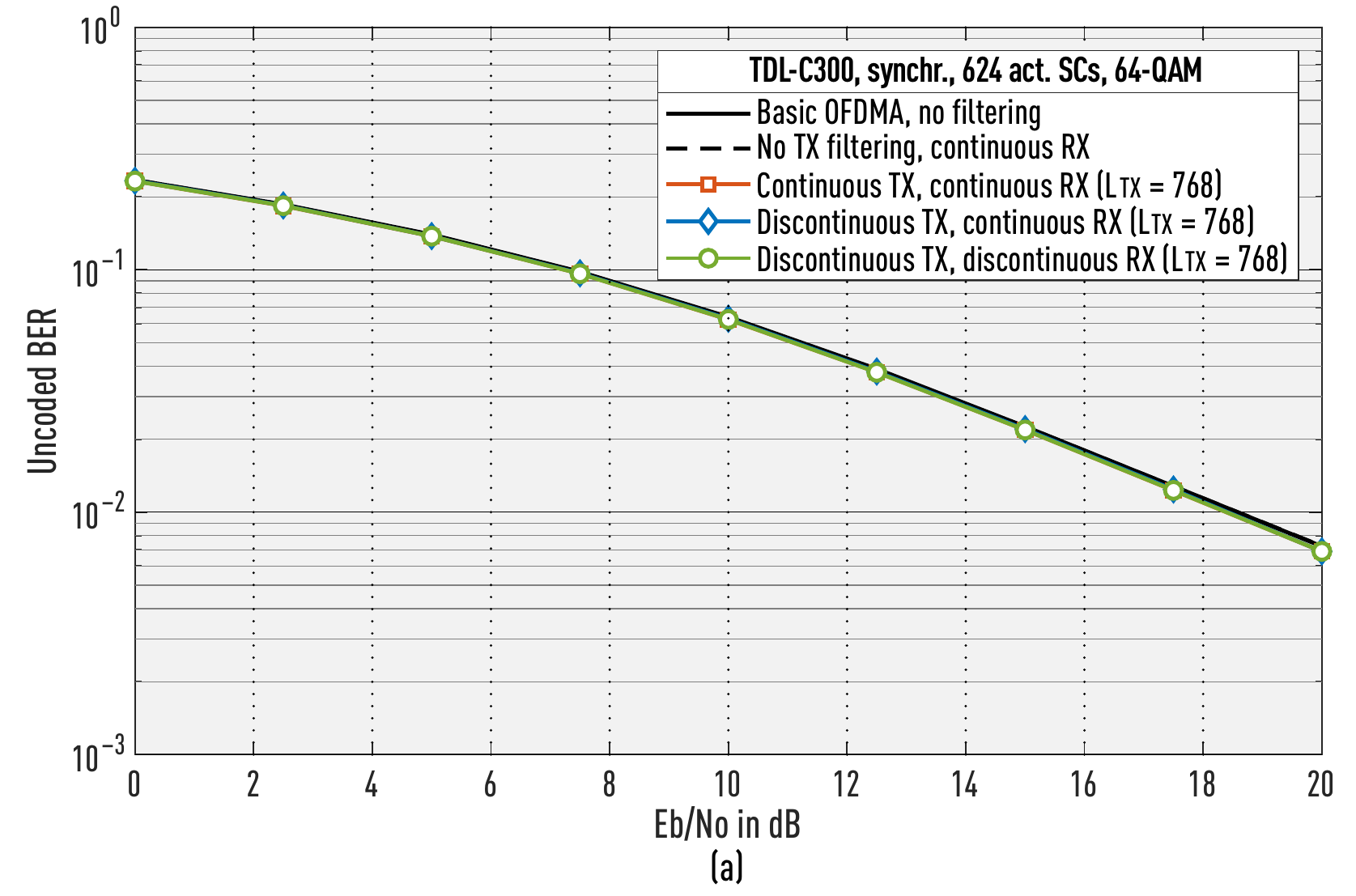}
  \includegraphics[width=0.49\textwidth]{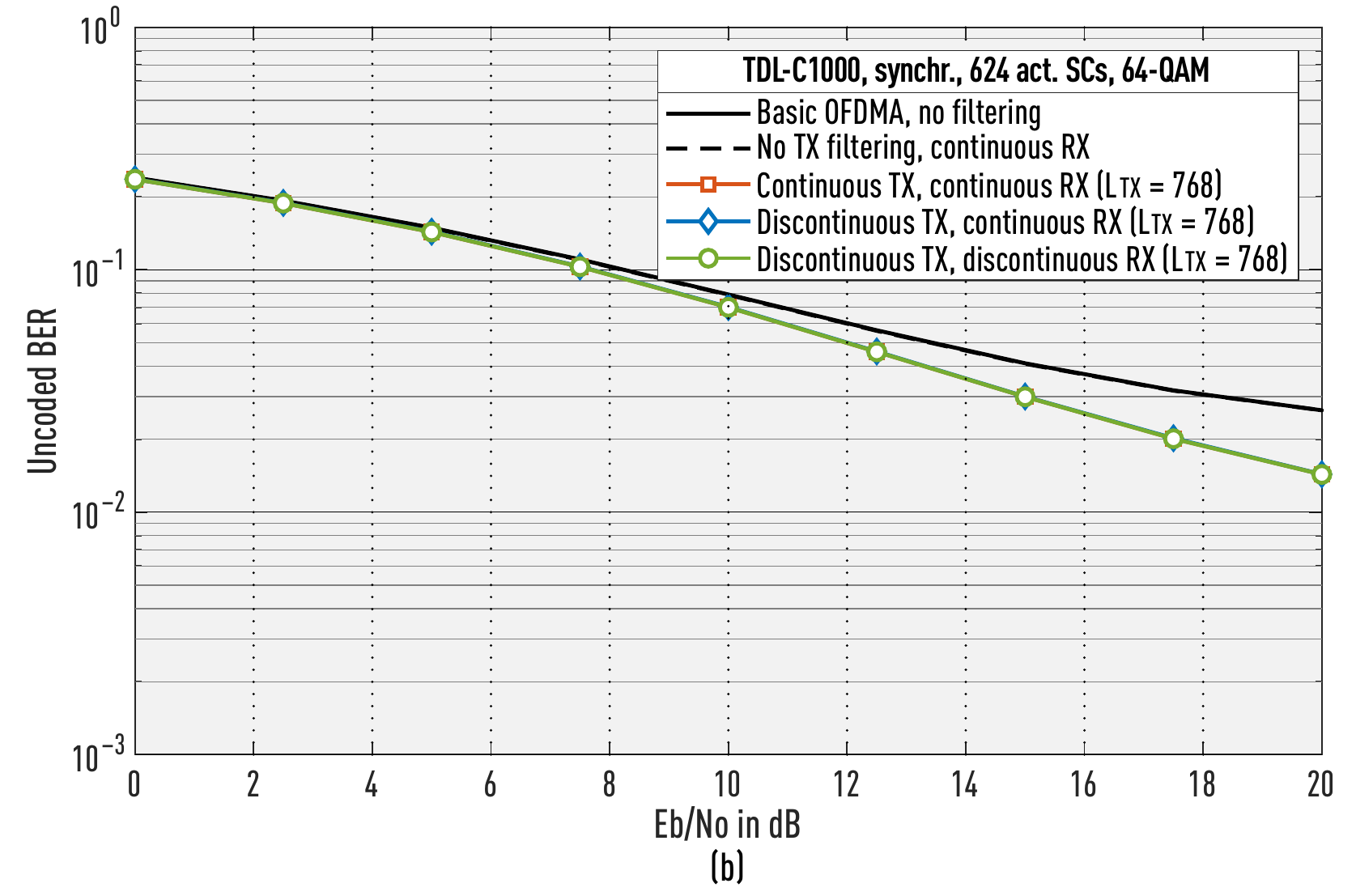}
  \caption{Performance simulation results with \ac{tdl}-C \SI{300}{ns} and \SI{1000}{ns} channels and 64-\ac{qam} modulation. (a) \ac{tdl}-C \SI{300}{ns} and (b) \SI{1000}{ns}. The number of \acp{prb} allocated for single subband is \num{52} (624 active \acp{sc} and 8 \acp{sc} for guardband). }
  \label{fig:simResTDL-C_wide}           
\end{figure*}

Fig.~\ref{fig:simResAWGN_624} and \ref{fig:simResTDL-C_wide} compare the simulated uncoded \ac{ber} performance of continuous and discontinuous subband filtering in \SI{10}{MHz} \ac{5g-nr} uplink scenario, with high-rate \ac{ifft} size of $N=1024$ and \ac{scs} of \SI{15}{kHz}. Single subband configuration is considered with 52 active PRBs of 12 subcarriers. In this case, eight non-active subcarriers are used as for transition bands. Table~\ref{tab:conf-wide} shows details of the considered scenarios and filtering configurations. The used channel models are \ac{awgn} and \ac{tdl}-C. Again two different values of the \ac{rms} channel delay spread, \SI{300}{ns} and \SI{1000}{ns}, are used for \ac{tdl}-C channel.

Fig.~\ref{fig:simResAWGN_624}(a) shows the uncoded \ac{ber} for \acs{qpsk}, \num{16}-\ac{qam}, and \num{64}-\ac{qam} modulations. As seen from this figure, both the continuous and discontinuous processing reach the theoretical \ac{ber} performance in \ac{awgn} channel. The \ac{psd}, as illustrated in Fig.~\ref{fig:simResAWGN_624}(b), is about \SI{5}{dB} lower at the out-of-band region for the continuous processing when compared with discontinuous processing. The average passband \ac{evm} for continuous and discontinuous processing are \SI{63.8}{dB} and \SI{63.4}{dB}, respectively.

The uncoded \ac{ber} performance in \ac{tdl}-C \SI{300}{ns} and \SI{1000}{ns} channels is shown in Figs.~\ref{fig:simResTDL-C_wide}(a) and \ref{fig:simResTDL-C_wide}(b), respectively. As seen from Fig.~\ref{fig:simResTDL-C_wide}(a), the performance of all the unfiltered and filtered schemes are approximately the same in \ac{tdl}-C channel with \SI{300}{ns} delay spread. However, for channel model with maximum delay spread exceeding the \ac{cp} duration as illustrated in Fig.~\ref{fig:simResTDL-C_wide}(b), the pulse shaping provided by the filtering gives slightly improved performance over the plain \ac{cp-ofdm} waveform. 

\subsection{Implementation Complexity} 
\label{sec:impl-compl-1}
\begin{figure*}[!ht]
  \centering
  \includegraphics[width=0.49\textwidth]{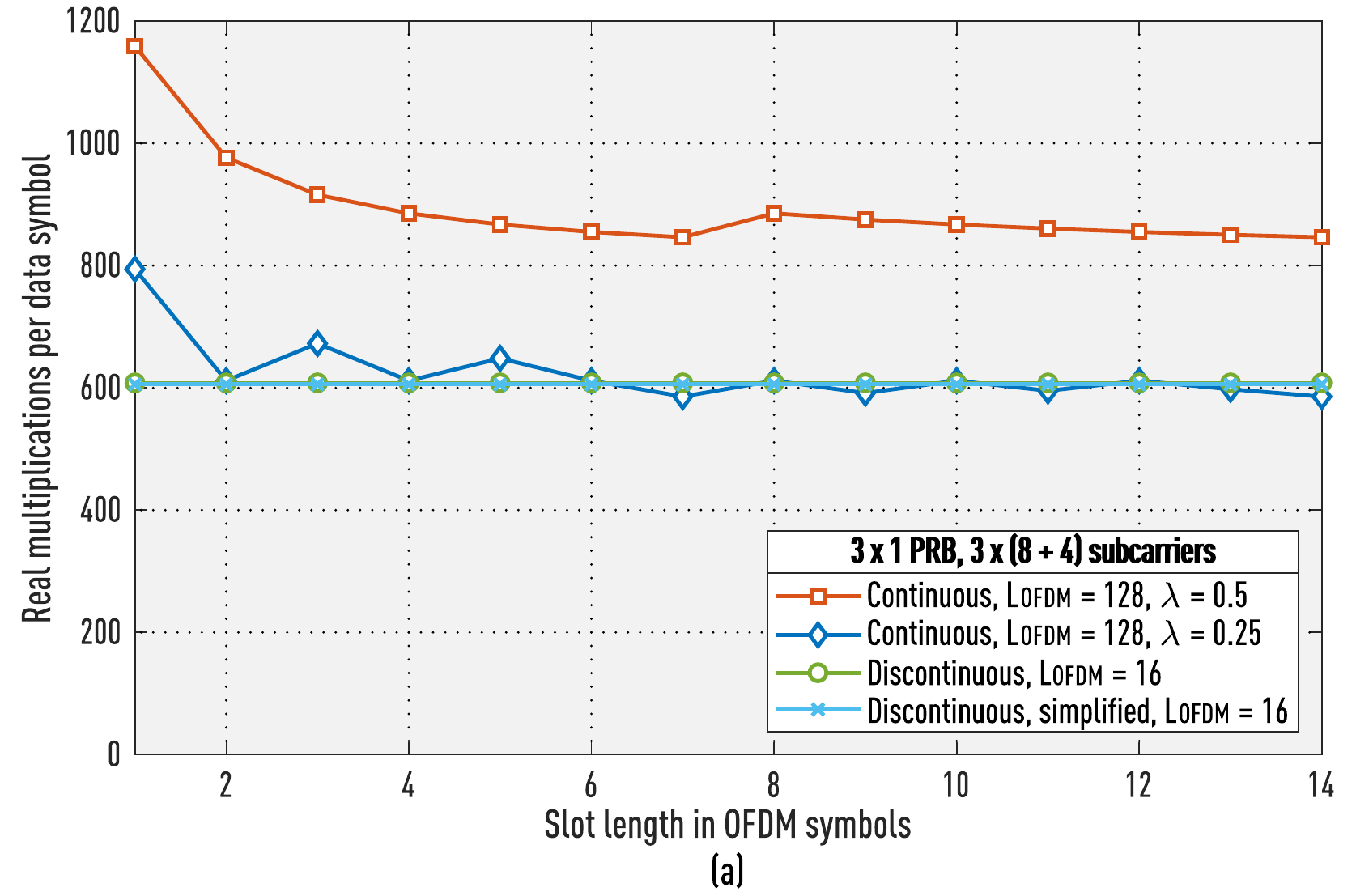}
  \includegraphics[width=0.49\textwidth]{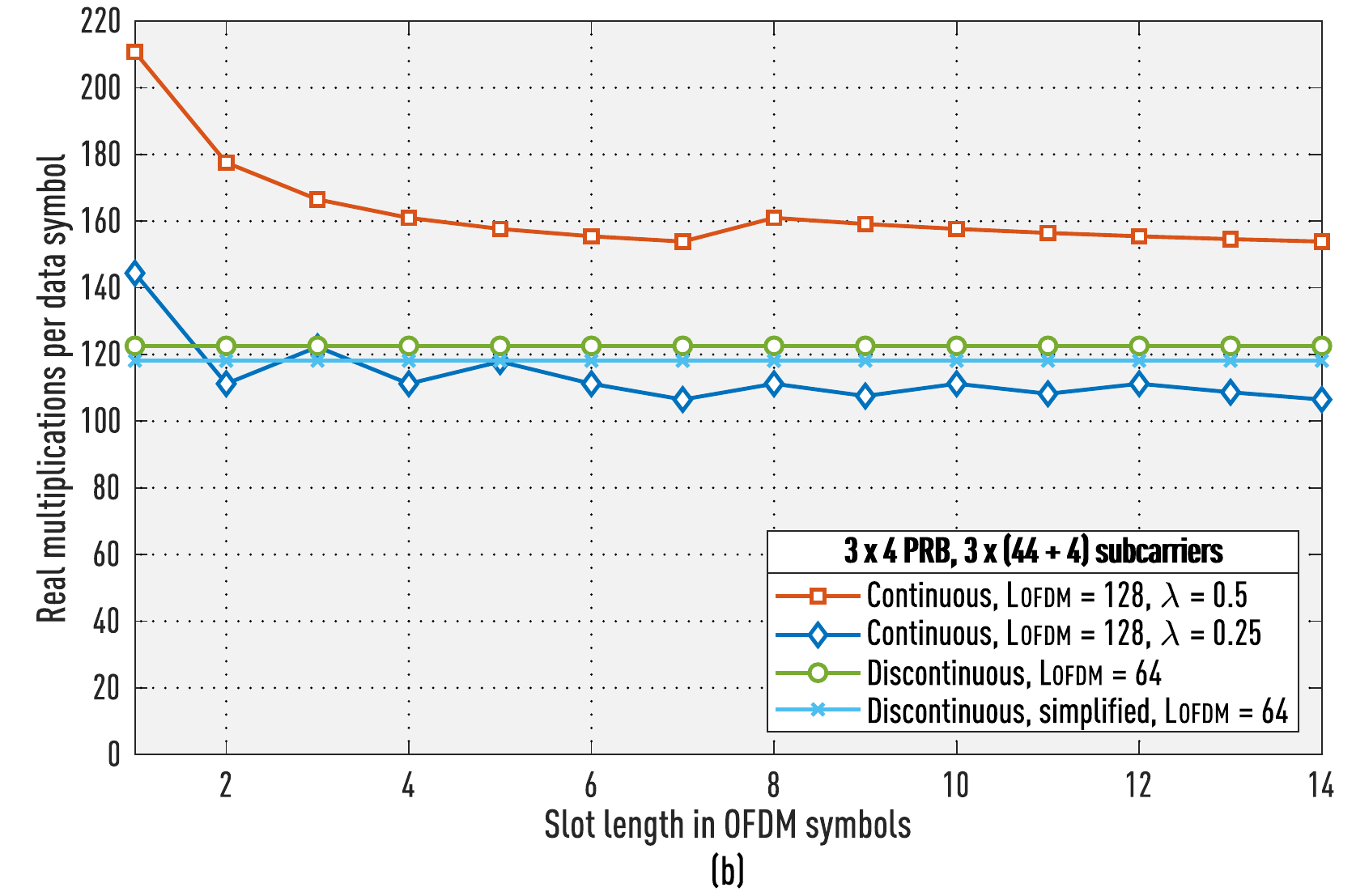}
  \caption{Computational complexity of continuous FC-F-OFDM RX processing with \SI{25}{\%} and \SI{50}{\%} overlaps and discontinuous FC-F-OFDM RX processing with \SI{50}{\%} overlap.
    (a) Three 1-PRB wide subbands.
    (b) Three 4-PRB wide subband. } 
  \label{fig:Complexity1}      
\end{figure*}

\begin{figure*}[!ht]            
  \centering   
  \includegraphics[width=0.49\textwidth]{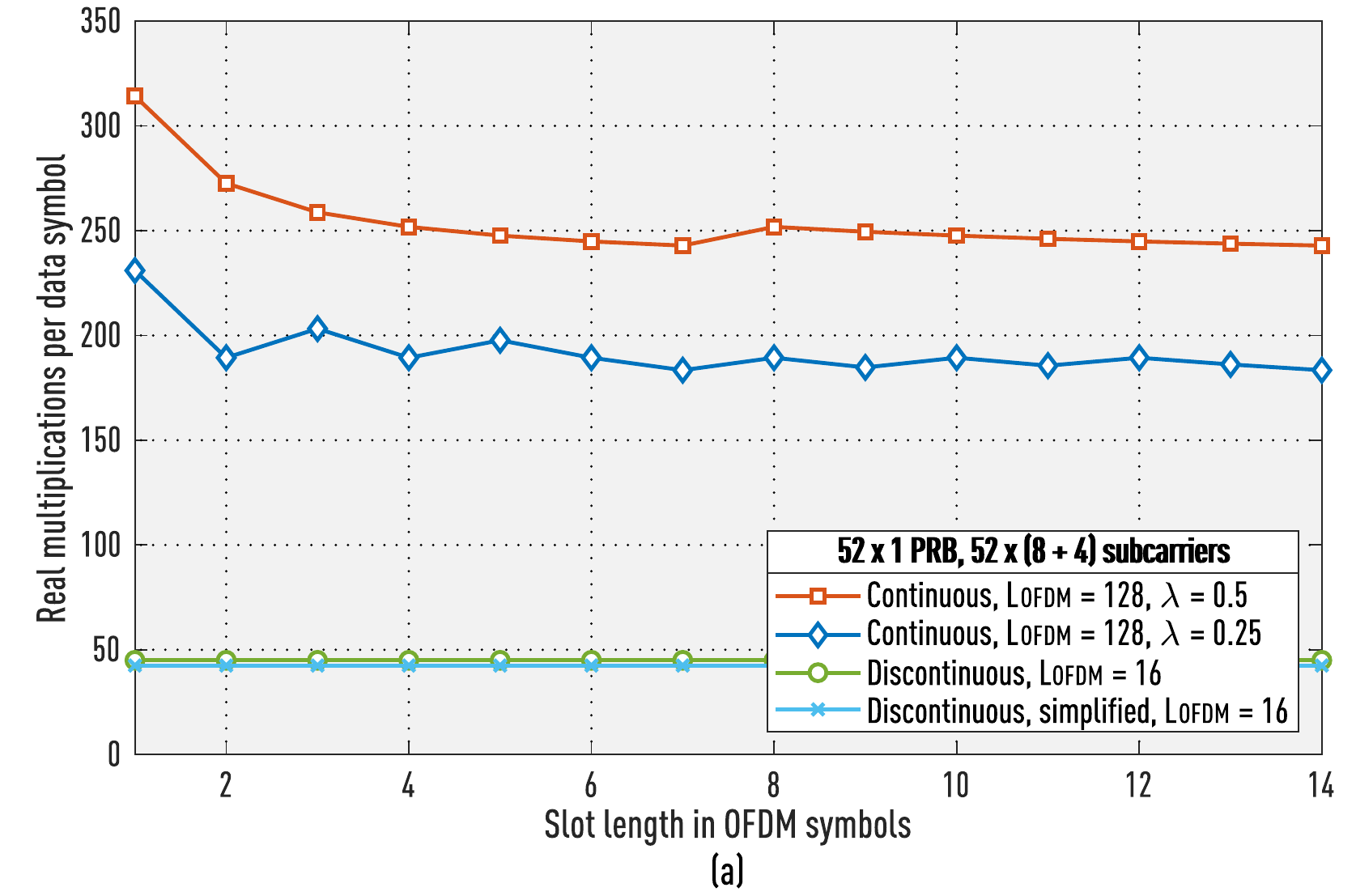}
  \includegraphics[width=0.49\textwidth]{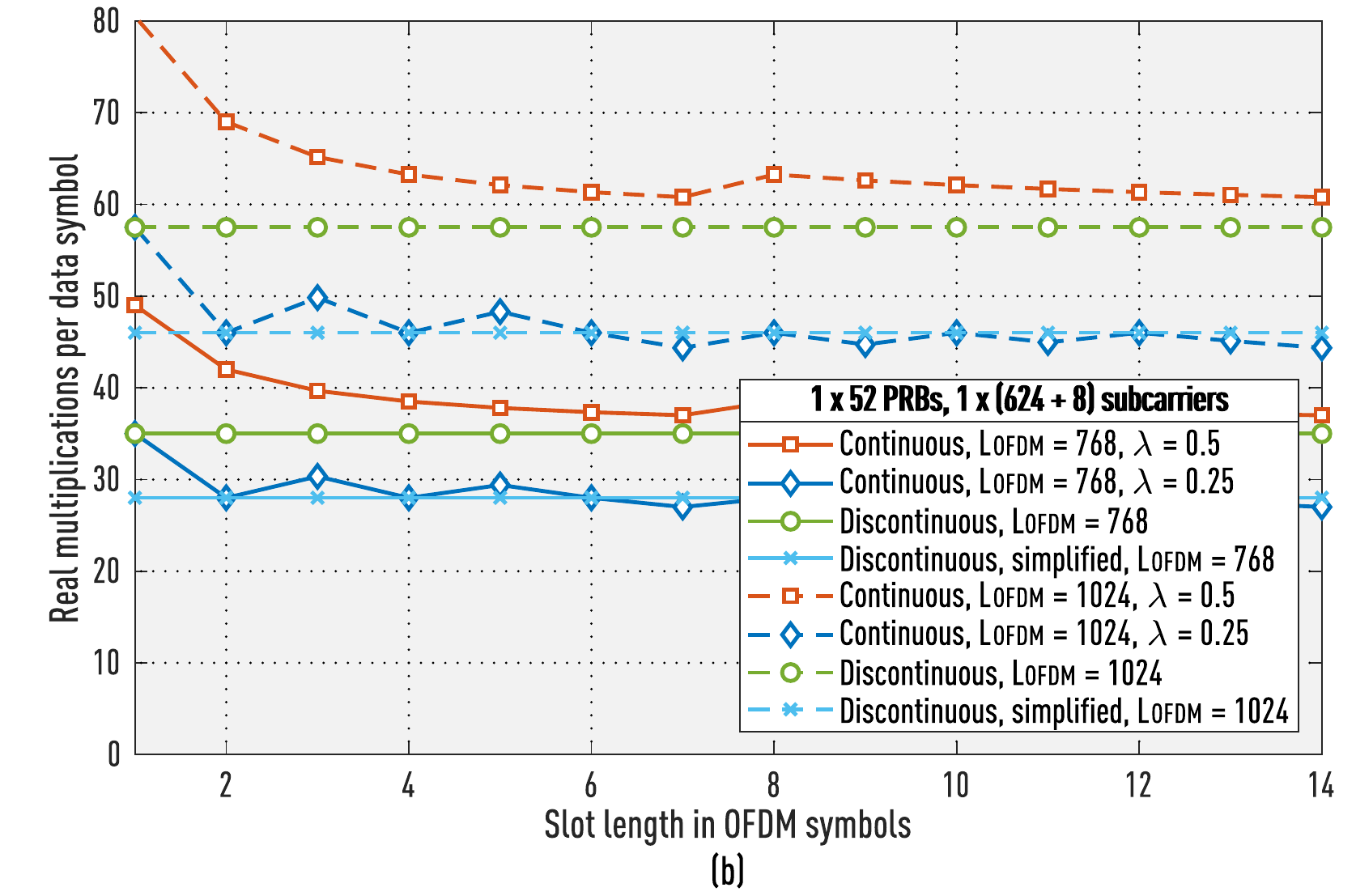}
  \caption{Computational complexity of continuous FC-F-OFDM RX processing with \SI{25}{\%} and \SI{50}{\%} overlaps and discontinuous FC-F-OFDM RX processing with \SI{50}{\%} overlap.
    (a) 52 1-PRB wide subbands.
    (b) Single 52-PRB wide subband.}
  \label{fig:Complexity2}      
\end{figure*}

Figs.~\ref{fig:Complexity1} and \ref{fig:Complexity2} compare the computational complexity of different \ac{fc}-based filtered \ac{ofdm} schemes for different subband configurations, again in the \SI{10}{MHz} \ac{5g-nr} or \ac{lte} case. The overlap factors of $\lambda=0.25$ and $\lambda=0.5$ are used for continuous processing, and $\lambda=0.5$ for discontinuous processing. The complexity is plotted as a function of the slot (\ac{tx} burst) length, in the range from 1 to 14 \ac{ofdm} symbols, using the number of real multiplications per transmitted \ac{qam} symbol as the complexity metric.

The short \ac{fc}-transform length $L_m$ is equal to the  \ac{ifft} length in \ac{ofdm} generation and it is selected as the smallest feasible power-of-two value. Notably, the smallest value of $L_m$ in continuous processing is \num{128}, while in discontinuous processing we can use $L_m=16$ for single PRB ($L_{\text{act},m}=12$ subcarriers) allocation and $L_m=64$ for four-PRB ($L_{\text{act},m}=48$ subcarriers) allocation. In addition to the 1-\ac{prb} and 4-\ac{prb} subband cases, also the fullband allocation with 52 \acp{prb} and $L_{\text{act},0}=624$ active subcarriers is included in the comparison. 

From Figs.~\ref{fig:Complexity1} and \ref{fig:Complexity2}, we can observe the following:
\begin{enumerate}
\item Discontinuous processing for all parameterizations and slot lengths provides lower complexity than continuous processing with \SI{50}{\%} overlap.  Discontinuous processing also provides constant complexity over different slot lengths.  With short slot lengths, the complexity of discontinuous schemes is significantly lower than that of continuous transmission with \SI{50}{\%} overlap.
\item With multiple relatively narrow subbands, this benefit is pronounced and significant also for higher slot lengths.  In these cases, the complexity of discontinuous processing is lower or similar to that of the continuous processing with \SI{25}{\%} overlap.
\item We remind that with \SI{50}{\%} overlap, the imperfections of \ac{fc} processing can be ignored, while the use of \SI{25}{\%} overlap degrades the performance with high \acp{mcs}.
\item The use of non-power-of-two short transform length ($L$) in building the \ac{cp-ofdm} symbols and \ac{fc} processing blocks may give significant complexity reduction in both continuous and discontinuous \ac{fc} processing, especially in the channel filter use case. This is shown for the full-band transmission case with $L=768$ instead of $L=1024$ (channel filtering example). This transform length can be efficiently implemented by three \acp{fft} of length \num{256} and some additional twiddle factors.
\item Discontinuous processing  allows to generate a single-{\prb} subband signal (e.g. 
for \ac{nb-iot})  by using  the \ac{fft} size of 16 for the \ac{ofdm} symbol generation and \ac{fc} processing at the sampling rate  of \SI{240}{kHz}, while prior art implementation require sampling rate of  \SI{1.92}{MHz}  with the \ac{fft} size of 128. 
\end{enumerate}       

The complexity evaluations for time-domain filtered \ac{ofdm} and \ac{wola} schemes and their relative complexity with respect to continuous \ac{fc} processing are given in \cite{J:Yli-Kaakinen:JSAC2017,J:Yli-Kaakinen:TSP2018}. 

\section{Conclusions} 
\label{sec:conclusions}
In this article, discontinuous symbol-synchronized \acf{fc} processing technique was proposed, with particular emphasis on the physical layer processing in 5G-NR and beyond mobile radio networks. The proposed processing approach was shown to offer various benefits over the basic continuous \ac{fc} scheme, specifically in terms of reduced complexity and latency as well as increased parametrization flexibility. The additional inband distortion effects, stemming from the proposed scheme, were found to have only a very minor impact on the link-level performance. The benefits are particularly important in specific application scenarios, like transmission of single or multiple narrow subbands, or in mini-slot type transmission, which is a core element in the ultra-reliable low-latency transmission service of \acf{5g-nr}.

Generally, discontinuous \ac{fc} processing can be regarded as an additional useful element in the toolbox for frequency-domain waveform processing algorithms, and it can be useful for various other signal processing applications as well. 

%
\begin{acronym}[FBMC/OQAM2]
\acro{3gpp}[3GPP]{third generation partnership project}
\acro{4g}[4G]{fourth generation} 
\acro{5g-nr}[5G-NR]{fifth-generation new radio}
\acro{5g}[5G]{fifth-generation}
\acro{F-ofdm}[F-OFDM]{subband filtered CP-OFDM}
\acro{aclr}[ACLR]{adjacent channel leakage ratio}
\acro{adsl}[ADSL]{asymmetric digital subscriber line}
\acro{afb}[AFB]{analysis filter bank} 
\acro{af}[AF]{amplify-and-forward} 
\acro{am/am}[AM/AM]{amplitude modulation/amplitude modulation \acroextra{[NL PA models]}}
\acro{am/pm}[AM/PM]{amplitude modulation/phase modulation \acroextra{[NL PA models]}} 
\acro{amr}[AMR]{adaptive multi-rate}
\acro{app}[APP]{a posteriori probability} 
\acro{ap}[AP]{access point} 
\acro{awgn}[AWGN]{additive white Gaussian noise}
\acro{b-pmr}[B-PMR]{broadband PMR} 
\acro{bcjr}[BCJR]{Bahl-Cocke-Jelinek-Raviv \acroextra{algorithm}}
\acro{ber}[BER]{bit error rate} 
\acro{bicm}[BICM]{bit-interleaved coded modulation} 
\acro{blast}[BLAST]{Bell Labs layered space time \acroextra{[code]}} 
\acro{bler}[BLER]{block error rate}
\acro{bpsk}[BPSK]{binary phase-shift keying} 
\acro{br}[BR]{bin resolution}
\acro{bs}[BS]{bin spacing}
\acro{cazac}[CAZAC]{constant amplitude zero auto-correlation}
\acro{cb-fmt}[CB-FMT]{cyclic block-filtered multitone}
\acro{ccc}[CCC]{common control channel}
\acro{ccdf}[CCDF]{complementary cumulative distribution function}
\acro{cdf}[CDF]{cumulative distribution function}
\acro{cdma}[CDMA]{code-division multiple access}
\acro{cfo}[CFO]{carrier frequency offset} 
\acro{cfr}[CFR]{channel frequency response} 
\acro{ch}[CH]{cluster head}
\acro{cir}[CIR]{channel impulse response} 
\acro{cma}[CMA]{constant modulus algorithm} 
\acro{cna}[CNA]{constant norm algorithm}
\acro{cp-ofdm}[CP-OFDM]{cyclic prefix orthogonal frequency-division multiplexing}
\acro{cpu}[CPU]{central processing unit} 
\acro{cp}[CP]{cyclic prefix} 
\acro{cqi}[CQI]{channel quality indicator} 
\acro{crlb}[CRLB]{Cram\'er-Rao lower bound}
\acro{crn}[CRN]{cognitive radio network}
\acro{crs}[CRS]{cell-specific reference signal}
\acro{cr}[CR]{cognitive radio} 
\acro{csir}[CSIR]{channel state information at the receiver}
\acro{csit}[CSIT]{channel state information at the transmitter}
\acro{csi}[CSI]{channel state information}
\acro{d2d}[D2D]{device-to-device} 
\acro{dc}[DC]{direct current}
\acro{dfe}[DFE]{decision feedback equalizer} 
\acro{dft-s-ofdm}[DFT-s-OFDM]{DFT-spread-OFDM}
\acro{dft}[DFT]{discrete Fourier transform} 
\acro{df}[DF]{decode-and-forward} 
\acro{dl}[DL]{downlink} 
\acro{dmo}[DMO]{direct mode operation}
\acro{dmrs}[DMRS]{demodulation reference signals}
\acro{dsa}[DSA]{dynamic spectrum access} 
\acro{dzt}[DZT]{discrete Zak transform} 
\acro{e-utra}[E-UTRA]{evolved UMTS terrestrial radio access}
\acro{ed}[ED]{energy detector}
\acro{egf}[EGF]{extended Gaussian function}
\acro{embb}[eMBB]{enhanced mobile broadband}
\acro{emse}[EMSE]{excess mean square error}
\acro{em}[EM]{expectation maximization} 
\acro{epa}[EPA]{extended pedestrian-A \acroextra{[channel model]}}
\acro{etsi}[ETSI]{European Telecommunications Standards Institute}
\acro{eva}[EVA]{extended vehicular-A \acroextra{[channel model]}}
\acro{evm}[EVM]{error vector magnitude}
\acro{f-ofdm}[f-OFDM]{filtered OFDM}
\acro{fb-sc}[FB-SC]{filterbank single-carrier} 
\acro{fbmc-coqam}[FBMC-COQAM]{filterbank multicarrier with circular offset-QAM}
\acro{fbmc/oqam}[FBMC/OQAM]{filter bank multicarrier with offset-QAM subcarrier modulation} 
\acro{fbmc}[FBMC]{filter bank multicarrier}
\acro{fb}[FB]{filter bank}
\acro{fc-f-ofdm}[FC-F-OFDM]{FC-based F-OFDM}
\acro{fc-fb}[FC-FB]{fast-convolution filter bank}
\acro{fc}[FC]{fast-convolution}
\acro{fdma}[FDMA]{frequency-division multiple access}
\acro{fd}[FD]{frequency-domain}
\acro{fec}[FEC]{forward error correction} 
\acro{fft}[FFT]{fast Fourier transform} 
\acro{fir}[FIR]{finite impulse response}
\acro{flo}[FLO]{frequency-limited orthogonal}
\acro{fmt}[FMT]{filtered multitone} 
\acro{fpga}[FPGA]{field programmable gate array} 
\acro{fs-fbmc}[FS-FBMC]{frequency sampled FBMC-OQAM} 
\acro{ft}[FT]{Fourier transform}
\acro{gb}[GB]{guard band}
\acro{gfdm}[GFDM]{generalized frequency-division multiplexing}
\acro{glrt}[GLRT]{generalized likelihood ratio test}
\acro{gmsk}[GMSK]{Gaussian minimum-shift keying}
\acro{hpa}[HPA]{high power ampliﬁer}
\acro{i/q}[I/Q]{in-phase/quadrature \acroextra{[complex data signal components]}} 
\acro{iam}[IAM]{interference approximation method}
\acro{ibe}[IBE]{in-band emission}
\acro{ibi}[IBI]{in-band interference}
\acro{ibo}[IBO]{input back-off}
\acro{ici}[ICI]{inter-carrier interference}
\acro{idft}[IDFT]{inverse discrete Fourier transform}
\acro{ifft}[IFFT]{inverse fast Fourier transform}
\acro{iid}[i.i.d.]{independent and identically distributed}
\acro{iota}[IOTA]{isotropic orthogonal transform algorithm}
\acro{iot}[IoT]{internet-of-things}
\acro{isi}[ISI]{inter-symbol interference}
\acro{itu-r}[ITU-R]{International Telecommunication Union Radiocommunication \acroextra{sector}}
\acro{itu}[ITU]{International Telecommunication Union}
\acro{kkt}[KKT]{Karush-K\"uhn-Tucker} 
\acro{kpi}[KPI]{key performance indicator} 
\acro{le}[LE]{linear equalizer}
\acro{llr}[LLR]{log-likelihood ratio} 
\acro{lmmse}[LMMSE]{linear minimum mean squared error} 
\acro{lms}[LMS]{least mean squares}
\acro{lpsv}[LPSV]{linear periodically shift variant} 
\acro{lptv}[LPTV]{linear periodically time-varying} 
\acro{ls}[LS]{least squares} 
\acro{lte-a}[LTE-A]{long-term evolution-advanced}
\acro{lte}[LTE]{long-term evolution}
\acro{lut}[LUT]{look-up table} 
\acro{mac}[MAC]{medium access control}
\acro{mai}[MAI]{multiple access interference}
\acro{map}[MAP]{maximum a posteriori} 
\acro{ma}[MA]{multiple access relay channel} 
\acro{mcm}[MCM]{multicarrier modulation}
\acro{mcs}[MCS]{modulation and coding scheme}
\acro{mc}[MC]{multicarrier}
\acro{mer}[MER]{message error rate} 
\acro{mf}[MF]{matched filter}
\acro{mgf}[MGF]{moment generating function}
\acro{mimo}[MIMO]{multiple-input multiple-output}
\acro{miso}[MISO]{multiple-input single-output}
\acro{mlse}[MLSE]{maximum likelihood sequence estimation} 
\acro{ml}[ML]{maximum likelihood} 
\acro{mma}[MMA]{multi-modulus algorithm} 
\acro{mmse}[MMSE]{minimum mean-squared error}
\acro{mmtc}[MTC]{massive machine-type communications}
\acro{mos}[MOS]{mean opinion score} 
\acro{mrc}[MRC]{maximum ratio combining} 
\acro{mse}[MSE]{mean-squared error} 
\acro{msk}[MSK]{minimum-shift keying} 
\acro{ms}[MS]{mobile station}
\acro{mtc}[MTC]{machine-type communications}
\acro{mud}[MUD]{multiuser detection} 
\acro{mui}[MUI]{multiuser interference}
\acro{music}[MUSIC]{multiple signal classification}
\acro{mu}[MU]{multi-user} 
\acro{nb-iot}[NB IoT]{narrow-band internet-of-things}
\acro{nbi}[NBI]{narrowband interference}
\acro{nb}[NB]{narrow-band}
\acro{nc-ofdm}[NC-OFDM]{non-contiguous OFDM}
\acro{nc}[NC]{non-contiguous}
\acro{nl}[NL]{non-linear} 
\acro{nmse}[NMSE]{normalized mean-squared error} 
\acro{npr}[NPR]{near perfect reconstruction}
\acro{nr}[NR]{new radio}
\acro{obo}[OBO]{output back-off} 
\acro{ofdm}[OFDM]{orthogonal frequency-division multiplexing} \acro{ofdma}[OFDMA]{orthogonal frequency-division multiple access} 
\acro{ofdp}[OFDP]{orthogonal finite duration pulse}
\acro{ola}[OLA]{overlap-and-add} 
\acro{ols}[OLS]{overlap-and-save}
\acro{omp}[OMP]{orthogonal matching pursuit}
\acro{oobem}[OOBEM]{out-of-band emission mask}
\acro{oob}[OOB]{out-of-band}
\acro{oqam}[OQAM]{offset quadrature amplitude modulation}
\acro{oqpsk}[OQPSK]{offset quadrature phase-shift keying}
\acro{osa}[OSA]{opportunistic spectrum access} 
\acro{pam}[PAM]{pulse amplitude modulation}
\acro{papr}[PAPR]{peak-to-average power ratio}
\acro{pa}[PA]{power amplifier} 
\acro{pci}[PCI]{perfect channel information}
\acro{per}[PER]{packet error rate} 
\acro{pf}[PF]{proportional fair} 
\acro{phy}[PHY]{physical layer}  
\acro{plc}[PLC]{power line communications}
\acro{pmr}[PMR]{professional (or private) mobile radio} 
\acro{ppdr}[PPDR]{public protection and disaster relief}
\acro{prb}[PRB]{physical resource block}
\acro{prose}[ProSe]{proximity services} 
\acro{pr}[PR]{perfect reconstruction}
\acro{psd}[PSD]{power spectral density} 
\acro{psk}[PSK]{phase-shift keying}
\acro{pswf}[PSWF]{prolate spheroidal wave function}
\acro{pts}[PTS]{partial transmit sequence}
\acro{ptt}[PTT]{push-to-talk} 
\acro{pucch}[PUCCH]{physical uplink control channel}
\acro{pusch}[PUSCH]{physical uplink shared channel}
\acro{pu}[PU]{primary user}
\acro{qam}[QAM]{quadrature amplitude modulation}
\acro{qoe}[QoE]{quality of experience} 
\acro{qos}[QoS]{quality of service}
\acro{qpsk}[QPSK]{quadrature phase-shift keying}
\acro{ram}[RAM]{random access memory} 
\acro{ran}[RAN]{radio access network}
\acro{rat}[RAT]{radio access technology} 
\acro{rbg}[RBG]{resource block group}
\acro{rb}[RB]{resource block}
\acro{rc}[RC]{raised cosine} 
\acro{rf}[RF]{radio frequency} 
\acro{rls}[RLS]{recursive least squares}  
\acro{rms}[RMS]{root-mean-squared \acroextra{[error]}}
\acro{roc}[ROC]{receiver operating characteristics}
\acro{rrc}[RRC]{square root raised cosine}
\acro{rrm}[RRM]{radio resource management}
\acro{rssi}[RSSI]{received signal strength indicator}
\acro{rx}[RX]{receiver}
\acro{sblr}[SBLR]{subband leakage ratio}
\acro{sc-fde}[SC-FDE]{single-carrier frequency-domain equalization} 
\acro{sc-fdma}[SC-FDMA]{single-carrier frequency-division multiple access} 
\acro{scs}[SCS]{subcarrier spacing}
\acro{subc}[SC]{subcarrier}
\acro{sc}[SC]{single-carrier}
\acro{sdma}[SDMA]{space-division multiple access}
\acro{sdm}[SDM]{space-division multiplexing} 
\acro{sdr}[SDR]{software defined radio}
\acro{sel}[SEL]{soft envelope limiter} 
\acro{ser}[SER]{symbol error rate} 
\acro{sfbc}[SFBC]{space frequency block code}
\acro{sfb}[SFB]{synthesis filter bank}
\acro{sic}[SIC]{successive interference cancellation}
\acro{simo}[SIMO]{single-input multiple-output}
\acro{sinr}[SINR]{signal-to-interference-plus-noise ratio}
\acro{sir}[SIR]{signal-to-interference ratio}
\acro{siso}[SISO]{single-input single-output}
\acro{slm}[SLM]{selected mapping}
\acro{slnr}[SLNR]{signal-to-leakage-plus-noise ratio}
\acro{sm}[SM]{spatial multiplexing}
\acro{sndr}[SNDR]{signal-to-noise-plus-distortion ratio}
\acro{snr}[SNR]{signal-to-noise ratio} 
\acro{softio}[SI-SO]{soft-input soft-output}
\acro{sqp}[SQP]{sequential quadratic programming}
\acro{sspa}[SSPA]{solid-state power amplifiers}
\acro{ss}[SS]{spectrum sensing} 
\acro{stbc}[STBC]{space time block code}
\acro{stbicm}[STBICM]{space-time bit-interleaved coded modulation}
\acro{stc}[STC]{space-time coding} 
\acro{sthp}[STHP]{spatial Tomlinson-Harashima precoder} 
\acro{su}[SU]{secondary users}
\acro{svd}[SVD]{singular value decomposition}
\acro{tbw}[TBW]{transition-band width}
\acro{tdd}[TDD]{time-division duplex}
\acro{tdl}[TDL]{tapped-delay line}
\acro{tdma}[TDMA]{time-division multiple access}
\acro{td}[TD]{time-domain}
\acro{teds}[TEDS]{TETRA enhanced data service}
\acro{tetra}[TETRA]{terrestrial trunked radio}
\acro{tfl}[TFL]{time-frequency localization} 
\acro{tgf}[TGF]{tight Gabor frame} 
\acro{tlo}[TLO]{time-limited orthogonal}
\acro{tmo}[TMO]{trunked mode operation} 
\acro{tmux}[TMUX]{transmultiplexer}
\acro{to}[TO]{tone offset}
\acro{tr}[TR]{tone reservation} 
\acro{tsg}[TSG]{technical specification group}
\acro{tx}[TX]{transmitter} 
\acro{ue}[UE]{user equipment} 
\acro{uf-ofdm}[UF-OFDM]{universal filtered OFDM}
\acro{ula}[ULA]{uniform linear array} 
\acro{ul}[UL]{uplink} 
\acro{urllc}[URLLC]{ultra-reliable low-latency communications}
\acro{v-blast}[V-BLAST]{vertical Bell Laboratories layered space-time  \acroextra{[code]}} 
\acro{veh-a}[Veh-A]{vehicular-A \acroextra{[channel model]}} 
\acro{veh-b}[Veh-B]{vehicular-B \acroextra{[channel model]}} 
\acro{wimax}[WiMAX]{worldwide interoperability for microwave access} 
\acro{wlan}[WLAN]{wireless local area network} 
\acro{wlf}[WLF]{widely linear filter}
\acro{wola}[WOLA]{weighted overlap-add} 
\acro{wola}[WOLA]{windowed overlap-and-add} 
\acro{zf}[ZF]{zero-forcing} 
\acro{zp}[ZP]{zero prefix}  
\acro{zt}[ZT]{Zak transform}
\end{acronym}

 

\bibliographystyle{IEEEtran}  
\bibliography{jour_short,conf_short,References} 

\ifhbonecolumn 
\else
\begin{IEEEbiography}[{\includegraphics[width=1in,height=1.25in,clip,keepaspectratio]{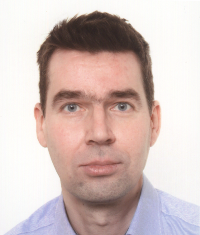}}]{Juha Yli-Kaakinen} received the degree of Diploma Engineer in electrical engineering and the Doctor of Technology (Hons.) degree from the Tampere University of Technology (TUT), Tampere, Finland, in 1998 and 2002, respectively.
  
Since 1995, he has held various research positions with TUT. His research interests are in digital signal processing, especially in digital filter and filter-bank optimization for communication systems and very large scale integration implementations.
\end{IEEEbiography}

\begin{IEEEbiography}[{\includegraphics[width=1in,height=1.25in,clip,keepaspectratio]%
{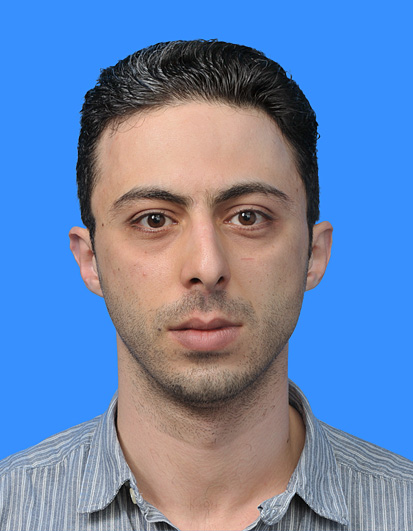}}]{AlaaEddin Loulou} received the bachelor's degree in electrical engineering from the Islamic University of Gaza, Gaza City, Palestine, in 2006, and the M.Sc. and D.Sc. degrees in communication engineering from the Tampere University, Finland, in 2013 and 2019. He has held various research positions with Tampere University.  Currently, he is working with Nokia Mobile Networks, Finland, as system on chip engineer.

His current research interests include enhanced orthogonal frequency-division multiplexing waveforms and advanced multicarrier schemes.
\end{IEEEbiography}

\begin{IEEEbiography}[{\includegraphics[width=1in,height=1.25in,clip,keepaspectratio]{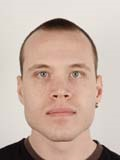}}]{Toni Levanen} received the M.Sc. and D.Sc. degrees from Tampere University of Technology (TUT), Finland, in 2007 and 2014, respectively. He is currently with the Department of Electrical Engineering, Tampere University. 

  In addition to his contributions in academic research, he has worked in industry on wide variety of development and research projects. His current research interests include physical layer design for 5G NR and beyond."
\end{IEEEbiography}

\begin{IEEEbiography}[{\includegraphics[width=1in,height=1.25in,clip,keepaspectratio]{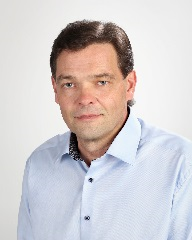}}]{Kari Pajukoski}%
received his B.S.E.E. degree from the Oulu University of Applied Sciences in 1992. He is a Fellow with the Nokia Bell Labs. He has a broad experience from cellular standardization, link and system simulation, and algorithm development for products. He has more than 100 issued US patents, from which more than 50 have been declared “standards essential patents”. He is author or co-author of more than 300 standards contributions and 30 publications, including conference proceedings, journal contributions, and book chapters. 
\end{IEEEbiography}

\begin{IEEEbiography}[{\includegraphics[width=1in,height=1.25in,clip,keepaspectratio]{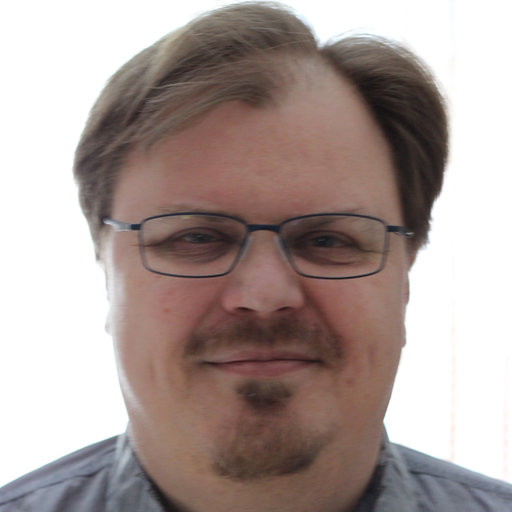}}]{Arto Palin} has long industrial experience in wireless technologies, covering cellular networks, broadcast systems and local area communications. He holds an MSc. (Tech.) degree from earlier Tampere University of Technology, and is currently working as Technical Leader at Nokia Mobile Networks, Finland, in the area of 5G SoC architectures.
\end{IEEEbiography}

\begin{IEEEbiography}[{\includegraphics[width=1in,height=1.25in,clip,keepaspectratio]{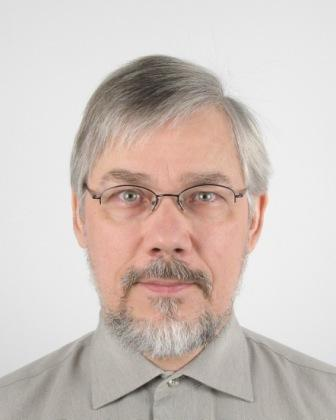}}]{Markku Renfors} (Life Fellow, IEEE) received the D.Tech. degree from the Tampere University of Technology (TUT), Tampere, Finland, in 1982.
Since 1992, he has been a Professor with the Department of Electronics and Communications Engineering, TUT, where he was the Head from 1992 to 2010. His research interests include filter-bank-based multicarrier systems and signal processing algorithms for flexible communications receivers and transmitters.

Dr.~Renfors was a corecipient of the Guillemin Cauer Award (together with T.~Saramäki) from the IEEE Circuits and Systems Society in 1987.
\end{IEEEbiography}

\begin{IEEEbiography}[{\includegraphics[width=1in,height=1.25in,clip,keepaspectratio]{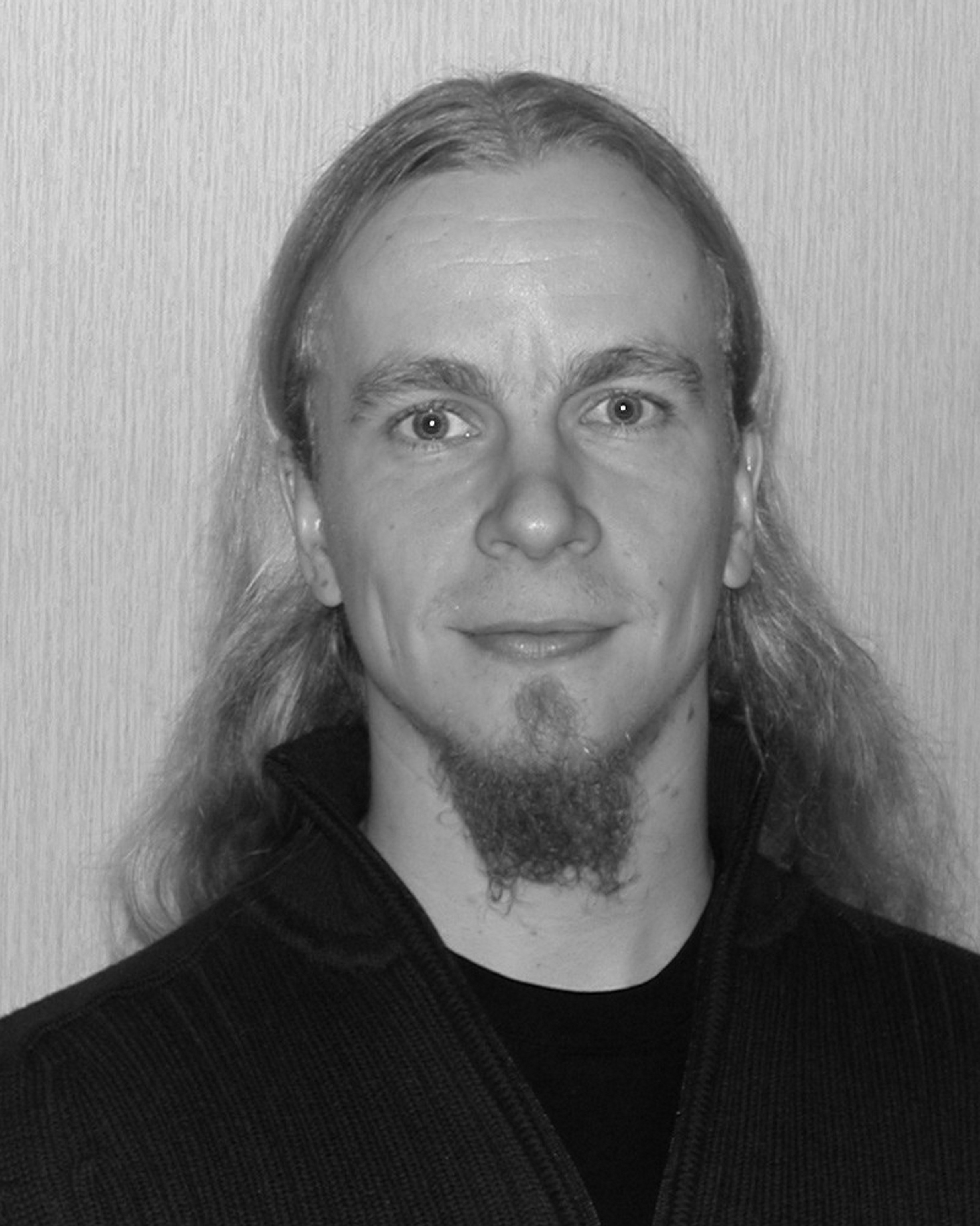}}]{Mikko Valkama} (Senior member, IEEE) received the D.Sc. (Tech.) degree (with honors) from Tampere University of Technology, Finland, in 2001. 

In 2003, he was with the Communications Systems and Signal Processing Institute at SDSU, San Diego, CA, as a visiting research fellow. Currently, he is a Full Professor and Department Head of Electrical Engineering at newly formed Tampere University (TAU), Finland. His general research interests include radio communications, radio localization, and radio-based sensing, with particular emphasis on 5G and beyond mobile radio networks.
\end{IEEEbiography}
\fi

\end{document}